\documentclass[utf8]{frontiersFPHY} 
\usepackage{url,hyperref,lineno,microtype,subcaption}
\usepackage[onehalfspacing]{setspace}
\usepackage{aas_macros}
\usepackage{graphicx}
\usepackage{color}
\newcommand{\la}{\lesssim}
\newcommand{\ga}{\gtrsim}

\newcommand{\sun}{\odot}
\newcommand{\alpacc}{$\alpha_\mathrm{acc}$}
\newcommand{\mdot}{$\dot{m}$}
\newcommand{\mseed}{$M_\mathrm{seed}$}

\newcommand{\msun}{M$_{\sun}$}
\newcommand{\rsun}{R$_{\sun}$}

\newcommand{\mj}{M$_\mathrm{J}$}

\newcommand{\tce}{$T_\mathrm{bce}$}
\newcommand{\teff}{$T_\mathrm{eff}$}
\newcommand{\tli}{$T(^7$Li$)$}
\newcommand{\ml}{$\alpha_\mathrm{ML}$}
\def\keyFont{\fontsize{8}{11}\helveticabold }
\def\firstAuthorLast{Tognelli {et~al.}} 
\def\Authors{E. Tognelli\,$^{1,2}$, S. Degl'Innocenti \,$^{1,3}$, P.G. Prada Moroni\,$^{1,3}$ L. Lamia$^{4,5,6}$, R.G. Pizzone$^{5}$, A. Tumino$^{5,7}$, C. Spitaleri$^{4,5}$ and A. Chiavassa$^{8}$}
\def\Address{$^{1}$Universit\`a di Pisa, Dipartimento di Fisica ``Enrico Fermi'', Largo Bruno Pontecorvo 3, Pisa, Italy\\
$^{2}$INAF, Osservatorio Astronomico d'Abruzzo, via Mentore Maggini, Teramo, Italy\\
$^{3}$INFN, Sezione di Pisa, Largo Bruno Pontecorvo 3, Pisa, Italy\\
$^{4}$Dipartimento di Fisica e Astronomia, Universit\'{a} di Catania, Catania, Italy\\  
$^{5}$INFN - Laboratori Nazionali del Sud, Catania, Italy\\ 
$^6$CSFNSM-Centro Siciliano di Fisica Nucleare e Struttura della Materia, Via Santa Sofia 64, Catania, Italy\\
$^{7}$Dipartimento di Ingegneria e Architettura, Universit\'{a} di Enna, Enna, Italy
$^{8}$Universit\'e C\^ote d'Azur, Observatoire de la C\^ote d'Azur, CNRS, Lagrange, CS 34229, Nice,  France
}
\def\corrAuthor{E. Tognelli, Universit\`a di Pisa, Dipartimento di Fisica ``Enrico Fermi'', Largo Pontecorvo 3, I-56127, Pisa, Italy}
\def\corrEmail{emanuele.tognelli@for.unipi.it}

\begin{document}
\onecolumn
\firstpage{1}

\title[Light element abundances in protostellar and pre-main sequence phase.]{Theoretical predictions of surface light element abundances in protostellar and pre-Main Sequence phase.}

\author[\firstAuthorLast ]{\Authors} 
\address{} 
\correspondance{} 

\extraAuth{}
\maketitle
\begin{abstract}
\section{}
Theoretical prediction of surface stellar abundances of light elements -- lithium, beryllium, and boron -- represents one of the most interesting open problems in astrophysics. As well known, several measurements of $^7$Li abundances in stellar atmospheres point out a disagreement between predictions and observations in different stellar evolutionary phases, rising doubts about the capability of present stellar models to precisely reproduce stellar envelope characteristics. The problem takes different aspects in the various evolutionary phases; the present analysis is restricted to protostellar and pre-Main Sequence phases. Light elements are burned at relatively low temperatures ($T$ from $\approx 2$ to $\approx 5$~million degrees) and thus in the early evolutionary stages of a star they are gradually destroyed at different depths of stellar interior mainly by (p, $\alpha$) burning reactions, in dependence on the stellar mass. Their surface abundances are strongly influenced by the nuclear cross sections, as well as by the extension toward the stellar interior of the convective envelope and by the temperature at its bottom, which depend on the characteristics of the star (mass and chemical composition) as well as on the energy transport in the convective stellar envelope. In recent years, a great effort has been made to improve the precision of light element burning cross sections. However, theoretical predictions surface light element abundance are challenging because they are also influenced by the uncertainties in the input physics adopted in the calculations as well as the efficiency of several standard and non-standard physical processes active in young stars (i.e. diffusion, radiative levitation, magnetic fields, rotation). Moreover, it is still not completely clear how much the previous protostellar evolution affects the pre-Main Sequence characteristics and thus the light element depletion. This paper presents the state-of-the-art of theoretical predictions for protostars and pre-Main Sequence stars and their light element surface abundances, discussing the role of (p, $\alpha$) nuclear reaction rates and other input physics on the stellar evolution and on the temporal evolution of the predicted surface abundances.
\tiny
\keyFont{ \section{Keywords:} Nuclear reactions, nucleosynthesis, abundances, Stars: pre-main sequence, Stars: evolution} 
\end{abstract}

\section{Introduction}
\label{sec:intro}
Light elements -- lithium, beryllium and boron (hereafter Li, Be and B) -- are burned at relatively low temperatures ($T$ from $\approx 2$ to $\approx 5$~million degrees) easy to reach in stellar interiors at the bottom of the convective envelope, even during the early pre-Main Sequence (pre-MS) evolution. Therefore, surface light elements are depleted if the mixing processes become efficient enough to bring them down to the destruction region. This property makes such elements very good tracers of the mixing efficiency in stellar envelopes whose theoretical treatment is still a difficult task in stellar physics. Due to the different burning temperatures,  the comparison between theory and observation for Li, Be and B, if possible, would be very useful to constrain theoretical models and in particular the extension of the convective envelope. The most of the observations concern the abundance of $^7$Li because, in most stars, surface $^6$Li is completely destroyed during the pre-MS phase and Be and B isotopic measurements are very problematic (e.g. \citet{cunha2010}, \citet{kaufer10}, \citet{delgado12}).

A huge amount of data for surface $^7$Li abundances are available both for disk, thick disk and halo field stars and for open clusters; however, the well known discrepancy between predictions and observations of this quantity in clusters or in the Sun (the so-called “lithium-problem”) is still an open question (see e.g.  \citet{xiong2002,piau02,sestito2003,deliyannis00,jeffries00,pinsonneault00,jeffries06, talon08a}). 

The theoretical prediction of surface light element abundances is complex because they are sensitive to both the input physics (i.e., equation of state, reaction rates, opacity, etc...) and chemical element abundances (i.e., initial abundance of deuterium, helium, metals, etc...)  adopted in stellar models, together with the assumed efficiency of microscopic diffusion and radiative acceleration (see e.g. \citet{piau02,burke04,richard2002,richard05,tognelli12,tognelli15b}). The situation is even more complicated because surface light element abundances seem to be affected by additional ``non standard'' physical processes, not routinely included in stellar evolutionary codes, as the possible presence of relevant magnetic fields and mass accretion processes in some young pre-MS stars (see e.g. \citet{baraffe10,macdonald2012,feiden13,somers14,somers2015}). Moreover, rotation-induced mixing, turbulent mixing, gravity waves and mass loss processes could play a role, though mainly for Main Sequence and more evolved stars, (see e.g. \citet{montalban2000,talon10,pace12,charbonnel13} and references therein).

The pre-MS is the first stellar phase where the star evolves as a fully formed object. To reach this evolutionary stage, the future star has to accrete mass, until its final value, in the previous "protostellar phase". The details of this phase, when matter of the protostellar cloud is still falling on the surface of the protostar, are complex and uncertain. The full understanding of how the protostellar accreting phase affects the predictions for pre-MS characteristics (and thus light element abundances) is still an open problem.
The inclusion of the protostellar accretion phase in evolutionary codes produces stars in the early pre-MS phase different from what expected in standard non accreting models, in which stars essentially contract along the Hayashi track. This eventually results in differences between standard and accreting models still visible during the whole pre-MS or the MS phase, with effects on both the structure and chemical composition of the stellar models (\citet{baraffe09,baraffe10,tognelli13b,kunitomo18,tognelli20}).

Light element burning cross sections are fundamental ingredients in the predictions of the time behaviour of light element stellar surface abundances. In recent years new values for (p,$\alpha$) reaction rates have been proposed, mainly estimated using the Trojan Horse Method, greatly improving the precision of these quantities.

The present review summarises the state-of-the-art of theoretical predictions for protostars and pre-MS stars and their light element surface abundances, in the light of recent improvements in the adopted input physics, updated reaction rates and description of the formation and evolution of pre-MS stars. 

The paper is structured as it follows. In Section 2 we qualitatively show the location of observed young pre-MS stars in the HR diagram and we compared it to the predictions of standard non accreting models. In Section 3 we give a brief overview of the main characteristics and evolutionary stages of a pre-MS model without protostellar accretion. In Section 4 we introduce the protostellar accretion phase, discussing the differences between spherical and disc accretion, along with the main parameters the determine the structure of an accreting protostar. In Section 5 we analyse the burning of light elements (Li, Be and B) in pre-MS stars and the predicted surface abundances during the pre-MS for stellar models of different masses without or with protostellar accretion phase. In section 6 we review the impact of updated cross sections for the burning of light elements and their impact in the predictions of surface abundances in pre-MS stellar models. In section 7 we summarize the main aspects highlighted in the review.

\section{Observational data of Pre-main sequence stars as test of theoretical models}
\label{PMSobservations}
This review is focused on theoretical predictions for protostellar/pre-MS models, which can be validated only through comparison with observational data. Given the difficulty in directly observing the stellar formation and the protostellar phase, only an investigation of the characteristics of very young pre-MS stars, close to the end of the protostellar phase, can indirectly give information on the previous accretion period. The availability of observations for very young pre-MS stars is thus fundamental. A great number of data is available for young pre-MS stars (ages $\sim 1$~Myr) with solar or slightly sub-solar metallicity; among these objects, some of them show a still detectable accretion disc, or protoplanetary disc, and very low accretion rates (see .g. \citet{hartmann96,muzerolle00,calvet05,muzerolle05,muzerolle05b,bae13,ingleby14}). Such residual accretion discs show clear footprints of a previous accretion phase. 

\begin{figure}[t]
\centering
\includegraphics[width=12cm]{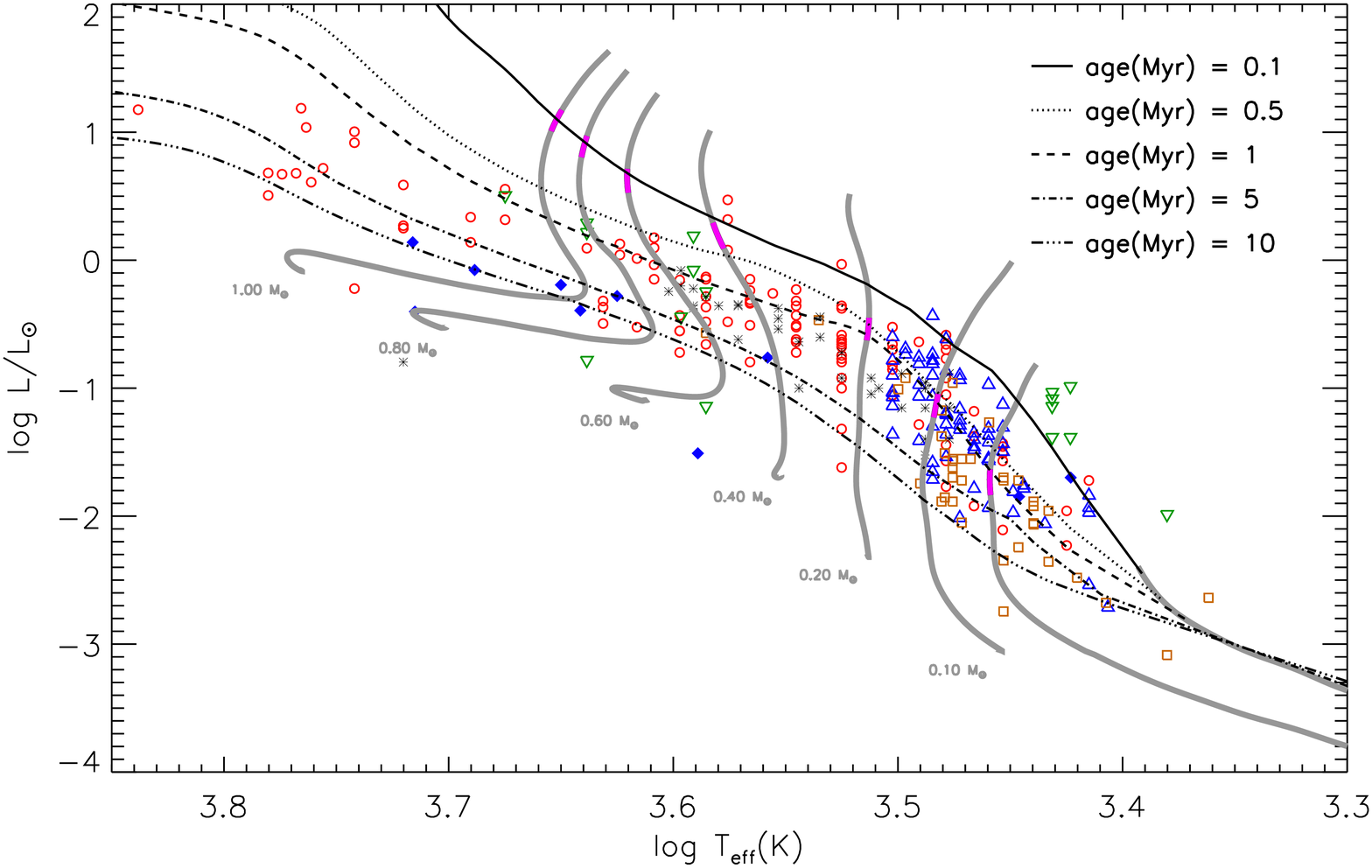}
\caption{HR diagram for young pre-MS stars extracted from the literature, compared with a set of standard evolutionary models and isochrones from the PISA database (\citet{tognelli11,tognelli18}.)}
\label{fig:dati}
\end{figure}
As an example, Fig.~\ref{fig:dati} shows a sample of young pre-MS stars compared to standard isochrones from 0.1 to 10~Myr and evolutionary tracks for masses in the range [0.01, 1.0]~\msun{} (\citet{tognelli11}). Such observed stars are fully formed, in the sense that the measured accretion rates are extremely small, thus, they have already reached their final mass. Thus they can be considered stars evolving as constant mass structures. This figure is intended to qualitatively show the position of observed young stars in a HR diagram; they are located in a region which theoretically corresponds to pre-MS models undergoing to a gravitational contraction (we will discuss this evolutionary stage in more details in Section~\ref{PMSevolution}). Standard stellar models generally agree with data for young stars in the colour-magnitude (CM) or in the HR diagram (see e.g. \citet{tognelli13,randich18}), as qualitatively shown in figure. We remark that stellar models should be able to populate such region of the HR diagram where young stars are observed. Thus, the simple comparison with observations of young associations/clusters (especially in the GAIA era) in the HR/CM diagram can put strong constraints on stellar evolution theoretical predictions (see e.g. \citet[][]{babusiaux18,randich18,bossini19}). This is a fundamental point  especially when accretion phases are taken into account (see Section~\ref{protostar}), helping in constraining free parameters adopted in model computations.

Other constraints are provided from pre-MS stars in double-lined eclipsing binary (EB) systems, whose masses and radii can be determined with high precision. In recent years, an increasing number of EB systems have been studied in detail, giving the possibility to check pre-MS model predictions against data (see e.g. \citet{mathieu07,gennaro12}).

Further constraints come from the measurements in low-mass stars of the surface abundance of lithium-7, which, being an element whose destruction rate is extremely sensitive to the temperature, can be used to test the temporal evolution of the pre-MS stellar structures. \citet{deliyannis00,charbonnel00,pinsonneault00,piau02,randich10,tognelli12}). 
These issues will be discussed in the present review. \\
\\
\\

\section{General characteristics of standard pre-main sequence evolution}
\label{PMSevolution}
\begin{figure}
    \centering
    \includegraphics[width=12cm]{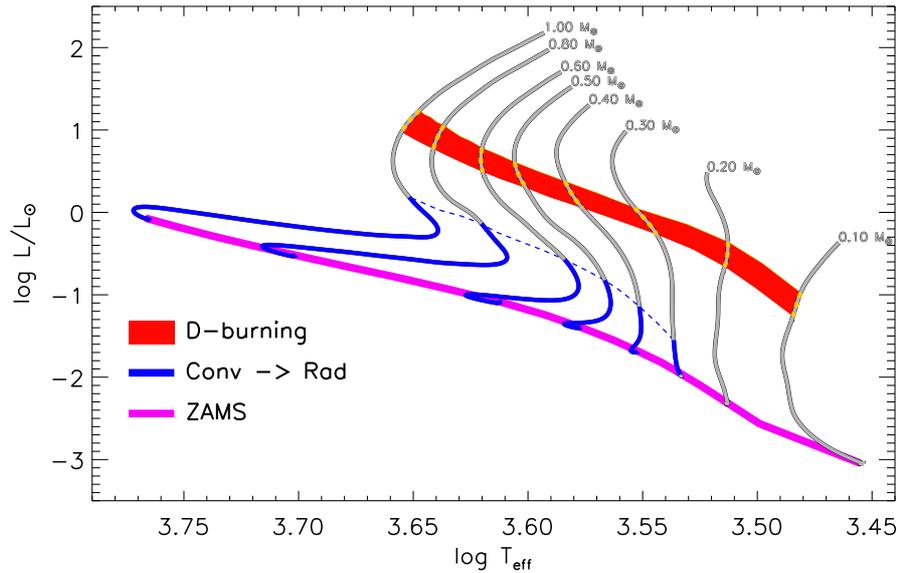}
    \caption{HR diagram for low-mass stars with indicated the main evolutionary stages during the pre-MS evolution: Hayashi track (fully convective star, grey line), partially convective star (blue line), locus of the Zero Age Main Sequence (ZAMS, magenta line). The locus corresponding to the deuterium burning is indicated by the red stripe.}
    \label{fig:std_pms}
\end{figure}
The pre-MS evolution starts at the end of the accretion phase and ends with the Zero Age Main Sequence, or simply ZAMS, position\footnote{The ZAMS corresponds to the phase when central hydrogen begins to be burned into helium with secondary burning elements at equilibrium and nuclear energy production fully supporting the star.}. The star is totally formed, and the mass can be considered constant at least for the whole pre-MS and Main Sequence (MS) evolution. The first consistent description of the pre-MS evolution was given by \citet{hayashi61} and \citet{hayashi63}; the basic idea is that a pre-MS star starts from a cold, expanse and luminous model. Due to the low temperatures, the opacity of the stellar matter is large and thus the radiative temperature gradient in the whole structure is larger than the adiabatic one. This leads to convective motions extended within the entire stellar structure; thus the star is fully mixed and chemically homogeneous. Figure~\ref{fig:std_pms} shows an example of the evolution of pre-MS solar metallicity low mass stars in the mass range [0.1, 1.0]~\msun{} computed using the PISA stellar evolutionary code (\citet{deglinnocenti08,dellomodarme12}), with the adopted input physics/parameters described in \citet{tognelli18,tognelli20}); the same figure also shows a qualitative representation of some of the main evolutionary stages characteristics of such mass range.

Due to their low temperatures, during these first stages of the standard pre-MS evolution, stars cannot produce the nuclear energy required to balance the surface energy losses by radiation and their evolution essentially consists in a gravitational contraction. The evolution time scale is thus given by the thermal (Kelvin-Helmholtz) time scale, which is the time of energy transport throughout the star. It is common to define  the Kelvin-Helmholtz time scale as the ratio between the total gravitational energy of the star and its luminosity $L$: 
\begin{equation}
\tau_{KH} = \frac{\beta}{2}\frac{GM^2}{R L}.
\end{equation}
The factor $\beta$ takes into account the density profile inside the star. For an (unrealistic) model of homogeneous and spherical star with a constant density, $\beta = 5/3$. 
The gravitational contraction leads to an internal temperature increase. We recall that for non-degenerate structures\footnote{When talking about \emph{degeneration} we refer to electron quantum degeneracy.} the central temperature, $T_c$, depends on the stellar mass $M$, the radius $R$ and the chemical composition (mean molecular weight $\mu$) in the following way:
\begin{equation}
T_c \propto \frac{\mu M}{R}.
\end{equation}
From the relation above a contraction naturally leads to a rise in $T_c$. Using this result, the Stephan-Boltzmann law ($L\propto R^2 $\teff$^4$) and the virial theorem, it can be shown that the luminosity of the star decreases following a simple power law, $L\propto t^{-2/3}$. 

The gravitational contraction is the only energy source until the central temperature reaches about $10^6$~K, when the deuterium burning reaction D(p,$\gamma$)$^3$He (D-burning) becomes efficient. Such a reaction generates the energy required to maintain the star stable on nuclear time scales, longer than the thermal one. This is guaranteed also by the steep dependence on the temperature of the energy generation rate, $\epsilon_{pD}$, ($\epsilon_{pD}\propto T_c^{12}$); such a dependence limits the $T_c$ increase, halting the gravitational contraction (because of the $T_c \propto 1/R$ relation). 

The ignition of the D-burning, due to the produced energy flux, maintains the star fully convective and deuterium is burnt in the whole star. The D-burning phase is shown in Fig.~\ref{fig:std_pms} as a the red stripe, which indicates the part of the Hayashi track where D-burning provides more than 10\% of the total stellar luminosity, for stars with different masses. 

The nuclear time scale of D-burning depends on the characteristics of the star, mainly on the mass. The luminosity of a star at the beginning D-burning phase increases with the stellar mass; this means that increasing the stellar mass the D-burning increases its efficiency to balance the higher energy losses at the stellar surface. Thus the rate of deuterium destruction increases with mass. The typical nuclear D-burning time scale for masses in the range 0.1 - 1~\msun{} varies between about $0.1$-2~Myr, depending on the mass; as an example the D-burning phase lasts about 1-2$\times 10^6$~yr for 0.1~\msun{} and about $10^5$~yr for a 1~\msun{} (see e.g. \citet{chabrier97,dantona97,tognelli11}).

The duration of the D-burning phase in pre-MS depends not only on the stellar mass but it is also proportional to the original stellar deuterium mass fraction abundance. Observations suggest that for disc stars a value of $X_D \approx 2\times 10^{-5}$ should be adopted (see e.g. the review by \citet{sembach10}); such a value is smaller than that predicted by the BBN ($X_D \approx 4 \times  10^{-5}$, see e.g. \citet{steigman07,pettini08,pitrou18,mossa2020}), as expected -- e.g. by galactic evolution models -- because deuterium is destroyed in stars. 

Once deuterium has been completely exhausted in the whole star a pure gravitational contraction phase starts again. As for the previous evolution the stellar luminosity is well approximated by the power law $L\propto t^{-2/3}$. This second gravitational contraction increases the temperature and density in the inner region of the star. Depending on the total mass, such a temperature increase could lead to a drop in the radiative opacity $\kappa_R$. For stars with $M\ga 0.3$~\msun, the internal opacity drop reduces the radiative gradient leading to the formation of a central radiative stable zone. As a consequence of this fact, the star leaves the Hayashi track in the HR diagram, shifting towards larger temperatures as the radiative core grows in mass, until the star efficiently ignites the central hydrogen burning (reaching the ZAMS). This part of the stellar evolution is traditionally called the \emph{Henyey track} and corresponds to the blue part of the evolutionary track in Fig.~\ref{fig:std_pms}. For $M< 0.3$~\msun, the temperature increase is not enough to produce such an opacity drop and the star continues its contraction along the Hayashi line. In this mass range, if the total mass is larger than approximately  $0.08$~\msun, the contraction continues until the central temperature is large enough to ignite central hydrogen burning, which becomes the main energy source of the star (see e.g. \citet{iben13}). On the other hand, if $M < 0.08$~\msun, during the contraction  the star become so dense that the pressure is dominated by the degenerate electron contribution; in such a configuration the pressure is only very slightly dependent on the temperature. Then the contraction slows down and the star (called brown dwarf) evolves along a cooling sequence which, in the HR diagram, follows a precise mass-radius relation.

This general picture describes the evolution of a pre-MS star in the standard case; theoretical calculations are started when the star is a fully formed object, chemically homogeneous at high luminosity (large radius) on the Hayashi line. However, it is well known that stars undergo a formation phase, the \emph{protostellar phase}, during which the mass is accreted from the protostellar cloud and/or from a disk to reach the final stellar mass. The inclusion of such a phase could, at least in principle, modify the standard theoretical picture. \\

\section{Protostellar accretion phase}
\label{protostar}

The stellar formation process starts with the collapse and the fragmentation of a molecular cloud that contracts forming denser cores which eventually become protostars and then stars. During this process, the protostellar mass progressively increases as the matter in the cloud falls onto the central dense object. The cloud collapse is a complex hydrodynamic problem, in which one has also to take into account cooling processes by molecules and dust. At a given time during the collapse a stable hydrostatic core forms, on which mass continues to fall, so that  the accretion treatment does not require anymore hydrodinamical models (e.g. \citet{stahler80,stahler801,stahler81,hartmann97,baraffe12}). 

Protostellar accretion has been analysed in the literature starting from the pioneering works by \citet{stahler80}, \citet{stahler88}, \citet{palla91}, \citet{palla92}, \citet{palla93}, \citet{hartmann97} and \citet{siess97}, to more recent works by \citet{baraffe09}, \citet{hosokawa09}, \citet{baraffe10}, \citet{tognelli13},   \citet{kunitomo17} and \citet{tognelli20}. Depending on the characteristics of the accretion assumed in the computations (chemical composition, magnetic fields, rotation, geometry...) the collapse of the cloud and the stellar formation can produce different evolution whose footprints are still visible in pre-MS stars. 
\\ 

\subsection{Cloud collapse and protostellar accretion}
\label{cloudcollapse}

The main phases of the protostellar evolution are briefly described below (for more details see \citet{larson69,larson72,larson03}). \\

\begin{itemize}
\item \textit{Isothermal collapse:} The protostellar cloud, during its first collapse (until the central density is lower than about $10^{-13}$~g~cm$^{-3}$) does not warm, because its density is too low to trap the energy produced by the contraction. When the density further increases above this limit the radiation can be partially trapped. 

\item \textit{Formation of the first Larson core:} the energy trapped inside the denser regions of the cloud prevent a further collapse of this region. A first temporarily hydrostatic core forms (with a mass of about 0.01~\msun{} and a radius of several AU) out of which the matter is still falling on the core. A transition region (shock front) develops close to the core surface where the matter settles and passes from supersonic to subsonic.

\item \textit{Second collapse:} the hydrostatic core contracts  as long as it radiates energy from its surface. So, although its mass is increasing due to mass accretion, its radius shrinks. The contraction of the core leads to a temperature rise, until the temperature of molecular hydrogen dissociation ($T\sim 2000~$K) is reached. Then contraction energy does not warm anymore the core but it is used to dissociate $H_2$, forcing the core to break the condition of hydrostatic equilibrium. At this stage, the core density and pressure increase.

\item \textit{Formation of the 2nd Larson core:} when $H_2$ is fully dissociated a further increase of density and pressure, due to contraction, while mass is still falling radially on the core, leads to a second, hydrostatic equilibrium for the central core with a mass of the order of $\sim 0.001$~\msun{} ($=1$~\mj, Jupiter mass) and a radius of about $1$~\rsun. From this moment on the central objects maintains its hydrostatic configuration while its mass increases.
\end{itemize}

The protostellar evolutionary phases listed above are quite general (for solar metallicity stars) and almost independent of the computation details. \citet{larson69} remarked that at some stages of the cloud evolution a hydrostatic central object (2nd Larson core) forms that can be considered the first protostellar core. The characteristics of this core (i.e. mass, radius, density and central temperature) appear to be barely sensitive to the adopted cloud initial conditions or to the adopted input physics (see e.g. \citet{masunaga00,machida2008,tomida13,vaytet2013}). Reasonable intervals for the mass, radius and temperature of the stable hydrostatic core are: mass range of 1~-~20~\mj, radius values of 0.5~-~10~\rsun{} and central temperature of 2-6$\times 10^{4}$~K. 
\begin{figure}[t]
\centering
\includegraphics[width=10cm]{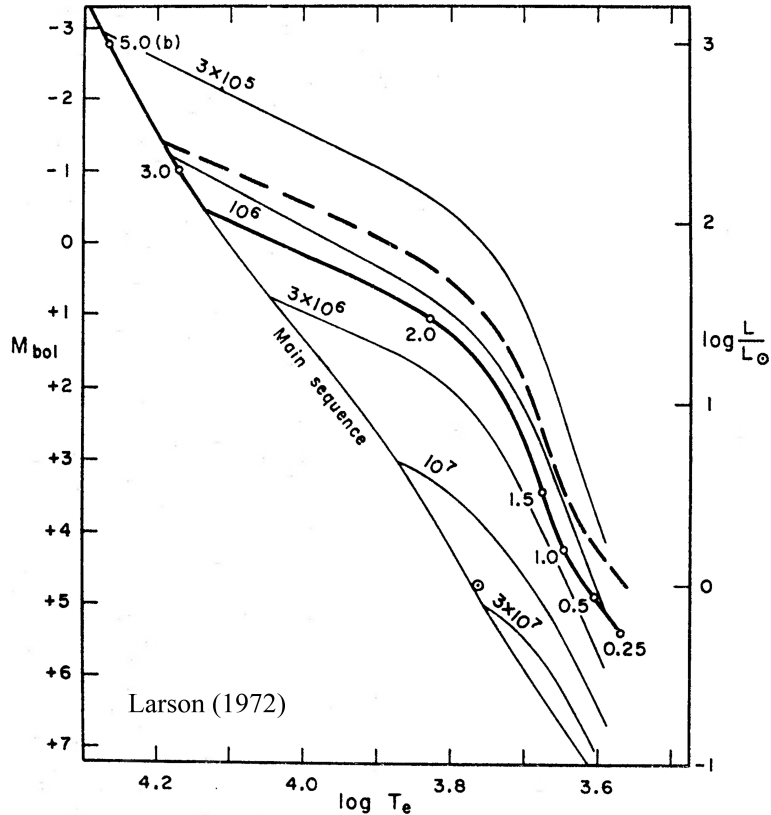}
\caption{Comparison between standard isochrones (thin solid lines) and the loci of the end of protostellar accreting sequence for two different initial temperatures of the cloud (10 K thick solid and 20 K thick dashed line). Circles mark the position of 0.25, 0.5, 1, 1.5, 2, 3, and 5~\msun{} models. Figure adapted from \citet{larson72}.}
\label{fig:larson72}
\end{figure}

In the HR diagram of Fig.~\ref{fig:larson72} the sequence that identifies the end of the protostellar accretion (when the star becomes visible) is compared to standard isochrones. Interestingly, the end of the protostellar evolution is very close to the position of the 1~Myr standard isochrone. \citet{larson69,larson72}, adopting selected accretion parameters, followed the subsequent evolution until the  the Hayashi track, finding that low mass stars ($M<1~$\msun) attain, after the protostellar accretion, characteristics similar to that of standard evolution along the Hayashi track. In contrast, as the mass increases ($M>2~$\msun), models skip the Hayashi line, ending the protostellar phase closer and closer to the MS position, where they join the "standard'' track.

It is worth to remark that theoretical models for the protostellar evolution cannot be easily checked with observations, as these accreting phases occur when the star is still embedded inside the cloud and thus the central core is largely masked by the matter around it.\\
\\
\\

\subsection{Protostellar accretion in hydrostatic stellar evolution codes}
\label{protostarhydrostatic}
As already discussed, hydrodynamic evolution of accreting stars is still a challenging task from the computational point of view. However, concerning the central protostar, it is not needed to employ a hydrodynamic code, as the protostar itself is in hydrostatic equilibrium after the formation of the 2nd Larson core. In this approximation, the central object can be described using a mono dimensional hydrostatic stellar evolutionary code (see e.g. \citet{siess97,stahler80,stahler801}). 

On the other hand, hydrodynamic models are needed to predict the structure of the envelope surrounding the protostar, which does not satisfy the hydrostatic conditions but it is essential to determine the characteristics of the accretion flow.  More precisely, the envelope gives information about the accretion rate, the percentage of the energy of the falling matter transferred to the star and the accretion geometry. Information about these quantities are needed inputs for hydrostatic protostellar models. Due to the still present uncertainty on hydrodynamic calculations, all the previous accretion parameters are affected by not negligible theoretical indeterminacy, as briefly summarised below. 
\begin{itemize}
\item \textit{Accretion rate.} The accretion rate (\mdot) defines the rate at which the star changes its mass; \mdot{} can vary by orders of magnitude during the accretion phase, passing from \mdot $< 10^{-6}$-$10^{-7}~$\msun/yr (quiescent accretion) to rapid and intense episodes of mass accretion (bursts) with \mdot$\sim 10^{-4}$-$10^{-3}~$\msun/yr, e.g. as observed in FU Ori stars, (\citet{hartmann96,hartmann16,audard14}).

\item \textit{Energy transferred by the accreted matter.} The matter falling to the star before reaching the stellar surface has a kinetic energy that can be estimated approximating the matter velocity to the free fall one. However, when it settles on the stellar surface, the kinetic energy has become equal to zero, so the kinetic energy has been converted into another energy form. It can be thermal energy carried inside the star (the accreted matter is hot), or the energy can be partially or totally radiated (photons) before the matter reaches the stellar surface (at the shock front). The fraction of the kinetic energy transferred to the protostar depends on the characteristics of the accretion flow (i.e. density, accretion geometry, accretion rate, see e.g. \citet{baraffe12}). 
\end{itemize}

The difficulty in treating simultaneously the protostar and the envelope evolution requires some simplifications, which mainly concern the geometry of the accreting protostar-envelope system:
\begin{itemize}
\item \textit{Spherical accretion} (see e.g. \citet{stahler80,stahler88,palla91,palla92,palla93,hosokawa09}). The star is supposed to be deeply embedded into the parental cloud and the matter falls on it almost radially. The whole stellar surface is subjected to the accretion and the energy radiated by the star can be reabsorbed by the envelope. The whole protostellar accretion occurs as a radial infall from a cloud that has mass enough to generate the star, at a fixed value of the accretion rate.

\item \textit{Disc accretion} (\citet{hartmann97,siess97,baraffe09,baraffe10,kunitomo17,tognelli13}). The matter falls from a boundary layer of a circumstellar disc and reaches the star via accretion streams. Most of the stellar surface is not subjected by the accretion and the star is free to radiate its energy, most of which is lost in space. The disc is assumed to be totally decoupled from the central star and it is not treated in the stellar evolution codes. The parameters that define the accretion (accretion rate, disk lifetime, accretion energy) are considered as external free parameters which can be obtained from detailed accretion disc evolution calculations (e.g. \citet{vorobyov10,baraffe12}).
\end{itemize}

The spherical accretion scenario is likely to describe the first stages of the formation of the protostar when it is still embedded within the cloud that retains an approximate spherical geometry. However, observations suggest that at some stage of protostellar evolution, the cloud collapses to a disc -- because of angular momentum conservation -- and that it is during the disc accretion that the star gains most of its final mass (see e.g. \citet{natta00,meyer07,watson07,machida10} and references therein). So, both scenarios are interesting and describe a part of the protostellar accretion.

The most important difference between spherical and disc accretion, which deeply affects the protostellar evolution, is the amount of energy retained by the accreted matter. Indeed, while in the spherical accretion it is possible to estimate the amount of energy retained by the accreted matter, in disc accretion this quantity is defined by a free parameter (\alpacc). The impact of \alpacc{} on the evolution is discussed in the next sections.\\
\\
\\

\subsection{Spherical and disc protostellar accretion}
\label{acccretiongeometry}
The spherical accretion scenario applies  to a star that is deeply embedded in a gas cloud. In this case, the evolution of the star and of the envelope have to be treated simultaneously. This allows (at least in principle) to have a consistent evaluation of the accretion rate and the amount of thermal energy that the accretion flows bring inside the star. Qualitatively, the energy emitted from the stellar surface is not free to escape into space since it has to interact with the matter around the star. Thus such an energy is partially reabsorbed by the matter in the envelope, and it eventually reaches the star. The effect of this process is that the star has a kind of external energy source that warms up the stellar surface. The injection of thermal energy from the accreted mass forces the star to expand or at least to compensate for the radius decrease caused by injection of mass.

\begin{figure}[t]
\centering
\includegraphics[width=8cm]{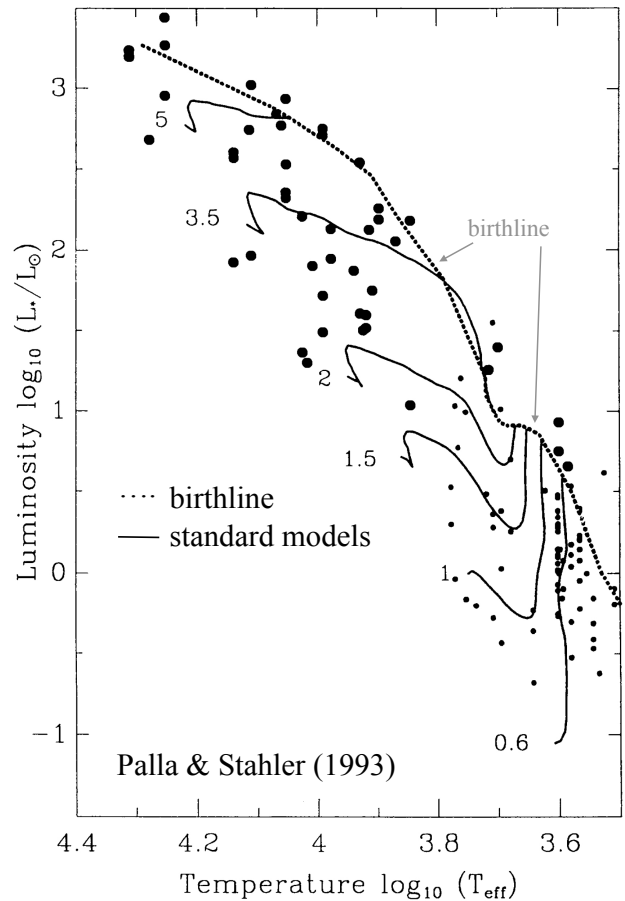}
\caption{Comparison between standard tracks (solid lines) and the birthline (dotted line). Dots represent T~Tauri and Herbig Ae/Be stars. Figure adapted from \citet{palla93}.}
\label{fig:stahler}
\end{figure}
The impact of spherical accretion on the formation of pre-MS stars has been largely analysed in the pioneering works by Larson (\citet{larson69,larson72}), Stahler and Palla (\citet{stahler80,stahler88,palla91,palla92,palla93}), and more recently also by Hosokawa and collaborators (\citet{hosokawa09,hosokawa09b,hosokawa10,hosokawa11}). One of the main results of such a spherical accretion scenario is that stars during the accretion phase remain bright and with large radii. Using a mild and constant accretion rate of $10^{-5}~$\msun /yr,  it is possible to obtain fully accreted stars in a region of the HR diagram that corresponds to the upper envelope of the locus where young pre-MS stars are observed (see Figure~\ref{fig:stahler}). This sequence was called ''\emph{birthline}'', that is the locus of stars with different masses where the accretion ends and the stars become optically visible (\citet{palla93}).

More recent sets of birthlines can be found in \citet{hosokawa09}: such accretion models was computed for different values of the accretion rate (from $10^{-6}$ to $10^{-3}~$\msun /yr), adopting a spherical protostellar accretion code (similar to that used by Stahler and Palla). They showed that increasing \mdot, the birthline moves towards larger luminosities and radii, thus still in full agreement with the observations. Moreover, since spherical accretion models produce low-mass stars (on the birthline) in a region that corresponds to the top of the Hayashi track of standard stellar models (see Figure~\ref{fig:stahler}), the differences between standard and spherical accreting models in pre-MS low-mass stars are negligible. This  validates the results of standard evolutionary tracks/isochrones (at least for ages higher than 1 Myr).

However, it is commonly accepted that stars do not accrete mass spherically during their entire protostellar phase; on the contrary they gain most of their mass from an accretion disc. This motivates the detailed study of protostellar accretion from a disk geometry. Differently from the spherical accretion, in the disc geometry the accretion streams cover only a very limited part of the stellar surface (few percent, see e.g. \citet{hartmann97}) and almost the whole star is free to radiate its energy into space. Another difference is that all the accretion parameters (i.e. accretion rate, fraction of energy inside the accreted matter, etc..) are treated as external parameters in disc accretion models.

In the disc accretion geometry, it is possible to follow also an analytic approach to analyse the main characteristics of the accreting star. Following the formalism presented in \citet{hartmann97}, it is possible to write a simple equation for the temporal evolution of the accreting star radius:
\begin{eqnarray}
\frac{\dot{R}}{R} = \frac{7}{3} \frac{R}{GM^2}\bigg[\beta_D - L_\mathrm{ph} + \bigg(\alpha_\mathrm{acc} - \frac{1}{7}\bigg)\frac{GM^2}{R}\frac{\dot{m}}{M}\bigg]
\label{eq:rdot}
\end{eqnarray}
where $M$ and $R$ are the stellar mass and radius, $\beta_D$ expresses the luminosity due to the deuterium burning (D-burning), $L_\mathrm{ph}$ is the luminosity of the stellar surface, \mdot{} is the mass accretion rate and $\alpha_\mathrm{acc}$ represents the fraction of the accretion energy deposed into the star (thermal energy of the accreted matter). 

Equation~(\ref{eq:rdot}) contains three terms: the first and second are the normal terms that define the evolution of the star with a surface radiative loss ($L_\mathrm{ph}$) with the inclusion of D-burning energy production $\beta_\mathrm{D}$, while the last term represents the accretion effect, which is proportional to \mdot. This term accounts for the fraction of the thermal energy of the accreted matter retained by the star, \alpacc. Such a parameter has to be specified as an external free parameter, ranging from 0 (no energy acquired by the star) to about 1 (or 1/2 in case of thin disc, see e.g. \citet{siess97}). From the same equation, it is also evident that $\alpha_\mathrm{acc} = 1/7 \equiv \alpha_\mathrm{acc,cr}$ defines a critical value; for $\alpha_\mathrm{acc} < \alpha_\mathrm{acc,cr}$ the third term is negative, and it contributes to the contraction of the star. For $\alpha_\mathrm{acc} > \alpha_\mathrm{acc,cr}$ the same term produces a radius expansion. It is common to refer to the case $\alpha_\mathrm{acc}\sim 0$ (or $\alpha_\mathrm{acc} \ll \alpha_\mathrm{acc,cr}$) as  \emph{cold disc accretion} and $\alpha_\mathrm{acc} > \alpha_\mathrm{acc,cr}$ as  \emph{hot disc accretion}.

Looking at eq.~(\ref{eq:rdot}) it is clear that a radius expansion requires a positive value of the right side of the equation, which can be obtained or via an efficient deuterium burning (large $\beta_D$) or via an efficient accretion energy transport into the protostar ($\alpha_\mathrm{acc} > \alpha_\mathrm{acc,cr}$). These two cases are discussed separately in the next two sections.\\
\\
\\

\subsection{D-burning during protostellar accretion}
\label{Dburning}
To check if the D-burning alone can produce a protostar with a large radius, in agreement with observations, we assume $\alpha_\mathrm{acc} = 0$. From eq.~(\ref{eq:rdot}), to produce a radius increase, D-burning has to supply the star with enough energy to counterbalance the radiative losses at the stellar surface plus the gravitational energy decrease caused by the mass ingestion. If this condition is not satisfied, the protostar contracts and the resulting model at the end of the protostellar phase has  a radius much smaller than that observed in young disk stars and expected in spherical accretion cases.
\begin{figure}[t]
\centering
\includegraphics[width=11cm]{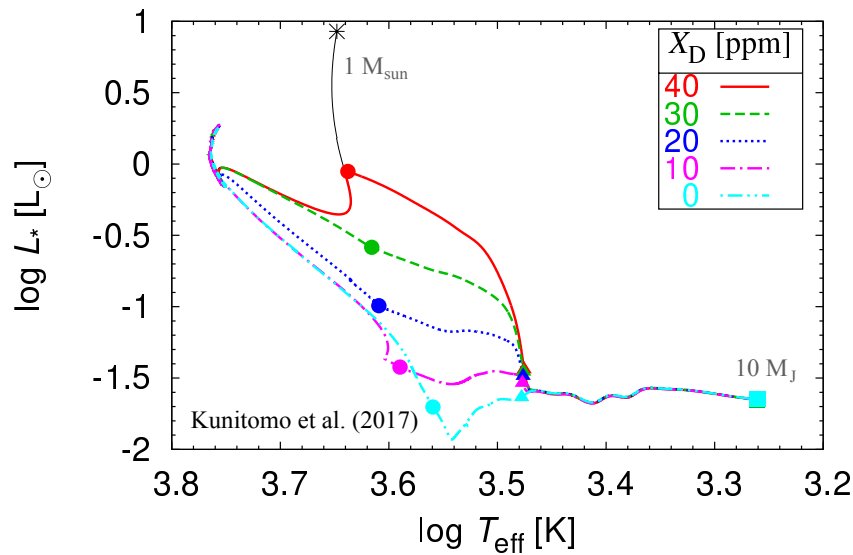}
\caption{Effect of different deuterium original abundances (see labels), $X_\mathrm{D}$, on the protostellar evolution of 1~\msun{} model. The protostellar accretion starts with a seed mass of 10~\mj, with a constant accretion rate of $10^{-5}$~\msun/yr. Filled circles indicate the end of the protostellar accretion phase and triangles the beginning of D-burning. Figure adapted from \citet{kunitomo17}.}
\label{fig:xd1}
\end{figure}
The dependency of the radius on original deuterium abundance $X_\mathrm{D}$ has been investigated in \citet{tognelli13} and more recently in \citet{kunitomo17}. In \citet{kunitomo17} the authors assumed for the 2nd Larson core mass the value $M_\mathrm{seed}=0.01$~\msun(=10~\mj). Figure~\ref{fig:xd1} shows a comparison between birthlines obtained assuming different values of $X_\mathrm{D}$, for a cold accretion scenario with \mseed=10~\mj. When no deuterium is taken into account in the stellar matter, the star inevitably contracts: in this model, the star ignites the hydrogen burning close to the end of the protostellar accretion, thus totally skipping the pre-MS evolution. The situation changes increasing the deuterium mass fraction abundance in the star. To partially reproduce the standard pre-MS evolution, a deuterium content of $X_D \approx 4\times 10^{-5}$ (i.e. 40 ppm) is required. If a more reliable deuterium content is adopted, $X_D \approx 2\times 10^{-5}$, the models with protostellar accretion converges to standard models only in the Henyey track; in this case, the evolution along the Hayashi track is missed contrarily to what observed in young clusters. We want to comment about the fact that the uncertainty on galactic deuterium mass fraction abundance is not larger than 10~ppm (see e.g. Fig.~2 and table~1 in \citet{sembach10}), thus an initial deuterium content of $X_D \approx 40$~ppm is an over estimation for disk stars. This fact seems to indicate that deuterium alone is not capable of maintaining the star bright enough to reconcile protostellar cold accretion models and the results obtained in a standard non accreting scenario.\\
\\
\\
\subsection{Accretion energy}
\label{accretionenergy}
\begin{figure}[t]
\centering
\includegraphics[width=12cm]{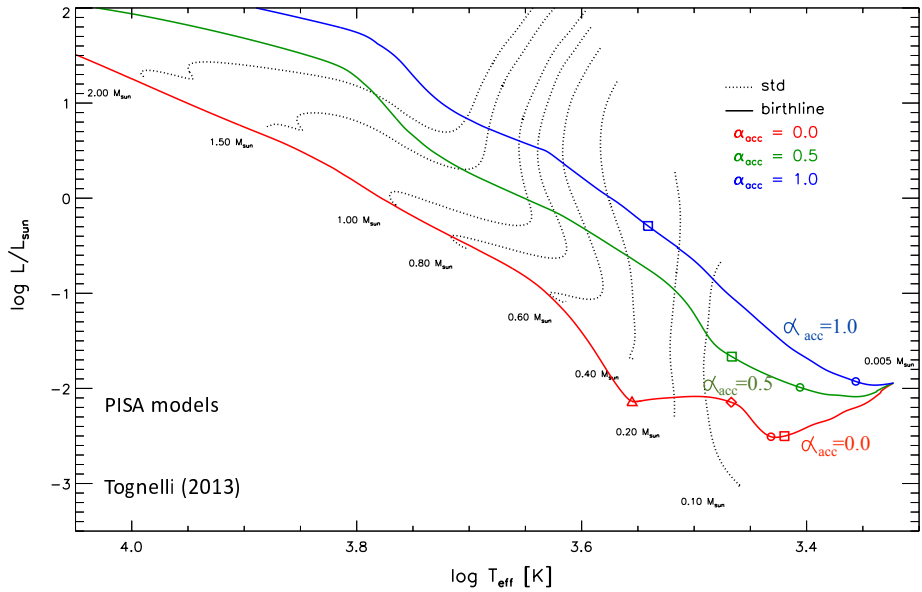}
\caption{Effect of \alpacc{} on the protostellar evolution for three values of \alpacc, 0 (cold accretion), 0.5 and 1 (hot accretion). Dotted lines are standard pre-MS tracks. Figure adapted from \citet{tognelli13}.}
\label{fig:accene}
\end{figure}

There is another natural way to obtain a radius expansion in protostars, which is assuming that the ingested matter retains part of its internal energy; this means to assume a value of $\alpha_\mathrm{acc} > \alpha_\mathrm{acc,cr}$. In \citet{hartmann97} it was shown that non-cold accretion models (\alpacc$>\alpha_\mathrm{acc,cr}$) can attain a radius expansion large enough to reproduce observed stars; in this case the disc accretion mimics  spherical-accretion birthline obtained by \citet{stahler88}. More recently, \citet{kunitomo17} analysed in more details the impact of \alpacc{} on the formation of a 1~\msun{} model, finding that the inclusion of a certain fraction of the total accretion energy (i.e. \alpacc$ \in [0,\,1]$) in the star is capable of maintaining the structure at large radii. Figure~\ref{fig:accene} shows the birthline computed in \citet{tognelli13} -- by means of the PISA stellar evolutionary code -- for solar metallicity stars using three values of \alpacc=0 (cold case), 0.5 and 1 (hot case), for a seed mass value of 5~\mj.
From Fig.~\ref{fig:accene}, it is evident that adopting a value of \alpacc$\ge 0.5$ models on the birthline are bright and intersect the standard evolutionary tracks in the standard Hayashi track, for all the selected mass range [0.1, 2.0]~\msun. Recently, \citet{tognelli20} have obtained similar results for metal poor models: they showed that even in the low metallicity case, the inclusion of accretion energy produces expanse objects that intersect the Hayashi track of standard non accreting models at the end of the protostellar accretion stage.\\
\\
\\

\subsection{Connecting the standard pre-MS and the protostellar accretion phase}
\label{protostarandPMS}
From the previous discussion, it emerges that, depending on the characteristics of the protostellar accretion, the protostar could end its first evolution with a structure similar or in some cases profoundly different from that obtained in a normal gravitational contraction along the Hayashi track. The largest discrepancy with standard pre-MS evolution occurs in the case of cold accretion starting from a seed of the order of few Jupiter masses, as in that case the classical Hayashi track is almost completely skipped (see e.g. \citet{baraffe09,tognelli20}). 

\begin{figure}[t]
\centering
\includegraphics[width=9cm]{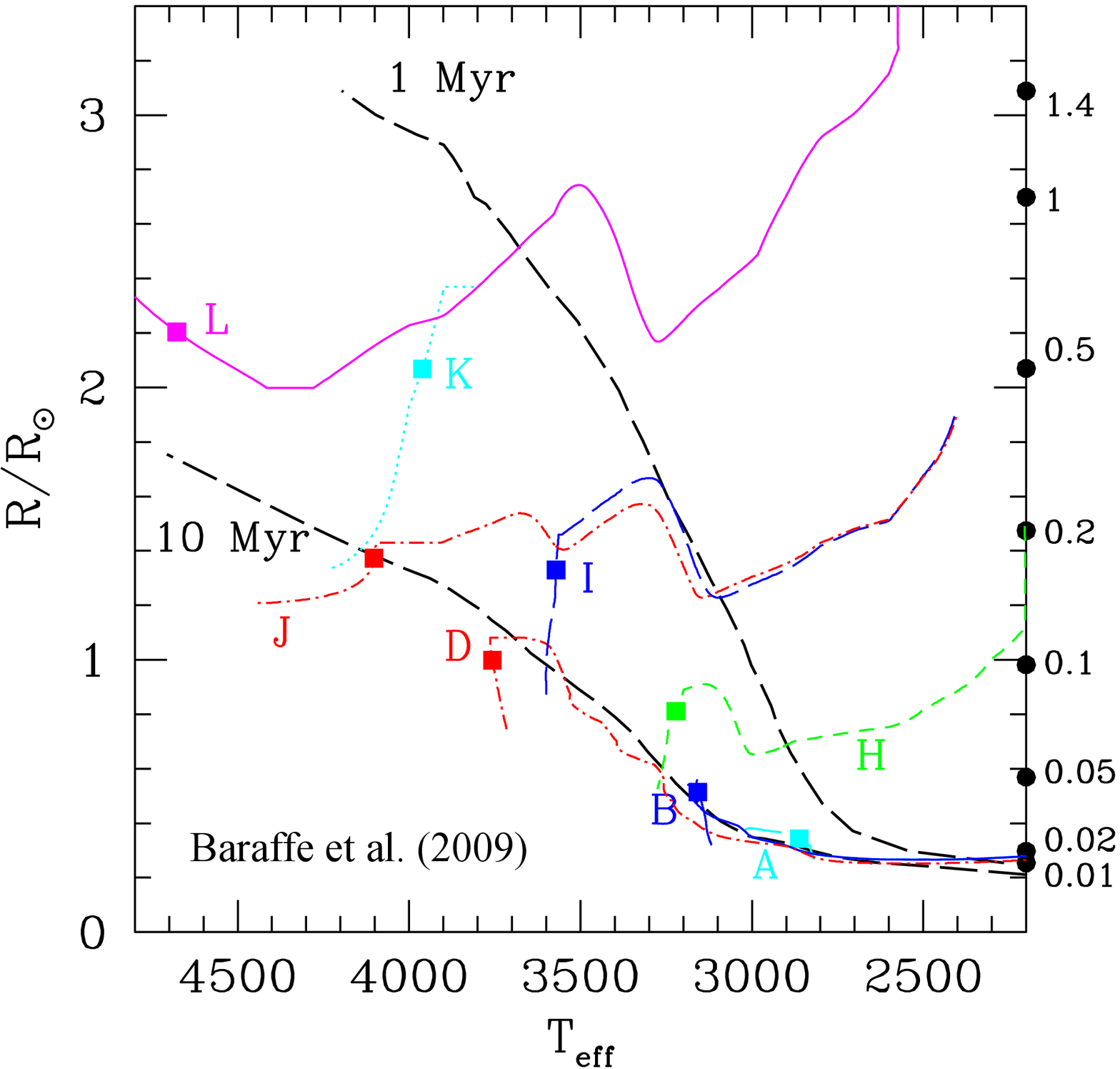}
\caption{Evolution in the HR diagram of protostellar models with different values of \mseed, compared with the standard 1 Myr and 10 Myr isochrones (black long dashed lines). The different letters indicate models with different seeds and final masses ($M_{fin}$), in particular: (A) \mseed= 1~\mj, $M_{fin}$ = 0.05~\msun, (B) \mseed= 1~\mj, $M_{fin}$=0.1~\msun, (D) \mseed= 1~\mj, $M_{fin}$=0.5~\msun, (H) \mseed = 10~\mj, $M_{fin}$ = 0.21~\msun, (I) \mseed = 50~\mj, $M_{fin}$ = 0.55~\msun, (J) $M_{fin}$=1.05~\msun, (K) \mseed = 100~\mj, $M_{fin}$ = 1.1~\msun, (L) \mseed=100~\mj, $M_{fin}$ =1.85~\msun. Filled squares represent the position of the 1~Myr model. Figure adapted from \citet{baraffe09}.}
\label{fig:bf09_tog20}
\end{figure}

Figure~\ref{fig:bf09_tog20} shows the evolution in the HR diagram of cold accretion models starting from different \mseed{} and ending with different final masses, as discussed in details in \citet{baraffe09}. It is difficult to reproduce the Hayashi track of pre-MS stars starting from \mseed{} of few Jupiter masses (i.e. cases A, B, D). Moreover, the position of the 1~Myr model (filled square) in accretion models is relatively far from the standard 1~Myr isochrone; in most of the cases, the position of pre-MS models with the inclusion of cold protostellar accretion at 1 Myr is very close to the standard non accreting 10 Myr isochrone,   witnessing the strong impact of cold accretion on pre-MS evolution. 

\begin{figure}[t]
\centering
\includegraphics[width=9cm]{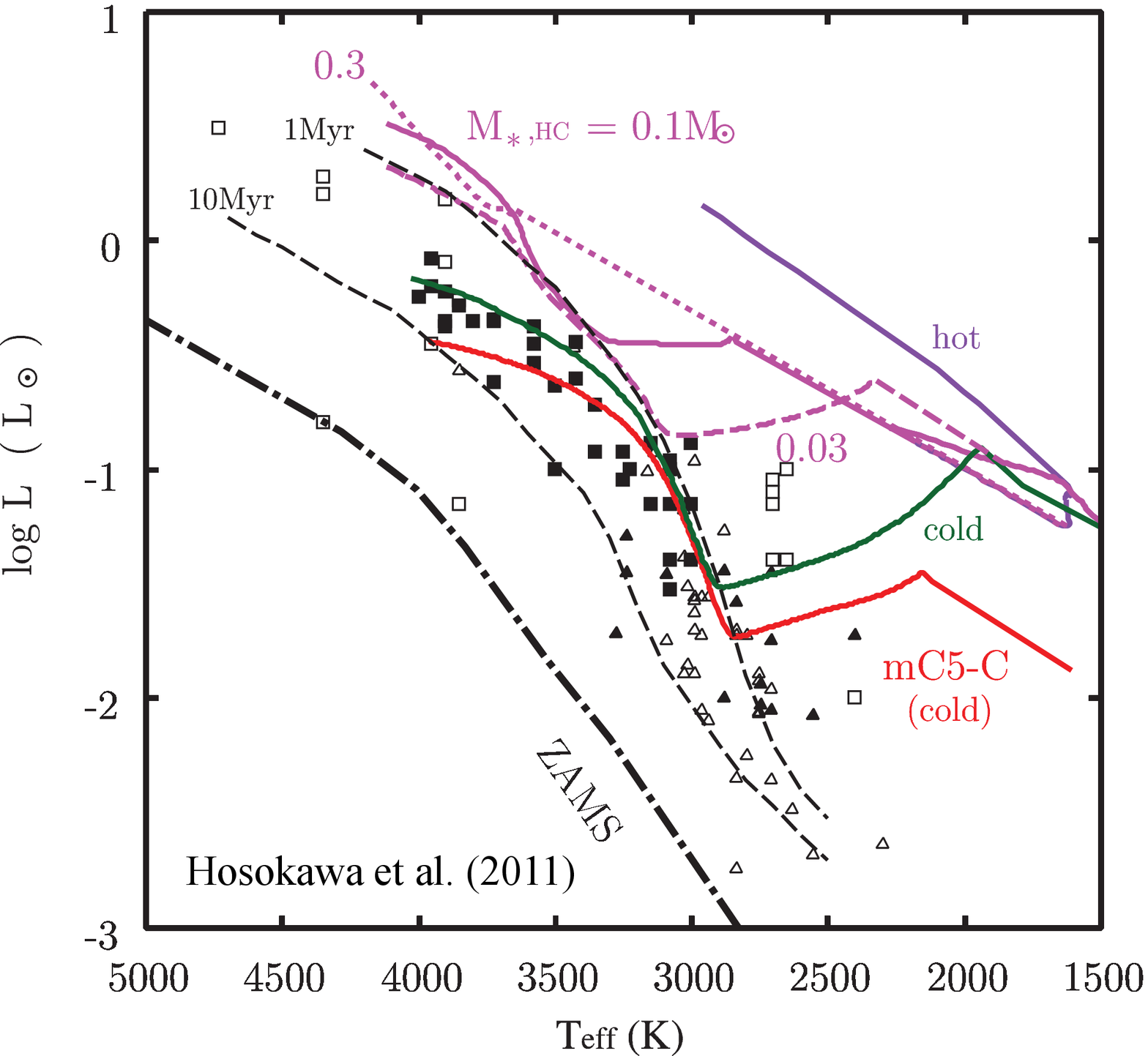}
\includegraphics[width=11cm]{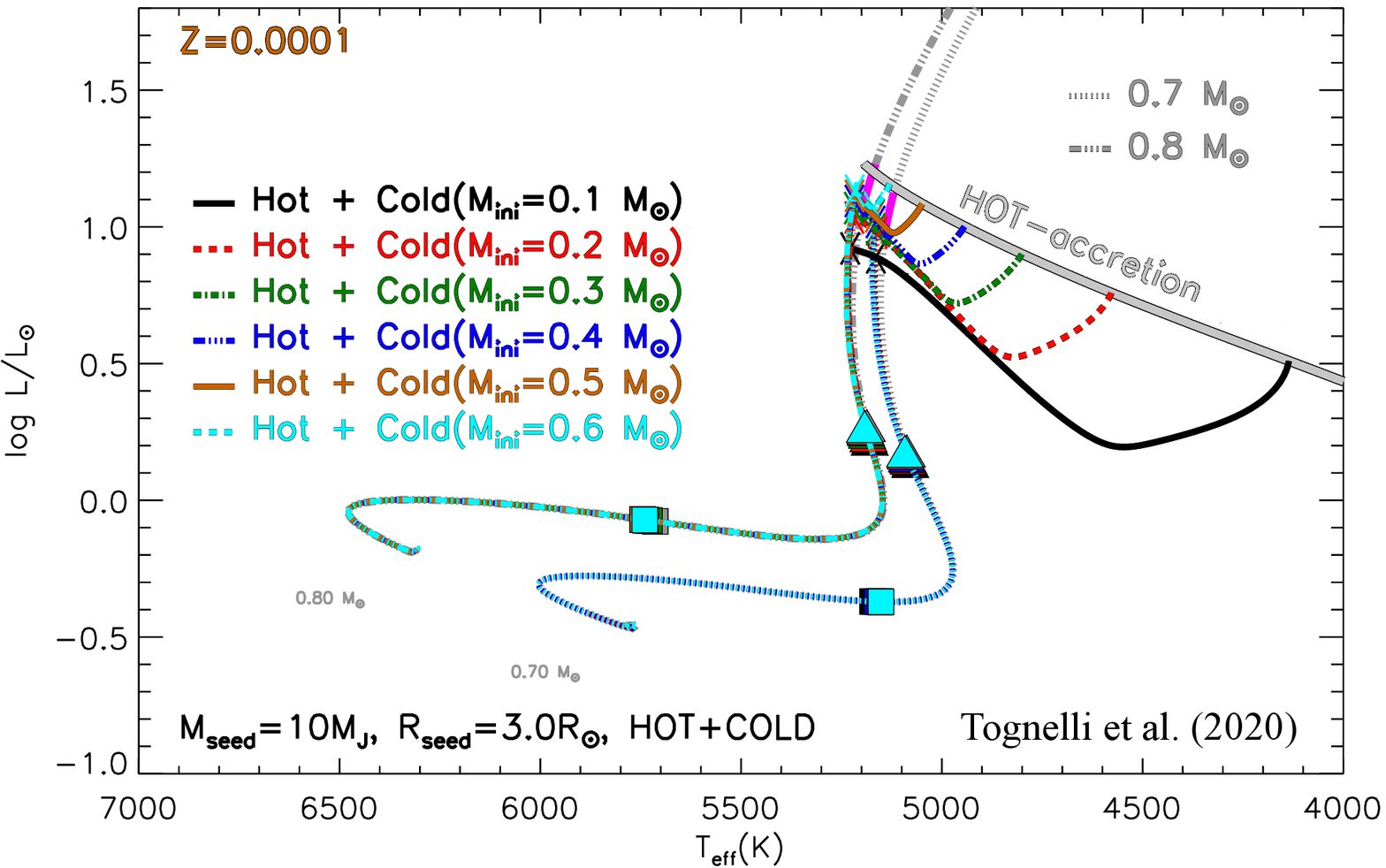}
\caption{Evolution in the HR diagram of purely and partially hot models computed with an accretion rate $\dot{m}=10^{-5}$~\msun{}/yr. Top panel: comparison between purely hot (solid purple line), purely cold (green and red solid lines) and hot+cold birthlines (magenta lines). The two purely cold cases differ for the seed radius, 3.7~\rsun{} (green line) and 1.5~\rsun{} (red line, mC5-C). The accretion switches from hot to cold at a given value of the mass, which is 0.03~\msun{} (magenta dashed line), 0.1~\msun{} (magenta solid line), and 0.3~\msun{} (magenta dotted line). Black lines are isochrones of 1 and 10 Myr (dashed lines) and the ZAMS (dot-dashed line) for standard non accreting models. Squares, triangles and circles represent observations of some young pre-MS stars. Figure adapted from \citet{hosokawa11}. Bottom panel: models at low metallicity (Z=0.0001) for a total final mass of 0.7 and 0.8~\msun{}. The accretion switches from hot to cold at a mass value 0.1, 0.2, 0.3, 0.4, 0.5, and 0.6~\msun{} as indicated in the labels. The thick grey line represents the hot birhtline ($\alpha_\mathrm{acc} = 1$). Figure adapted from \citet{tognelli20}.}
\label{fig:hot_cold}
\end{figure}

As discussed, it is likely that stars first accrete in a spherical hot scenario and then, at a given stage, switch to a disk-like accretion. In this case the transition from hot to cold accretion occurs for some value of the protostellar mass (possibly dependent on the amount of mass available in the cloud/disk). This mixed scenario has been investigated firstly by \citet{hosokawa11} to show that the protostar remains bright enough to end the protostellar phase close to a Hayashi track. Top panel of Fig.~\ref{fig:hot_cold} shows the models by \citet{hosokawa11}. The purely hot accretion scenario (purple solid line), which corresponds to a hot birthline obtained assuming a spherical accretion, attains large luminosities and radii well above the standard 1~Myr isochrone. Figure shows also the results of models where the accretion switches from hot to cold at a given value of the stellar mass, namely 0.03~\msun{} (magenta dashed line), 0.1~\msun{} (magenta solid line) and at 0.3~\msun{} (magenta dotted line). It is interesting to notice that in all cases, the birthline is still quite luminous, being very close to the 1~Myr isochrone. Similar results have been obtained for metal poor models by \citet{tognelli20} (bottom panel of Fig.~\ref{fig:hot_cold}).

\citet{baraffe09} and \citet{baraffe12} investigated also the possibility to produce bright objects using an episodic accretion. The basic idea behind the models is that during intense bursts mass accretion phases the protostar can accrete matter in the hot-accretion configuration ($\alpha_\mathrm{acc} > \alpha_\mathrm{acc,cr}$, see the Appendix in \citet{baraffe12}), to switch back to cold accretion at the end of each burst. The authors showed that in this case it's still possible to produce models that end their protostellar accretion close to the standard position of the Hayashi track, to reproduce data (see also \citet{tognelli20} for metal poor protostars). 

What emerges from the previous analysis is that, if one assumes masses and radii typical of the 2nd Larson core, cold models cannot produce the observed bright stars in young clusters, but it is required the presence of hot accretion phases. Thus, the results seem quite comfortable: in most hot disk or spherical geometry, the protostellar accretion leads to pre-MS models with characteristics similar to those predicted in standard pre-MS evolution. More importantly, the position in the HR diagram of such models is in agreement with observational data. On the contrary, for the accretion parameters leading to a final mass model different to that of the standard one, as in the cold accretion scenario, the position in the HR diagram is in disagreement with disk star observations, rising doubts about the validity  of such models.\\
\\
\\

\section{Light elements surface abundances and nuclear burning during the pre-MS phase}
\label{elementabundances}
Lithium, together with beryllium and boron, belong to the class of light elements burnt in pre-MS, because of their relatively low nuclear destruction temperature (between 2-5 million degree). The threshold values for the burning temperature depend mainly on the considered element, on the stellar mass (density and evolutionary stage) and slightly on the chemical composition of the star (in particular on helium and metals abundances). 
For pre-MS solar metallicity stars in the mass interval [0.08, 1.0]~\msun,  the ranges of burning temperatures for the different elements approximately are: 2.4-3.5$\times 10^6$~K ($T(^{6,7}$Li)), 3.5-4.0$\times 10^6$~K ($T(^9$Be)) and 4.2-5.0$\times 10^6$~K ($T(^{10,11}$B)). In the literature the temperatures given for the burning are sometimes slightly different form the values reported here; the reason is that usually authors do not take into account that stars with different masses ignites these elements at slightly different temperatures because the nuclear burning rates also depends (even if at a lower level) on the density in the region where the burning occurs. Moreover, such temperatures can be different for MS or pre-MS stars with the same mass because of the different time scales in which light elements are destroyed. In MS stars the evolutionary time scale is much longer than that in pre-MS (for the same mass and chemical composition) consequently a smaller burning rate due to a smaller threshold temperature at the bottom of the convective envelope, is compensated by the longer time during which that element is destroyed. As a result the burning of light elements in MS can efficiently occur even at thresholds temperature smaller than those required in pre-MS.
\begin{figure}[t]
    \centering
    \includegraphics[width=12cm]{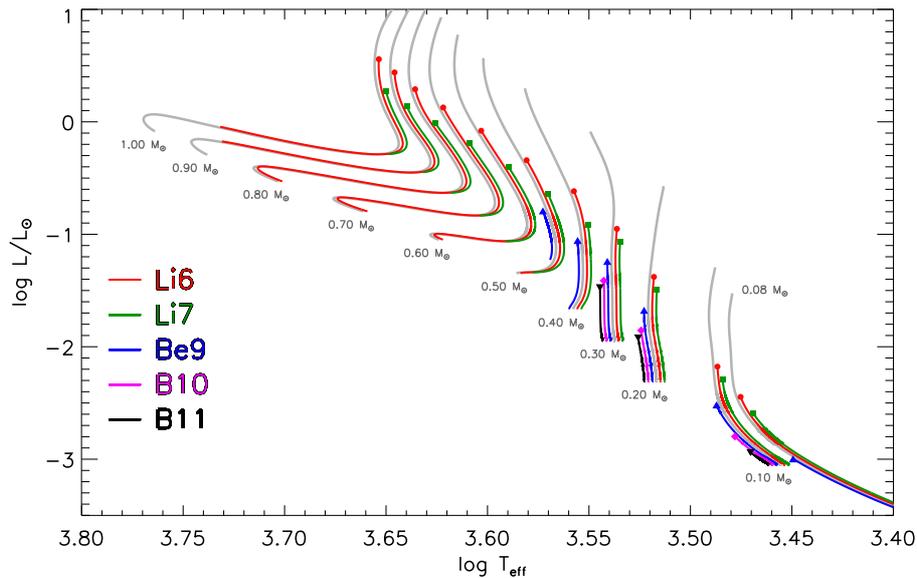}
    \caption{Evolutionary tracks for solar metallicity in the HR diagram with indicated the regions where surface light element abundances decrease due to burning (where the temperature at the bottom of the convective envelope is higher than the burning temperature of Li, Be and B). The stellar models have been computed using the PISA evolutionary code with the same input parameters described in \citet{lamia15}.}
    \label{fig:ele_burning}
\end{figure}
Due to the differences in their burning temperatures in pre-MS, Li, Be and B are gradually destroyed at different depths inside the stellar interior and at different ages, in dependence on the stellar mass. As an example, Fig.~\ref{fig:ele_burning} shows the portion along the evolutionary track where surface Li, Be and B are burnt at the bottom of the convective envelope in a set of solar chemical composition stars in the mass range [0.08, 1.0]~\msun. It is interesting to notice that while Li is burnt (at the bottom of the convective envelope) in the whole selected mass range, surface $^9$Be burning occurs only for masses between about 0.08 and 0.5~\msun, while B is burnt in an even smaller mass range (about 0.1 - 0.3~\msun). 

The abundance of light elements at the stellar surface are strongly influenced by the nuclear burning as well as by the inwards extension of the convective envelope and by the temperature at its bottom. Consequently, the comparison between theory and observation for Li, Be and B surface abundances are useful to constrain theoretical models and in particular the convective envelope depth.

From the observational point of view, most of the data for light elements concern the abundance of $^7$Li whose line (at 670.779~nm) can be safely resolved even in cold stars, as witnessed by the huge amount of data for stars in clusters or isolated stars at different metallicities (see e.g. \citet{sestito05,delgadomena14,delgadomena15,gomez18} and references therein).

$^6$Li burns at a lower temperature  with respect to $^7$Li, consequently it is almost completely destroyed when $^7$Li burning becomes efficient. Thus a potential detection of observable amounts of $^6$Li in stellar atmospheres would constrain the destruction of the less fragile $^7$Li (\citet{copi1997}). Since the depth of the convective zone increases with metallicity, $^6$Li is almost completely depleted in high metallicity disk stars, as in the Sun (see e.g. \citet{asplund09}) and it is below the detection level also for most thick disk and halo stars (see e.g. \citet{Spite10}). The possible abundance of $^6$Li below the limit of detection also for halo stars could be explained by the fact that the amount of $^6$Li formed by the standard Big Bang and by the cosmic rays is supposed to be very low. Moreover, a very small $^6$Li abundance in these stars would be very difficult to detect, in particular because the lines (doublets) of $^6$Li and $^7$Li are overlapping (see also discussion in Sec.~\ref{Liatmouncertainties}).

Beryllium and boron measurements are more problematic than $^7$Li observations. $^9$Be abundance is measured using near-UV lines, only in stars with \teff $\ga 5000$~K, which corresponds in pre-MS to a mass range where Be is expected to be preserved and not destroyed (see e.g. \citet{garcia95,santos04,randich07,smiljanic10,delgado12,lamia15}).

The abundance of the boron isotopes is even more difficult to measure than Be because the boron lines fall mainly in the UV part of the spectra where the Earth atmosphere is not transparent. In addition, for disk metallicity stars, B lines suffer strong blending problems (see, e.g., \citet{cunha2010}). Similarly to Be, B abundance are available in a mass range where B is expected to be not burnt in standard models. Despite the observational difficulties Be and B surface abundances data are available for some stars even at low metallicities (see e.g. \citet{boesgaard05}, \citet{lodders09}, \citet{primas09}, \citet{tan09}, \citet{boesgaard11}). In the observed stars, the ratio $^{11}$B/$^{10}$B seems to be of the order of four, in agreement with solar values and meteorite results, even if it is very difficult to spectroscopically discriminate among the boron isotopes (see, e.g. \citet{proffitt99},  \citet{prantzos12}). Be and B surface abundances have been also measured in the Sun where, as expected, they are not burned (see e.g. \citet{asplund05}, \citet{asplund09}, \citet{lodders09}, \citet{lodders10}).

The temperatures for light elements burning can be reached in stellar interiors during the pre-MS evolution of stars with masses larger than about 0.05-0.1~\msun{} (depending on the requested burning threshold temperature). We recall that at the beginning of the pre-MS evolutions stars are, independently of their mass, fully convective. Thus, if a nuclear burning occurs at this evolutionary stage, the burning affects the chemical abundance in the whole star, from the centre to the surface. However, as the star contracts and warms up, the opacity decreases at the stellar centre and stars with $M\ga 0.3$~\msun{} develop a radiative core. From this moment on the chemical evolution of the surface is (during the pre-MS) decoupled from that of the centre if the bottom of the convective envelope does not reach a region deep and hot enough to process --via nuclear burning-- the surface matter. Thus, for partially convective pre-MS stars, a condition to have a partial surface depletion of a specific element is that the bottom of the convective envelope reaches a temperature high enough to make nuclear burning efficient and then recedes toward the external parts of the star at lower temperatures.
\begin{figure}[t]
\centering
\includegraphics[width=12cm]{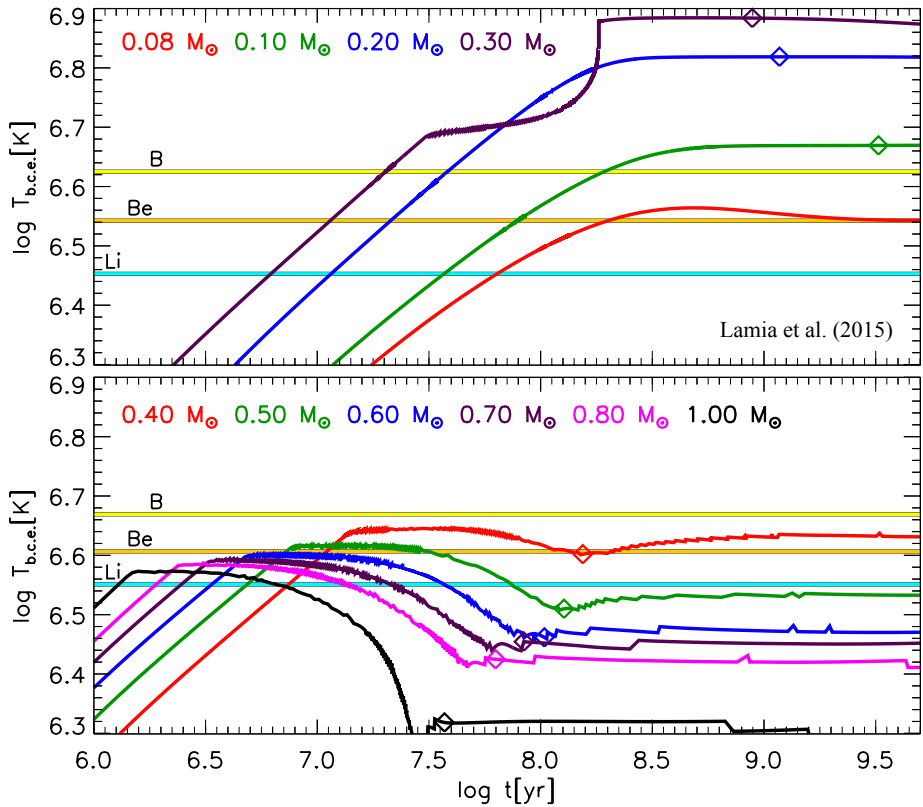}
\caption{Temporal evolution of the temperature at the bottom of the convective envelope for masses in the range [0.08,1.0]~\msun. The threshold temperature required to ignite Li, Be and B burning are indicated as coloured horizontal lines. The ZAMS position is marked by a diamond. The not regular behaviour of the 0.3~\msun{} model at $\log t(yr) \sim 7.5-8.3$ is caused by the formation of a transient convective core before the ZAMS (figure adapted from \citet{lamia15}).}
\label{fig:Tce}
\end{figure}

Figure~\ref{fig:Tce} shows the temporal evolution of the temperature at the bottom of the convective envelope, \tce{}, for stars with different masses between 0.08 and 1.0~\msun{} at solar metallicity, with indicated the approximate values for the Li, Be and B burning temperatures. In fully convective stars \tce{} coincides with the central temperature, T$_c$. When stars are fully convective \tce{} progressively increases until the star reaches the ZAMS, while in stars that develop a radiative core \tce{} stops increasing when the radiative core forms and \tce{} slowly decreases as the radiative core grows in mass. This has a direct impact on the interval of time during which the surface light elements depletion occurs. Considering e.g. $^7$Li abundance in fully convective stars (i.e. M$\le 0.3$~\msun), \tce{} overcomes \tli{} at young ages (i.e. about 50~Myr for 0.1~\msun{} and 5~Myr for 0.3~\msun); then surface Li burning continues during the whole pre-MS and MS phase. Since \tce{} continuously increases, the burning efficiency increases too. On the other hand, in partially convective stars \tce{} reaches a maximum and then decreases as the star evolves towards the ZAMS. For $M\ga 0.5$~\msun, \tce{} decreases below \tli{} at some point during the pre-MS, thus halting the lithium burning at the bottom of the convective envelope. From this moment on, surface lithium abundance remains constant during the pre-MS phase. Figure~\ref{fig:Tce} also shows that increasing the mass of the star the time interval during which surface lithium is destroyed is shorter and the maximum value of \tce{} reduces too; this is due to the fact that increasing the mass the convective envelope becomes thinner. This indicates that, increasing the mass, surface lithium is destroyed progressively less efficiently. 

The situation is similar for the other light elements; clearly one has to take into account the different burning temperatures, so that the mass range in which Be and B are destroyed at the base of the convective envelope is different from that in which lithium is burned. As an example, for solar composition models, Be can be burnt at the bottom of the convective envelope in the mass interval $0.08\la M$/\msun$\la 0.5$. On the other hand B in destroyed only in the mass range $0.1\la M$/\msun$\la 0.3$.

\begin{figure}[t]
    \centering
    \includegraphics[width=12cm]{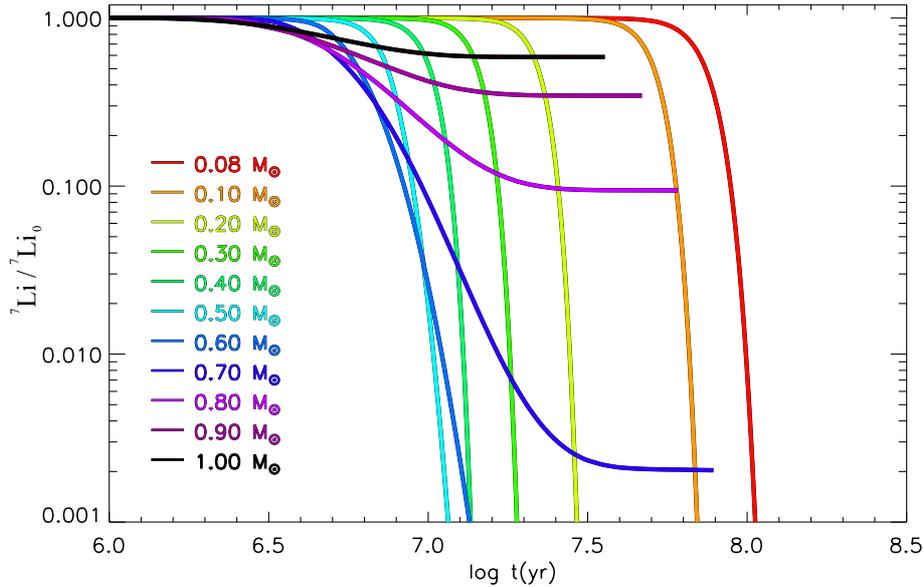}
    \caption{Temporal evolution of surface lithium abundance (normalized to the initial one) during the pre-MS for stars in the mass range [0.08, 1.0]~\msun and solar chemical composition. Stellar models have been computed using the same input parameters described in \citet{lamia15}.}
    \label{fig:li_age}
\end{figure}

Figure~\ref{fig:li_age} gives an example of the $^7$Li surface abundance time behaviour predicted for stars in the mass range [0.08, 1.0]~\msun; it is important to notice that light element surface abundances depend not only on the capability of \tce{} to overcome the threshold temperature for the considered element, but also on the duration of the burning and on the difference between the threshold and \tce. In particular, this last quantity is very important because the burning rate of light elements is proportional to $T^a$ with $a\approx 20$ for lithium, $a\approx 23$ for beryllium and $a\approx 25$ for boron. 

Referring to Fig.~\ref{fig:li_age}, in fully convective stars, i.e. $M\la 0.3$~\msun, at a fixed age, the surface lithium depletion progressively increases increasing the stellar mass. For partially convective models, this behaviour breaks up and, at a fixed age, the lithium depletion decreases as the mass increases. This is clearly visible in the figure comparing e.g. the predicted lithium abundance at about $\log t(yr)= 7.5$ for 0.7, 0.8, 0.9, and 1.0~\msun{} models. The amount of residual surface lithium increases with the mass, as the consequence of the decrease of \tce{} (in time) when the radiative core forms and grows.

Figures~\ref{fig:Tce} and \ref{fig:li_age} refer to a standard evolution, where the star is fully formed at the top of the Hayashi track. The situation can be different if  protostellar accretion is taken into account, in particular in those cases where the star at the end of the protostellar phase is compact and faint, which corresponds essentially to the case of cold accretion models. This could affect light element burning in two different ways: 1) in principle for some possible values of the accretion parameters it could be possible the burning of light elements (most likely lithium) during the protostellar phase 2) accretion can change the pre-MS stellar characteristics with respect to those already predicted by the standard scenario so that the light element burning efficiency is changed too.
We will discuss the effect of the protostellar accretion on the surface chemical composition in Section~\ref{Liprotostar}. \\
\\
\\

\subsection{Surface lithium abundance in open clusters}
\label{Liinopen}
Many questions are still open about the large discrepancies between the predicted and observed surface lithium abundance in young clusters, where standard models tend to underestimate the surface abundance at a given age (see e.g. \citet{dantona00,dantona03,jeffries06,tognelli12} and references therein). Moreover, the presence of a large scatter in the observed Li abundance among stars with similar \teff{} in young clusters poses questions about the possible mechanisms producing different amounts of lithium depletion in stars with the same mass, age and chemical composition (\citet{king00,jeffries00,clarke04,xiong06,king10}). 

\begin{figure}[t]
\centering
\includegraphics[width=10cm]{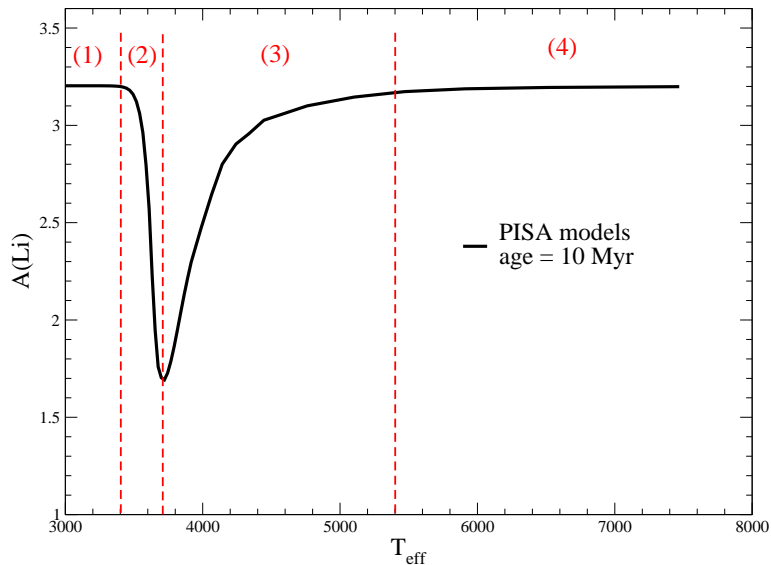}
\caption{Theoretical surface lithium abundance predicted for solar metallicity stars at 10~Myr, as obtained in standard evolutionary models using the PISA evolutionary code.}
\label{fig:ex_li_pattern}
\end{figure}

It is worth noticing that, qualitatively,  standard models (without accretion) are capable to produce a pattern of lithium vs mass (or \teff) similar to that observed in young clusters. This pattern can be divided into three regions, and, referring to Fig.~\ref{fig:ex_li_pattern}, it can be summarised as it follows:
\begin{itemize}
\item  Starting from a certain value of the effective temperature (that depends on the cluster age) the surface lithium content, at a given cluster age, increases with the \teff{} (or the stellar mass), until it reaches a plateau corresponding to stars that do not deplete Li (hot stars). Regions (3)-(4) in Fig.~\ref{fig:ex_li_pattern} correspond to partially convective models of increasing mass. As previously discussed, the more massive is the star, the thinner is the convective envelope and, in turn, the less efficient is the surface Li depletion. The plateau corresponds to stars with a convective envelope so thin that \tce$<$ \tli. 

\item For lower masses (and thus \teff{}), e.g. in regions (1)-(2), stars are fully convective and lithium is burned in the whole star.  At a fixed age the lithium burning efficiency increases with the stellar mass and lithium surface abundance rapidly changes  varying the stellar mass.

\item In region (1), reducing the mass (or \teff), one approaches the minimum mass that reaches the Li burning temperature in fully convective stars. Below this minimum mass the surface lithium abundance is constant and equal to the original value.
\end{itemize}
\quad\\

\subsection{Lithium abundance evolution during protostellar accretion}
\label{Liprotostar}
As discussed in the previous section, the inclusion of the protostellar accretion phase could (in dependence of the adopted accretion parameters) drastically alter the evolution of a pre-MS star. In this section we briefly review the main effects of the protostellar accretion phase on the surface lithium abundance during the protostellar and pre-MS evolution as a function of the different possible accretion parameters.

This problem was first analysed by \citet{baraffe10}, who showed that the inclusion of protostellar accretion in solar metallicity stars with different input parameters can lead to a variety of cases for which the resulting lithium abundance (in pre-MS or in MS) is different from what expected in standard pre-MS evolution (see also \citet{baraffe12,tognelli13,kunitomo18}). We recall that accretion models depend on many parameters, but the main quantities that strongly affect the pre-MS evolution are the seed mass and the accretion energy deposed into the star. The general picture that emerges is that in cold accretion models lithium is efficiently destroyed  during the protostellar accretion or at the very beginning of the pre-MS phase. Thus these stars should show a very low surface lithium content.

A detailed analysis of the effect of the protostellar accretion on surface lithium abundance for different subsolar metallicities ($Z=0.005$, $Z=0.001$, and $Z=0.0001$) was discussed in \citet{tognelli20}. We also performed some tests to verify that what obtained for sub-solar metallicity is still valid at solar metallicities. 

The results by \citet{tognelli20}, similarly to what already obtained by \citet{baraffe10}, show that two scenarios can be found:
\begin{itemize}
\item  \emph{Pure cold accretion case}. The accretion leads to stellar structures at the end of the protostellar phase different with respect to standard non accreting models, affecting also the lithium burning efficiency. If the seed mass is of the order of few Jupiter masses, the models result to be so compact and hot that start to efficiently deplete lithium before the end of the accretion phase. The level of depletion is mainly determined by the seed mass and it is only slightly affected by the other accretion parameters (accretion rates, initial radius). After the protostellar phase, for masses larger than about 0.1-0.2~\msun, lithium is completely destroyed in an extremely hot and fully convective pre-MS structure. This prediction is in complete disagreement with observations available for young clusters, where $M\approx 0.8$-1~\msun{} stars show no relevant lithium depletion (see Section~\ref{Liuncertainties} for more details). Moreover, such accreting models are in disagreement with the observed position of very young disk pre-MS stars in the HR diagram. The disagreement between data and accretion models is partially mitigated if a larger seed mass is adopted (of the order of 10~\mj). In this case it is possible to reduce the level of lithium depletion in very low mass stars (i.e. $M\la 0.3$~\msun), but not for stars close to 1~\msun{} where lithium is totally depleted in pre-MS.

\item \emph{Hot accretion case}. In Section~\ref{protostarandPMS} we showed that if stars accrete part of their mass during an hot accretion phase (during which the protostar is maintained at large radius by the accretion energy), the star at the end of the accretion phase is more similar to a standard evolutionary models. In this case, protostars are relatively  cold and they do not deplete an appreciable amount of Li. Then, when the star enters the pre-MS the residual lithium is essentially equal to the original one, as predicted by models without accretion and from this moment on the lithium evolution proceeds as in standard stellar evolutionary models.
\end{itemize}

These two scenarios embrace many other possible solutions, obtained by modify/tuning the accretion parameters and the accretion history to produce, at least in principle, intermediate scenarios. However, a fine tuning of the accretion parameters that depends also on the stellar mass is unlikely and could produce artificial results (\citet{tognelli20}). The two extreme scenarios highlight an important point. The expected Li abundance is strictly connected to the protostellar evolution. Stars that due to the inclusion of the protostellar accretion skip the Hayashi track (i.e. pure cold accretion) undergo to an efficient lithium burning during the protostellar phase, in disagreement with standard predictions. This kind of models are excluded, at least for disk metallicities, by observational data.

The possible effects of accretion on stellar characteristics and Li temporal evolution could also be linked to the question of the luminosity spread observed in star forming regions. The problem consists in the fact that stars with the same \teff{} and the same chemical composition show different luminosities (see e.g. \citet{hillenbrand09,jeffries09b,dario10,dario10a}). A possible dependence on the protostellar accretion of such a luminosity spread was analysed by \citet{baraffe09}; the adoption of a different accretion history during the protostellar phase can strongly affect the luminosity and \teff{} of a star at the end of the protostellar phase, as already discussed in previous sections. If this is the case, faint stars, which experienced cold accretion, should show a clear lower lithium content than bright ones. In other words, such a luminosity spread should directly reflect in a surface lithium content spread. This point deserves to be further investigated to clearly confirm or exclude the presence of a correlation between lithium content and luminosity in star forming regions. \\
\\
\\

\subsection{Lithium in old metal poor stars}
\label{LilowZstars}
An interesting aspect to be discussed about lithium evolution is the cosmological lithium problem. Halo stars show a lithium plateau for \teff$ > 5900$ K and [Fe/H]~$<-1.5$, the so called Spite plateau (\citet{spite82a,spite82}), with a constant logarithmic lithium abundance\footnote{$A(Li)$ indicates an observational notation for abundances, where $A($Li$) = \log N_\mathrm{Li}/N_\mathrm{H} +12$, with $N$ being the number of particles of a given specie.} $A($Li$) = 2.1$-$2.4$ (\citet{cp05,asplund06,melendez10,sbordone10}), and references therein). From the theoretical point of view, stars with such temperatures and metallicities are expected to preserve their initial lithium content, moreover galactic enrichment due to cosmic rays and spallation processes should be totally negligible at such low metallicities. Thus the Spite plateau is expected to represent the primordial lithium content produced during the big bang nucleosynthesis (BBN). 

However, BBN predicts a primordial lithium content of $A($Li$) = 2.67$-$2.75$, (see e.g. \citet{cyburt16,pitrou18}). This estimate depends on the density of baryons, which is related to the fluctuations of the cosmic microwave background measured with WMAP and Planck satellites. The BBN predictions for the primordial lithium abundance are thus 0.3-0.6~dex larger than the Spite plateau value. This discrepancy constitutes the so called "cosmological lithium problem''. Several attempts to introduce new physics (exotic particles) or to review the reaction rates during the BBN have been performed, but without any firm conclusion (see e.g. \citet{fields11,pizzone14,gpp16,coc17,damone18,lamia19}). 

Similarly, on the stellar evolution side, the problem has been analysed to find a possible mechanism to deplete the same lithium amount  for the stars with different masses and metallicities which populate the Spite plateau. Diffusion has been investigated as a possible solution, as it slowly brings surface lithium below the convective region (\citet{richard05}). This process acts on timescales typical of the MS evolution, but its efficiency depends on the stellar mass and thus in the mass range corresponding to the Spite Plateau the effect of diffusion increases with \teff. Thus no Spite plateau would be possible without tuning the diffusion efficiency. Also turbulent mixing could produce an effect similar to that of pure diffusion, on similar time scales, but also in this case an ad hoc tuning is required (see e.g. \citet{richard05,spite12} and references therein). Also mass loss coupled to diffusion and turbulent mixing can be tuned to produce a constant lithium depletion along the Spite plateau (\citet{swenson95,vauclair95}), but, again, there is the need for a fine tuning of the parameters.

Another possibility is that lithium depletion occurs during the pre-MS. In this case, \citet{fu15} suggested that a certain level of undershooting\footnote{The term "undershooting" indicates a sink of the external convection toward the stellar interior larger than the one predicted in classical models i.e. by the Schwarzschild criterion.} at the bottom of the convective envelope of pre-MS stars could increase the depletion of surface lithium. In addition, a residual matter accretion, regulated by the stellar luminosity, could provide gas with pristine chemical composition (and thus lithium to the star), obtaining in pre-MS the depletion level required to produce the Spite plateau. However, in such models, MS diffusion must be inhibited to avoid a \teff{} (or mass) dependent depletion on MS time scales. 

Recently \citet{tognelli20} analysed the possibility to produce a constant lithium depletion in pre-MS taking into account in the models the protostellar evolution with different accretion parameters. As discussed in Sec.s~\ref{protostarandPMS} and \ref{Liprotostar}, depending on the scenario adopted during the protostellar evolution, stars at the beginning of the pre-MS can be profoundly different from the ones evolved starting from the Hayashi track. The reason is that the protostellar phase can deeply affect the thermal structure of a star. As a result, it is possible to induce a lithium depletion in pre-MS or even during the protostellar phase, but it requires the adoption of a fine tuning of the parameters that govern the the stellar mass accretion (see e.g. \citet{fu15,tognelli20}). Moreover, as already discussed, the models that show a significant Li depletion, follow a pre-MS evolution in the HR diagram that is different to that observed for high metallicity pre-MS stars. The lack of Galactic very young and metal poor stellar systems, in which one could observe pre-MS stars with Spite plateau metallicities, avoid the possibility to restrict the range of valid accretion parameters and reach firm conclusions.

To conclude, the proposed mechanisms could in principle alleviate the cosmological lithium problem, but the weakness of all these suggested solutions is that a fine tuning of the parameters is still required to produce a constant lithium depletion reproducing the Spite plateau. \\
\\
\\

\subsection{Uncertainties on predicted surface lithium abundance}
\label{Liuncertainties}
The predicted depletion of surface lithium abundance (and, in general, of the light element surface abundances) is affected by the uncertainties on the input physics adopted in stellar models and on the assumed chemical composition, that influence the extension of convective envelope and temperature structure of the star. In particular, also the uncertainty on the nuclear burning cross sections play a role in the light element abundance predictions. In the literature there are several attempts to estimate the impact of the uncertainties on the parameters/input physics adopted in stellar models on the predictions of lithium abundance in low and very-low mass stars (see e.g. \citet{piau02,burke04,tognelli12,tognelli15b}). The quantities that mainly affects lithium, as analysed in the quoted papers, are the following (see e.g. Chapter~3 in \citet{tognelli13}):

\begin{itemize}
\item \emph{Radiative opacity and equation of state}. The extension of the convective envelope is determined by the Schwarzschild criterion which simply states that a region is convective if the radiative temperature gradient ($\nabla_{rad} \equiv (d\log T/ d\log P)_{rad}$) is larger than the adiabatic one. The radiative gradient is proportional to the Rosseland mean radiative opacity $\kappa_R$, thus  change in $\kappa_R$ directly affects the position of the convective unstable  boundary. Generally an uncertainty on $\kappa_R$ of about $\pm 5\%$ is assumed in the computations (\citet{badnell05,blancard2012,mondet2015}). Similarly an uncertainty on the adiabatic gradient, $\nabla_{ad}$, thus in the equation of state, modifies the position of the bottom of the convective envelope.  An increase in $\kappa_R$ or a decrease of $\nabla_{ad}$ lead to an extension of the convective envelope that can reach deeper and hotter layers, increasing the efficiency of surface lithium burning (see e.g. \citet{tognelli13}). The variation of surface lithium abundance, due to the change in the equation of state or radiative opacity, strongly depends on the selected mass range and age. However, in those models that efficiently deplete lithium (e.g. for 0.7 and 0.8~\msun) a variation in lithium abundance of approximately 0.1-0.2~dex (due to equation of state uncertainty) and 0.4-0.5~dex (due to opacity error) can be obtained. In the worst cases (i.e. $M\sim $0.6~\msun) a variation of $5\%$ of the radiative opacity can lead to a difference of $\sim 0.8$~dex in the predicted lithium content. The effect on surface lithium of the equation of state or radiative opacity reduces as the mass increases. 

\item \emph{Outer boundary conditions}. The outer boundary conditions are the pressure $P_{atm}$ and temperature $T_{atm}$ at the bottom of the atmosphere. These quantity deeply affect the temperature profile in the convective envelope thus modifying also its depth. The uncertainty on $(P_{atm},\,T_{atm})$ it is not provided by stellar atmospheric calculations, but one can test the effect on the stellar characteristics of the adoption of different atmospheric models available in the literature (see e.g. \citet{tognelli11}). As said above, the effect on lithium depends on the mass/age of the models; a typical variation of 0.3-0.5~dex is expected, which reduces as the mass increases. 

\item \emph{Mixing length parameter}. The convection efficiency in super-adiabatic regimes in 1D stellar evolution codes commonly relies on mixing-length theory (MLT) \citet{bohm58}. In this formalism, the scale on which the heat is efficiently transported by convection is defined as $\ell = \alpha_\mathrm{ML} \times H_p$, where $H_p$ is the local pressure scale height and $\alpha_\mathrm{ML}$ is the mixing length parameter, a free parameter to be calibrated. The extension of the convective envelope and thus the temperature at its bottom and the the surface lithium abundance are strongly affected by the adopted $\alpha_\mathrm{ML}$. The adoption of different values of this quantity within plausible ranges can produce a variation of surface lithium abundance as large as 1 order of magnitude in those stars where the external envelope is largely super adiabatic and where lithium is efficiently destroyed at the bottom of the convective envelope (i.e. for masses in the range [0.5, 1.0]~\msun{} (see e.g. \citet{piau02} and \citet{tognelli12}). 

\item \emph{Nuclear cross section}. The error on the cross section for the $^7$Li(p, $\alpha)^4$He reaction directly affects the rate at which lithium is destroyed and thus its temporal evolution. Since the energy released by such reactions is inconsequential for the stellar structure, the only effect is on the surface lithium content at a fixed age. In Section~\ref{reactionandelements} we will discuss this point in more detail. 

\item \emph{Chemical composition}. The initial abundance of helium ($Y$) and metals ($Z$) in the star is not known, but it can be estimated from the observed [Fe/H], assuming for metal rich stars the same relative abundance of metals of the Sun, while for metal poor galactic stars a suitable alpha-enhancement must be introduced \footnote{With \emph{alpha enhancement} one indicates an enhancement of the relative abundance of $\alpha$ elements (C, O, Ne, Mg, Si, S, Ar and Ca) with respect to the solar composition. It is generally expressed as: [$\alpha$/Fe]=$\log ($N$_{\alpha}$/N$_{Fe})_\mathrm{star}$ - $\log ($N$_{\alpha}$/N$_{Fe})_{\odot}$.}. The conversion of [Fe/H] into $Y$ and $Z$ depends on the assumed values of : 1) the primordial helium mass fraction produced in the BBN ($Y_p$), 2) the metal-to-helium enrichment ratio ($\Delta Y / \Delta Z$), 3) the metal-to-hydrogen ratio in the Sun ($(Z/X)_{\odot}$), 4) and the [$\alpha$/Fe] (alpha-enhancement) for metal poor stars (\citet{gennaro12,tognelli12,tognelli15b}). The observational error on [Fe/H] has thus to be combined with the uncertainties of such quantities, to estimate the final global uncertainty on the initial helium and metal mass fraction to be used in the computation of stellar models (see e.g. \citet{gennaro12,tognelli12,tognelli15b}); for solar chemical composition, the uncertainty on $Y$ and $Z$ are estimated to be of the order of 4-5\% for $Y$ and about 20\% for $Z$ (\citet{tognelli15b}). The variations of $Y$ and $Z$ have a strong impact on the lithium burning because they change both the extension of the convective envelope and the temperature inside a star (see e.g. \citet{piau02}, \citet{tognelli13} and \citet{tognelli15b}); the uncertainty on the chemical composition can produce a variation of the surface lithium abundance up to one order of magnitude, especially in stars with $M\la 0.7$~\msun. The effect reduces at larger masses. 
\end{itemize}

\begin{figure}[t]
\centering
\includegraphics[width=8.5cm]{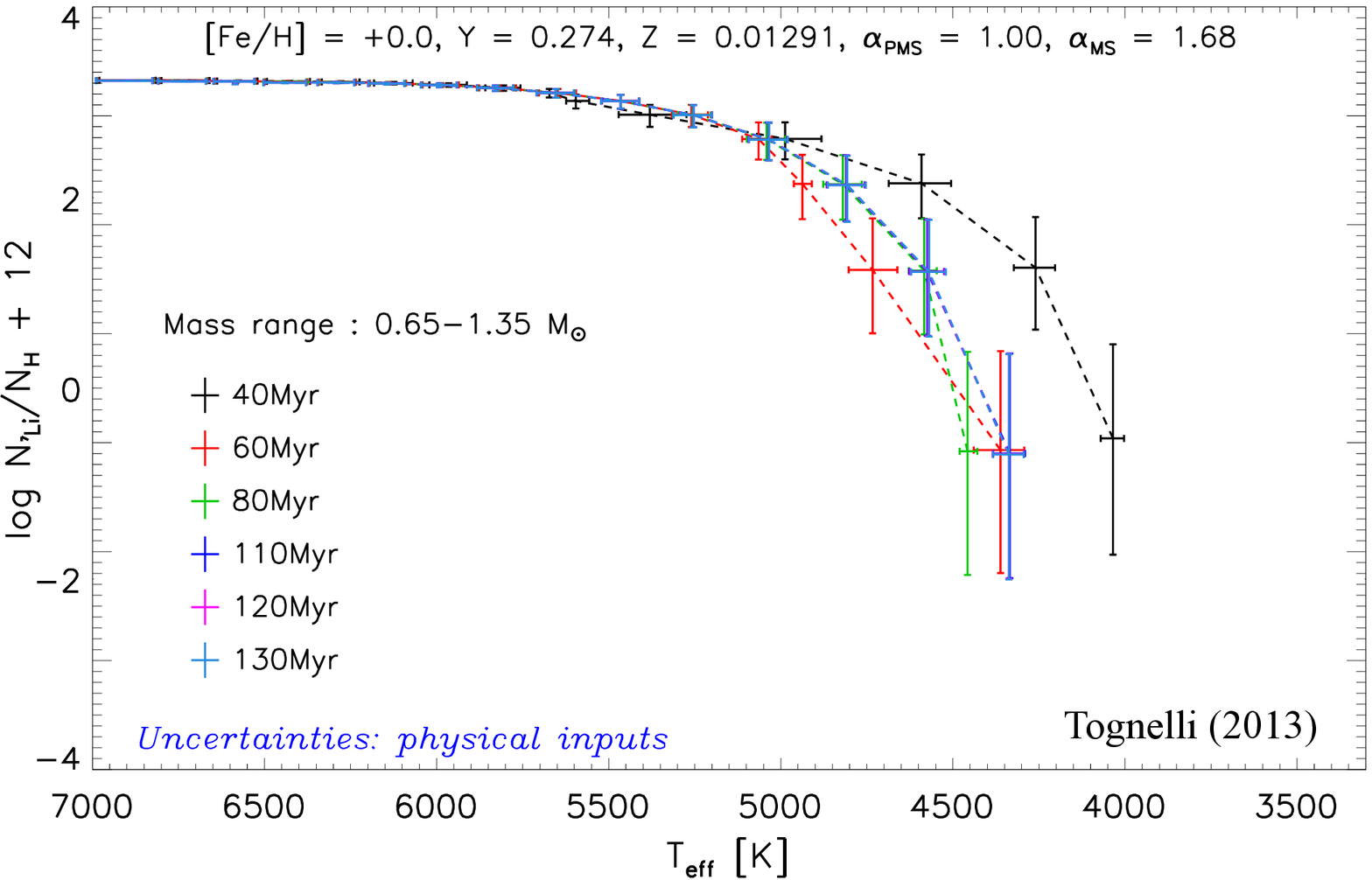}
\includegraphics[width=8.5cm]{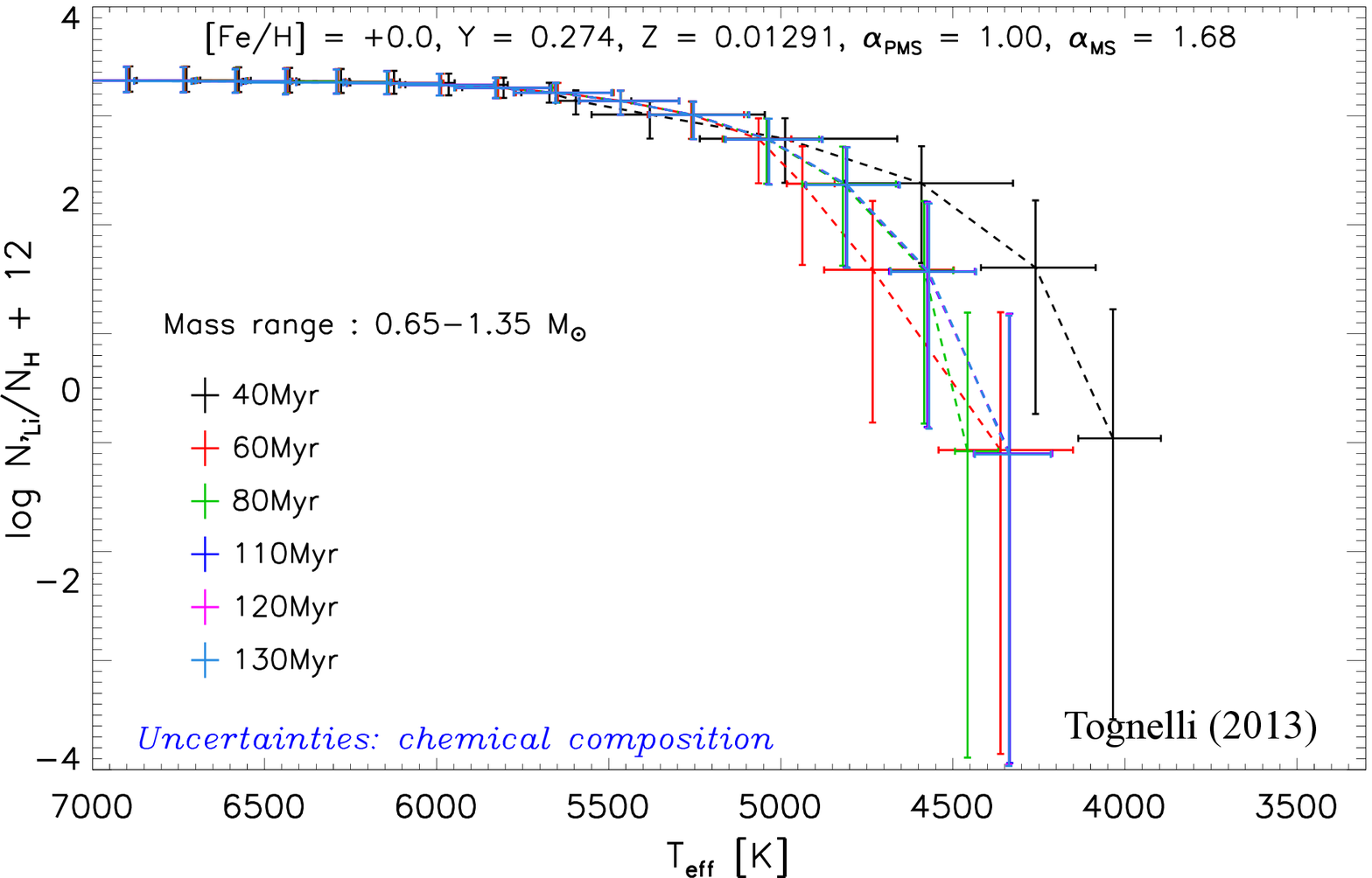}
\caption{Uncertainties on surface lithium abundance and effective temperature due to the errors on adopted input physics (left panel) and chemical composition (right panel). Figure adapted from \citet{tognelli13}.}
\label{fig:unc_li}
\end{figure}

\citet{tognelli13} quantitatively evaluated the impact on the predicted surface lithium abundance of the uncertainties in the input physics and in the initial chemical composition discussed above, calculating upper and lower limits of surface $^7$Li in stellar models. Figure~\ref{fig:unc_li} shows the estimated upper/lower limits (plotted as error bars) of surface lithium abundance and effective temperature, due to the contribution of the input physics uncertainties (top panel) and chemical composition indeterminacy (bottom panel). Stars with different masses at different ages typical of young clusters are shown (for more details on the procedure adopted to obtain these limits see \citet{tognelli12} and \citet{tognelli13}). The errors on the present input physics and the typical uncertainties on the adopted chemical composition have a drastic impact on the predicted surface lithium abundance, which can vary by more than 1 order of magnitude, especially for stars with \teff$\la 4700$~K.

In standard models, the only possibility to deplete surface lithium in pre-MS is via convective mixing. If the bottom of the convective envelope reaches a region hot enough to burn lithium, then the surface lithium decreases in time. The level of depletion depends on a key parameter in convective stars, which is the efficiency of convective energy transport. A more efficient convective transport produces hotter stars that consequently experience a  more efficient lithium burning. The opposite occurs if the convection efficiency reduces. 

A precise physical treatment of external convection would require three-dimensional hydrodynamic models which have been improved in recent years, but only for limited regions of the star corresponding mainly to the atmospheric regions (see e.g. \citet{nordlund09}, \citet{collet11}, \citet{freytag12}, \citet{magic13}, \citet{trampedach13}, \citet{trampedach14}, \citet{trampedach15}, \citet{pratt16}, and references therein). These codes are state-of-the-art (magneto) hydrodynamic code that solves the time-dependent hydrodynamic equations for mass, momentum, and energy conservation, coupled with the 3D radiative transfer equation, in order to correctly account for the interaction between the radiation field and the plasma. However, hydrodynamic calculations still cannot cover the wide range of physics quantities needed to model the Galactic stellar populations. Moreover, their results cannot be easily adopted in stellar evolutionary codes, although attempts to implement approximations directly based on 3D simulations in 1D stellar models exist in the literature (e.g. \cite{lydon92},  \citet{ludwig99}, \citet{arnett15,arnett18}).  The commonly adopted procedure to treat the convection efficiency in super-adiabatic regimes in 1D stellar evolution codes relies on mixing-length theory (MLT) \citet{bohm58}, where convection efficiency depends on the free parameter $\alpha_\mathrm{ML}$. A variation of such a parameter can produce a large effect on the surface lithium abundance at a given age in stars with a super-adiabatic envelope. This effect is particularly important in stars with masses larger than about 0.5-0.6~\msun. It has been shown in the literature that models with a reduced convection efficiency (\ml $< $ \ml$_{\odot}$, solar calibrated value) attains a better agreement with data for both young clusters and binary stars (\citet{piau02,dantona03,tognelli12}). 
 
\begin{figure}
\centering
\includegraphics[width=0.92\linewidth]{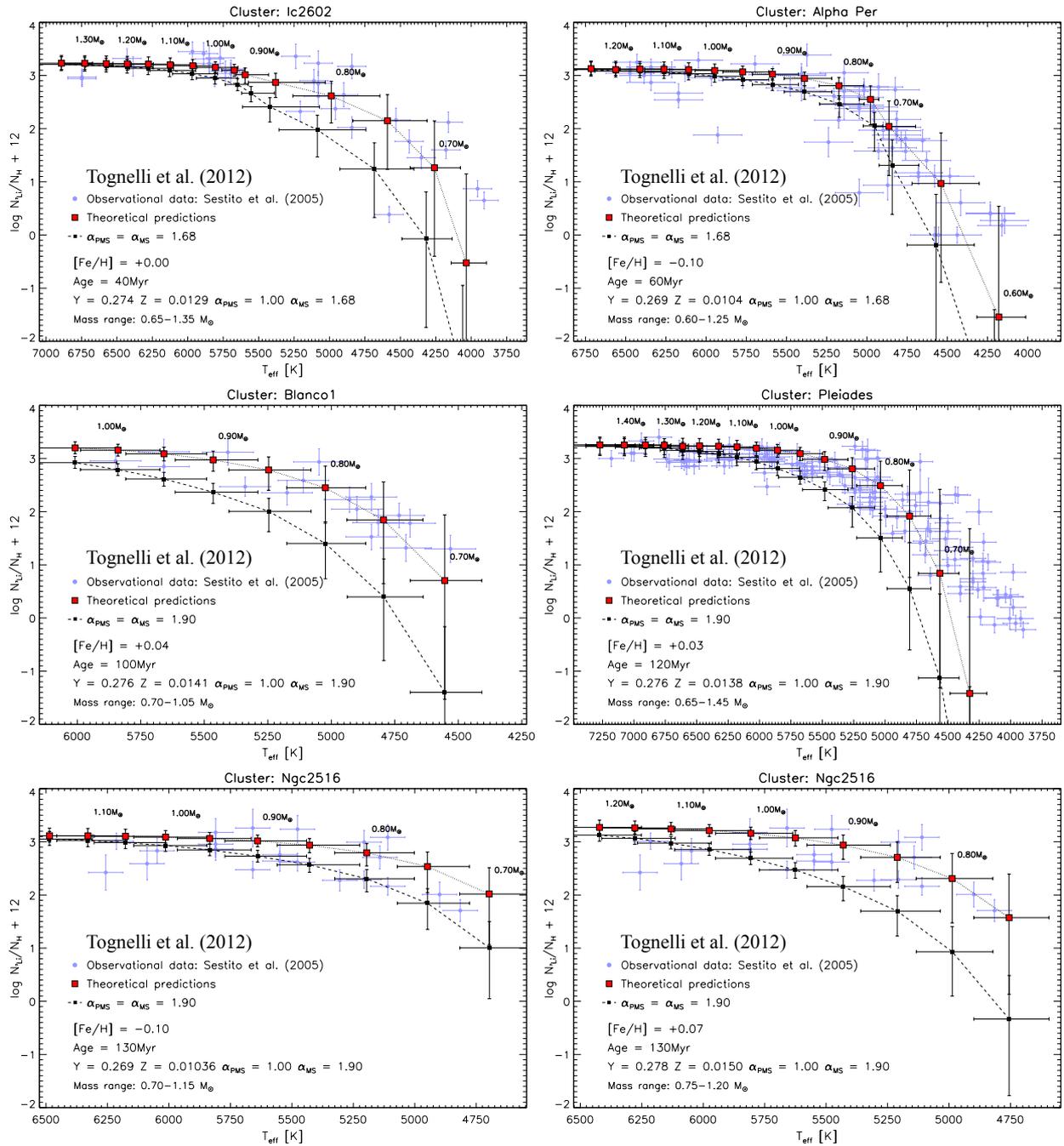}
\caption{Comparison between data and theoretical model predictions for surface lithium in young clusters. Models with the same mixing length parameter in pre-MS and MS phases, calibrated with MS stars, are shown as dashed lines while models with calibrated convection efficiency in MS and artificially reduced mixing length parameter (\ml=1) in pre-MS are shown as continuous lines. In the case of NGC2516, the models were computed using two different values of [Fe/H], [Fe/H] = -0.10 (bottom left panel) and [Fe/H] = $+0.07$ (bottom right panel), as reported in the literature. Figure adapted from \citet{tognelli12}.}
\label{fig:tog12_li}
\end{figure}

Figure~\ref{fig:tog12_li} shows the results obtained by \citet{tognelli12} where the observed surface lithium abundance in five young open clusters (IC2602, $\alpha$~Per, Blanco1, Pleiades and NGC2516) has been compared to theoretical predictions obtained adopting two different values for the mixing length parameter during the Pre-MS phase: one calibrated on MS stars to reproduce their colours and the other corresponding to a much less efficient convective energy transport. The best value to be used in pre-MS has been estimated in order to reproduce the $A($Li$)$ vs \teff{} pattern. The availability of young clusters  to perform such an analysis (ages below 100-150~Myr) is mandatory to avoid possible effects due to MS non-standard mixing processes which act on timescales of the order of $\sim$~Gyr.
Referring to Fig.~\ref{fig:tog12_li}, it is evident that in all cases, the adoption of a constant value of \ml{} (calibrated on MS stars) produces a lithium depletion much larger than observed. On the other hand it is possible to tune \ml{} during the pre-MS to reproduce the observed lithium pattern and the most important result is that the derived \ml{} in pre-MS is independent of the cluster age and on the stellar mass. The authors derived a value of \ml$_\mathrm{,PMS} = 1$. We mention that the reduction of the efficiency of superadiabatic convection in stellar envelope has been put forward as a plausible mechanism to explain the discrepancies in the radius observed and predicted in pre-MS binary systems. A value of \ml$\sim 1$ has been suggested to explain the radius in young binary systems (see e.g. \citet{gennaro12}). 

\begin{figure}[t]
\centering
\includegraphics[width=0.96\linewidth]{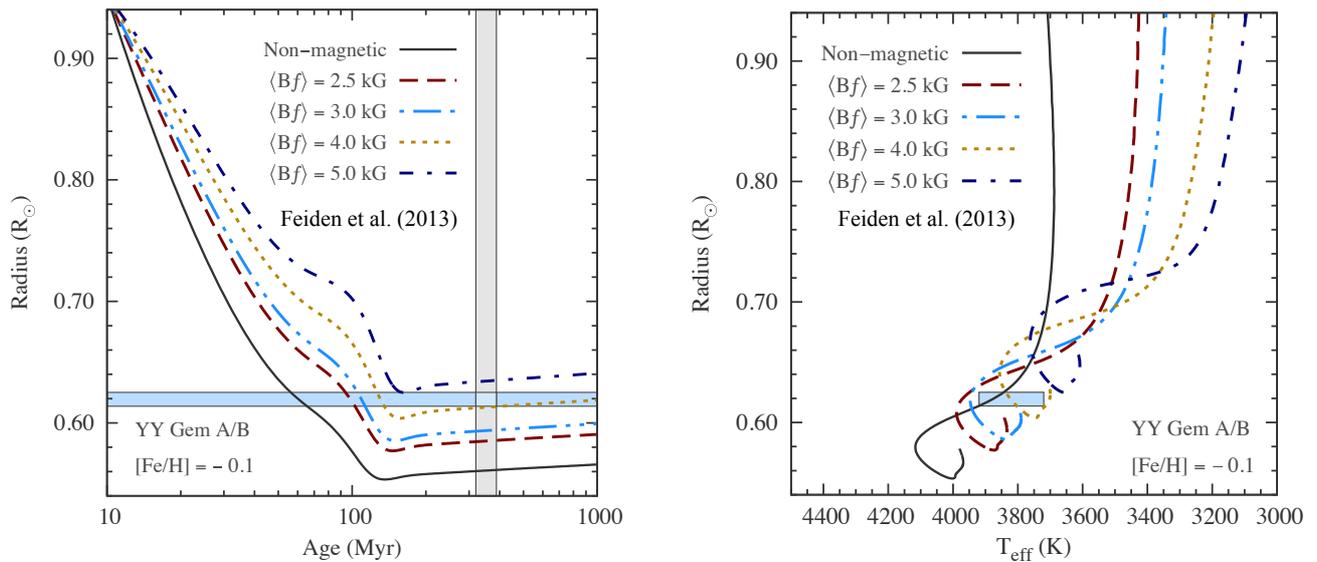}
\caption{Effect of magnetic fields (for the labelled surface magnetic field strength $<$B$f>$) on the radius evolution of a partially convective pre-MS star ($M=0.599$~\msun{}). The radius is compared to that of the observed YY Gem binary system (horizontal and vertical stripes). Left panel: radius vs age. Right panel: stellar radius vs \teff{} for magnetic and non magnetic models. Figure adapted from \citet{feiden13}.}
\label{fig:feid13}
\end{figure}

As said, \ml{} is a free parameter which reflects the present not precise knowledge of external convection efficiency, thus one should find a physical reason for its variation during the evolution of a given star. A reduced convection efficiency could be motivated by the attempt in 1D evolutionary codes to mimic the main effects of some non-standard mechanisms active in young stars, such as the presence of a not negligible magnetic field (especially in the convective region, see e.g.  \citet{ventura98}). To this regard \citet{feiden13} found that the inclusion of a magnetic field in partially convective stars  produces a radius expansion caused by the inhibition of convection efficiency in the convective envelope (see also \citet{feiden2012a}). Figure~\ref{fig:feid13} shows the results of their work on evolutionary models computed with and without the inclusion of a magnetic field. The models are compared to the characteristics of YY~Gem binary system (both stars have masses of about 0.6~\msun, \citet{torres2002}). Such a system exhibit evidences of a relatively strong magnetic field (surface spots, X-ray, gyrosynchrotron radio emissions, flaring events). Standard models underestimate the radius of the components by about 8\%, a difference that can be erased if a magnetic field of 4-5~KG is included in the computations. The stronger is the magnetic field the larger the radius of the star at the same age (left panel). In the radius vs \teff{} plane, the inclusion of a magnetic field produces a cooler and brighter star (see right panel of Fig.~\ref{fig:feid13}). \citet{feiden13} also showed that it is possible to reproduce the main effects of a magnetic field in 1D non-magnetic stellar models by using a properly tuned value of the mixing length parameter. To do this an $\alpha_{ML}$ value lower than the solar calibrated one (i.e. close to unity) should be adopted.

\begin{figure}
\centering
\includegraphics[width=10cm]{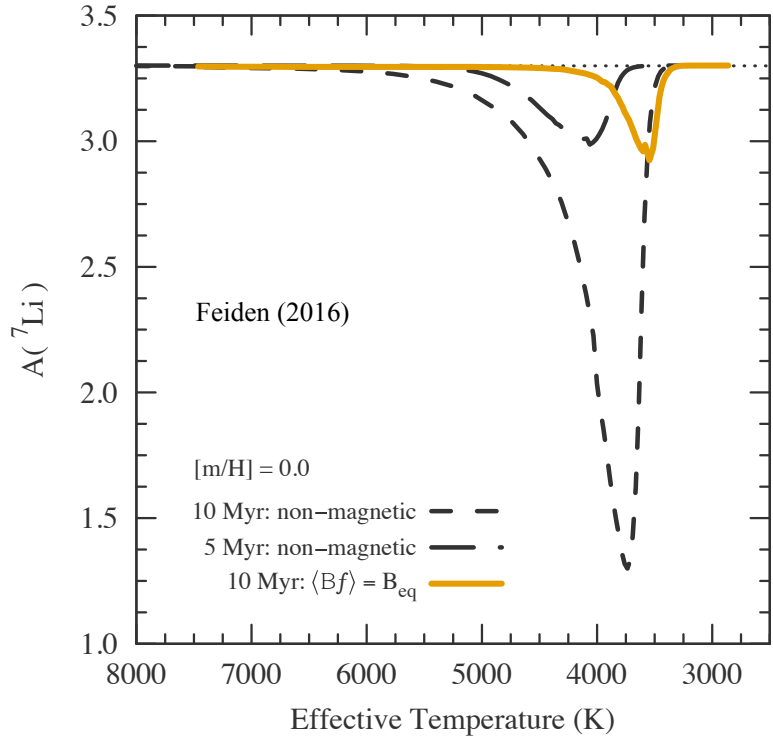}
\caption{Effect on the surface lithium abundance of the presence of a magnetic field (B$_\mathrm{eq}$, see text). The surface lithium abundance behaviour as a function of the effective temperature is shown for standard (non magnetic) models of 5 and 10 Myr compared with a model of 10 Myr in which a magnetic field is included. Figure adapted from \citet{feiden16}.}
\label{fig:feid16}
\end{figure}

The presence of a magnetic field makes the star cooler and modifies the temperature stratification inside the star: this has a direct impact on surface lithium burning efficiency (\citet{feiden16}). Figure~\ref{fig:feid16} shows the comparison between the expected surface lithium abundance as a function of \teff{} in standard non magnetic models of 5 and 10~Myr compared to a model of 10~Myr in which a magnetic field is included (the magnetic field strength B$_\mathrm{eq}$ shown in figure varies with the mass but in the mass range 0.1-1~\msun{} it is of the order of 2-2.6~KG). The inclusion of the magnetic field has a strong impact on the resulting lithium abundance, drastically reducing the level of depletion and thus pointing in the direction to improve the agreement between data and model predictions for pre-MS stars.


\begin{figure}[t]
\centering
\includegraphics[width=10cm]{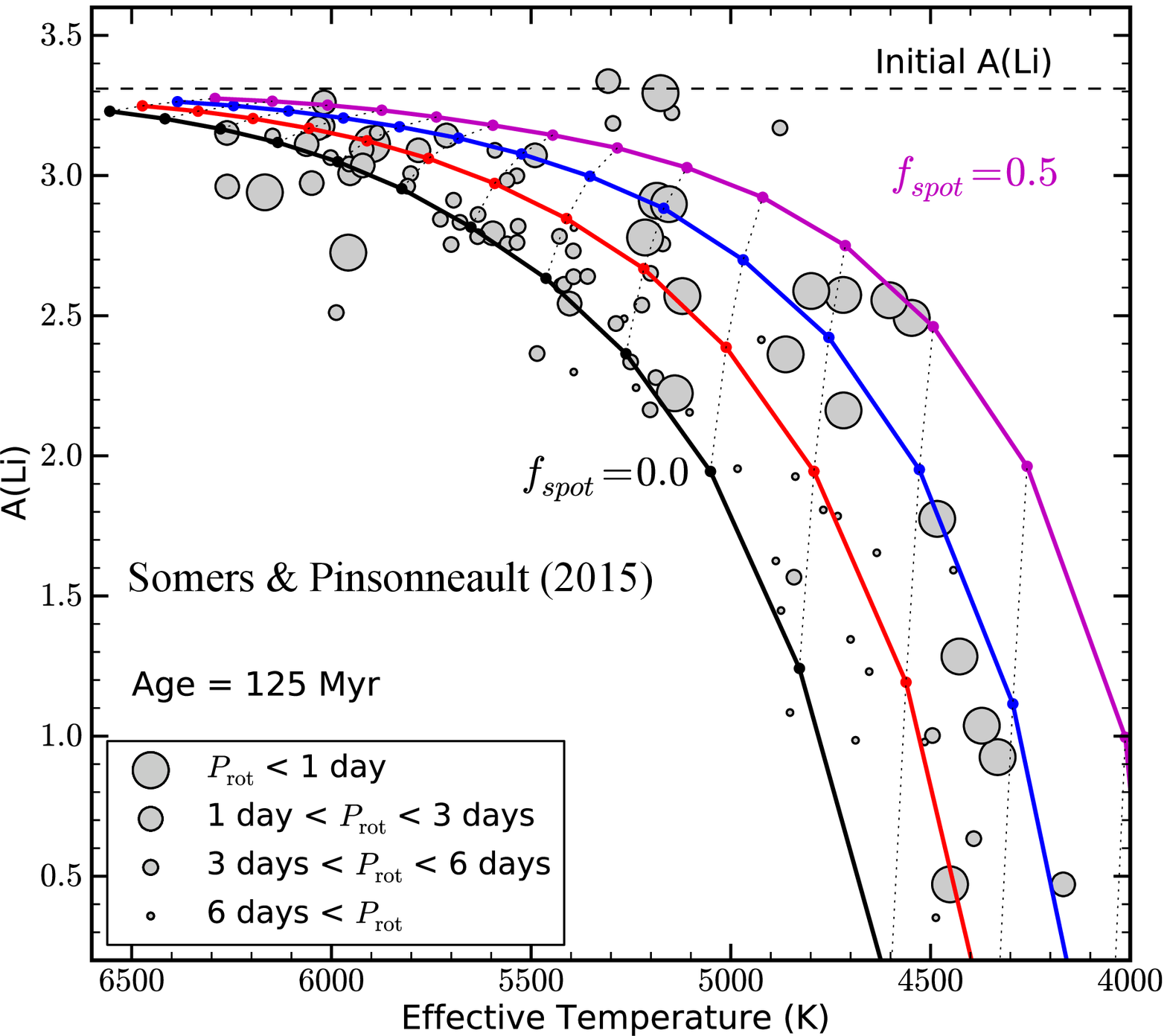}
\caption{Comparison between Pleiades data and predicted surface lithium abundance obtained in models that account for surface spots for several values of the spot coverage fraction $f_\mathrm{spot}$. Dotted lines connect models with the same mass and different values of $f_\mathrm{spot}$. The dimension of the circles is proportional to the rotation velocity of the star. Figure adapted from \citet{somers2015}.}
\label{fig:som15}
\end{figure}

Another aspect related to the presence of magnetic field in the star is the possibility to include in stellar models a surface spots coverage fraction (see e.g. \citet{jackson14}     and \citet{somers2015}). The effect of the spots is to reduce the outgoing flux at the stellar surface producing a radius inflation and a decrease of the stellar effective temperature. Such an effect goes in the same direction of an artificially decreased convection efficiency and, as expected, leads to a cooler envelope and to a less efficient lithium burning. Figure~\ref{fig:som15} shows an application of stellar models with surface spots to the surface lithium abundance observed in the Pleiades cluster (see \citet{somers2015}). Standard models predict a level of pre-MS lithium depletion larger than that observed. The agreement can be restored assuming that a certain fraction of the stellar surface is covered by spots; increasing the coverage fraction, the models are cooler and thus the surface lithium depletion is reduced.
It's important to notice that the presence of magnetic fields of different strength or a different spot coverage fraction in stars with similar mass can introduce a star-to-star scatter in the surface lithium abundance. This could partially answer another important open question about young clusters, i.e. which is the cause of a spread in the lithium abundance, measured in stars with similar effective temperature. The extent of such spread is much larger than the quoted uncertainties, so it represents a real spread (see e.g. \citet{xiong06}). The inclusion of a different surface spots coverage or different magnetic fields strength could produce stars with similar effective temperatures (but different total masses) thus leading to an apparent dispersion in the lithium abundance.

Additional mechanisms that can alter the level of lithium burning in stars have been analysed in the literature. An induced extra mixing due to the presence of rotation, gravity waves, diffusion, or mass loss has been put forward to reproduce the surface lithium abundance pattern typical of older stars (ages $\ga 500$~Myr). However, such mechanisms are not relevant for the evolution of young pre-MS stars and thus we will not discuss them in this context (see e.g. \citet{charbonnel13}, \citet{eggenberger12}, \citet{landin06}) \\

\subsection{Uncertainties on atmospheric models for surface lithium abundance determination}
\label{Liatmouncertainties}

The determination of surface element abundances involves the interpretation of the observed absorption lines through atmospheric models as accurate as possible. However modelling stellar atmospheres is a difficult task and the uncertainties on the measurements of surface element abundance clearly affects the comparison between theory and observations. Here we limit to briefly discuss the main difficulties in modelling realistic stellar atmospheres, the interested reader can find more details in other reviews (see e.g. \citet{paula2019}).

The photosphere of low mass stars is covered with a complex and stochastic pattern -- associated with convective heat transport -- of downflowing cooler plasma and bright areas where hot plasma rises, the so called granulation (\citet{nordlund09}). As already discussed, convection is a difficult process to understand, because it is non-local and three-dimensional, involving non-linear interactions over many disparate length scales. In recent years it has become possible to use numerical three-dimensional (3D) radiative hydrodynamical (RHD) codes to study stellar convection in atmosphere such as {\sc Stagger Code} (\citet{collet11,nordlund09}) and CO5BOLD (\citet{freytag12}). Nowadays, the use of large grids of simulations covering a substantial range of values of effective temperature and surface gravity for stars in different regions of the HR diagram (\citet{ludwig09,magic13,trampedach13}) have proven that the convection-related surface structures have different size, depth, and temporal variations, depending on the stellar type (\citet{tremblay13,beeck13,magic14}). Moreover, the related activity (in addition to other phenomena such as magnetic spots, rotation, dust, etc.) has an impact in stellar parameter determination (\citet{bigot11,creevey12,chiavassa12}), radial velocity (\citet{allende13,chiavassa18}), chemical abundances estimates (\citet{asplund09,caffau11}), and photometric colours (\citet{chiavassa18,bonifacio18}). 

Chemical abundance ratios inferred from spectra of cool stars is based on the understanding of limitations and uncertainties of spectroscopic analyses. In this context, radiation transfer in the atmospheres of late-type stars generally takes place under non-local thermodynamic equilibrium (NLTE) conditions, rather than the idealised LTE (\citet{asplund05b}). 
The full 3D NLTE treatment would require to compute NLTE radiative transfer inside radiative hydrodynamical simulations and coupling it to the equations of gas movements. In these simulations the computational cost is dominated by the radiative transfer calculations which can be greatly  reduced by adopting an approximated solution based on the opacity binning or multi-group method (\citet{nordlund82}). However, introducing NLTE calculations at this stage would largely increase the computation time making very complicated to obtain a relaxed simulation. This is why 3D NLTE radiative transfer calculations are only affordable in a post-processing manner, i.e., each 3D RHD simulation is computed in LTE and then the so called $<$3D$>$ models are computed averaging multiple snapshots of 3D RHD simulations over regions of equal optical depth and over the time series (\citet[e.g.,][]{asplund04,caffau09,lind17,nordlander17,magic13,amarsi18}). This approach offers a middle-ground between full 3D NLTE and 1D NLTE, by accounting for NLTE in model atoms of arbitrary size, and through the use of time-independent 1D structures derived from the full 3D hydrodynamic simulations (\citet{bergemann17}). Using this method, \citet{wang2021} derived a new 3D NLTE solar abundance of A(Li) = 0.96 $\pm$ 0.05, which is 0.09 dex lower than the commonly used value and provided a grids of synthetic spectra and abundance corrections publicly available. Eventually, it has also become possible to undertake large samples of observations from disk and halo stars with this 3D NLTE analysis (\citet{amarsi19,bergemann19}). Unluckily at present 3D atmospheric calculations are not still available for Pre-Main Sequence atmospheres. 
         
The measurement of surface lithium abundances constitutes an important example of efforts undertaken in this field. In Sec.\ref{elementabundances} we mentioned that the stellar surface abundance of $^6$Li is expected to be negligible, moreover its identification is very difficult. The presence of $^6$Li in metal-poor halo stars can only be derived from the asymmetry in the red wing of the $^7$Li doublet line at 670.8~nm. Several authors attempted to detect $^6$Li using 1D hydrostatic models and assuming LTE for a number of metal-poor stars with [Fe/H], lower than $-2$ dex (\citet{cayrel99,asplund06}). \citet{cayrel07} pointed out that the intrinsic line asymmetry -- due to the stellar surface convection -- in the $^7$Li doublet would be almost indistinguishable from the asymmetry produced by a weak $^6$Li blend on a (presumed) symmetric $^7$Li profile. 

\begin{figure}
\centering
\includegraphics[width=0.48\linewidth]{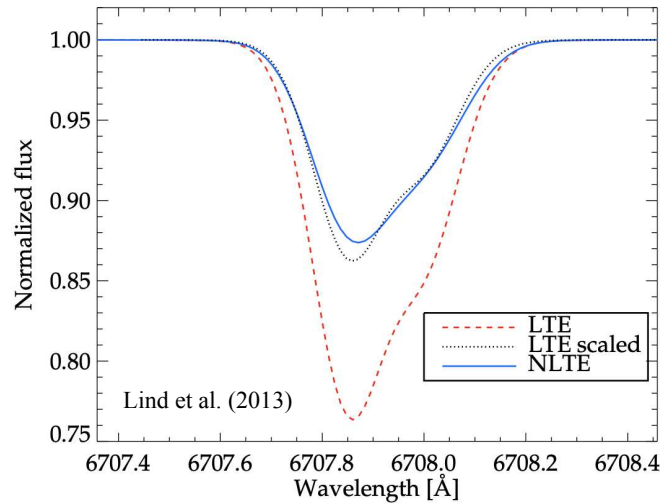}
\caption{Example of synthetic stellar surface profiles of the $^7$Li doublet line at 6708.0 \AA , computed from 3D RHD simulation of metal-poor stars with an LTE (dashed) and NLTE (solid) approach. The $^7$Li-abundance is the same. The LTE line profile interpolated to meet the same equivalent width as the NLTE line is also displayed (dotted). An additional absorption appears in the red wing relative to the blue for NLTE models due to strongly differential effects in granules and inter-granular regions (\citet{lind13}).}
\label{fig:atmospheres}
\end{figure}

The total line strength of the Li resonance line determines the $^7$Li-abundance and the shape of the line profile determines the isotopic ratio due to the shift between $^6$Li and $^7$Li isotopic components. Thus it's critical to resolve the strongly differential NLTE effects on the granules and inter-granular regions (Fig.~\ref{fig:atmospheres}), because they have a preferential influence over the blue- and red-shifted part of the line profile, respectively (\citet{lind13}). To investigate this aspect, \citet{steffen12} and \citet{lind13} used a 3D NLTE treatment with 3D RHD simulation snapshots carried out with CO5BOLD and {\sc Stagger Code} , respectively. They re-analysed the Li feature in some metal-poor stars and were not able to confirm the previous claimed detection of $^6$Li. However, they pointed out that a full understanding of 3D NLTE line formation is necessary to make correct measurements of $^6$Li, even though from their studies they could give only upper limits for the isotopic ratio $^6$Li/$^7$Li. In particular, the 3D NLTE approach is important to characterise the calibration lines, to decrease the observational error.
Eventually, a very recent publication by \citet{gonzalez19}, confirms the non detection of $^6$Li for a very metal poor binary star ([Fe/H]$\sim-3.7$ dex), finding an upper limit for the isotopic ratio of $^6$Li/$^7$Li$~< ~10\%$. 
\\

\section{Effects of light element burning cross sections on pre-Main Sequence characteristics and on light element surface abundances}
\label{reactionandelements}
The predicted temporal evolution of light elements is affected by the stellar evolutionary stage and by the model structure, which depends on the input physics adopted in the computations. One of the key ingredients is the adopted light element burning cross section, as derived from measurements of indirect/direct processes. Thus, it is worth discussing how the recent determination of such cross sections at energies relevant for stellar evolution have changed the prediction of surface light element abundances in low mass stars.

For stellar calculation the reaction rate of a two body process can be written in the following way (see e.g \citet{rolfsrodney}),
\begin{equation}
    N_A <\sigma v>_b = \sqrt{\frac{8}{\pi \mu}}\frac{N_A}{(K_B T)^{\frac{3}{2}}} \int_0^{+\infty} \sigma(E)_b \, E \, e^{-\frac{E}{K_BT}}\, dE \,\, (\mathrm{cm}^3\,\,\mathrm{mol}^{-1}\,\mathrm{s}^{-1})
    \label{eq:rate}
\end{equation}
where $\sigma(E)$ is the cross section of the process, the subscript $b$ means that the reaction rate is for two bare nuclei (i.e. without any electron screening effect), $T$ is the temperature in Kelvin (K). In stellar evolution calculations, the energy at which the process occurs are generally so small that it is convenient to write the cross section in terms of another quantity called the astrophysical factor $S(E)$ defined as it follows,
\begin{equation}
    S(E)_b = E\, \sigma(E)_b\, e^{2\pi\eta(E)}
\end{equation}
with $\eta(E)$ the Sommerfeld parameter related to the tunnel effect of two interacting charged particles, that can be written as:
\begin{equation}
    2\pi \eta(E) = \frac{Z_1\,Z_2\,e^2}{2\varepsilon_0 \hbar} \sqrt{\frac{\mu}{2E_\mathrm{cm}}} = 31.3998\,Z_1\, Z_2\,\sqrt{\frac{A_\mu}{E_\mathrm{cm}(\mathrm{KeV})}}
\end{equation}
where $\mu = m_1 m_2 /(m_1+m_2)$ is the reduced mass and $A_\mu$ is the same quantity but expressed in atomic mass units (amu), $E_\mathrm{cm}(\mathrm{KeV})$ is the energy in the centre of mass expressed in KeV. 
Using this quantity, eq.~(\ref{eq:rate}) assumes the following form,
\begin{equation}
    N_A <\sigma v>_b = \sqrt{\frac{8}{\pi \mu}}\frac{N_A}{K_B T^{\frac{3}{2}}} \int_0^{+\infty} S(E)_b\, e^{-2\pi\eta(E) - \frac{E}{K_BT}}\, dE \,\, (\mathrm{cm}^3\,\,\mathrm{mol}^{-1}\,\mathrm{s}^{-1})
    \label{eq:rate1}
\end{equation}
For many application in stellar astrophysics, it is possible to expand the astrophysical factor around a specific value of the energy, thus obtaining,
\begin{eqnarray}
S(E) \approx S(E_0) \bigg(1 + \frac{S'(E_0)}{S(E_0)} (E-E_0) + \frac{1}{2}\frac{S''(E_0)}{S(E_0)} (E-E_0)^2 + \dots\bigg )
\end{eqnarray}
The quantity $E_0$ is also known as the Gamow peak energy, and it corresponds to the energy where the exponential quantity inside the integral in eq.~(\ref{eq:rate1}) has its maximum value. $E_0$ is defined in the following way (\citet{rolfsrodney}),
\begin{equation}
E_0 = 1.22 (A_\mu Z_1^2 Z_2^2T_6^2)^{1/3}\,\, (\mathrm{KeV})
\end{equation}
with $T_6$ the temperature expressed in million kelvin.

The expansion of $S(E)$ given above depends on the temperature at which the considered reaction occurs (thorough $E_0$), which in turn depends on the stellar mass. However, at low energy typical of reactions in stars, $S(E)$ varies slowly with the energy, thus it is convenient to expand $S(E)$ around $E\approx 0$: in this case the reaction rate can be evaluated knowing $S(0)$ and its derivatives (usually it is enough to have $S'(0)$ and $S''(0)$).

Light element (p,$\alpha$) reaction rates have been recently revised through the indirect  Trojan Horse Method (THM, see e.g. \citet{baur86,spitaleri03,spitaleri16,spitaleri19} and references therein), which is particularly useful to measure cross sections at astrophysical energies by-passing extrapolation procedure, often affected by systematic uncertainties due, for instance, to electron screening effects. THM allows to measure the astrophysically relevant cross sections in correspondence, or very close, to the Gamow peak without experiencing the lowering of the signal-to-noise ratio due to the presence of the Coulomb barrier between the interacting charged particles. Moreover in the last years THM was successfully applied to reactions induced by unstable beams \cite{pizzone16,lamia19} as well as neutron induced reactions which may play a role also in the context of light element nucleosynthesis and BBN. In particular the $^3$He(n,p)$^3$H was studied at the astrophysically relevant energies (see \cite{Pizzone20} and references therein). THM was also applied to reactions between heavier systems, which are relevant in the late stellar evolutionary stages (\citet{tumino2018}). We will discuss the effects of these improvements and of the present errors on nuclear cross sections on the light elements surface abundance calculations in pre-Main Sequence stars.
\\

\subsection{Effects of deuterium burning cross sections on pre-MS evolution}
As discussed in Sec.s~\ref{PMSevolution}, \ref{acccretiongeometry} and \ref{Dburning}, deuterium burning plays a crucial role in the first stages of pre-MS or protostellar evolution. The value of the cross section of the p(D,$\gamma$)$^3$He process in stellar conditions has been reported by several authors (see \citet{adelberger11} for a review) both from measurements and theoretical calculations along with its uncertainty. \citet{adelberger11} redetermined the best value for the astrophysical factor $S(E)$ for such a reaction at zero energy ($S(0)$) and the uncertainty on it, concluding that the current uncertainty on $S(0)$ for such burning reaction is $\approx 7\%$. Recently, \citet{mossa2020} redetermined the D+p cross sections at energies typical of the BBN (between 32-263 KeV) -- thus larger than those used in stellar calculations -- estimating an uncertainty of about $3\%$.

We tested the impact on pre-MS evolution of the D+p cross section variation, using the uncertainty given by \citet{adelberger11} at stellar energies, which is $\pm7\%$. Such a variation of the deuterium burning reaction rate produces a negligible effect on the stellar structure evolution. The negligible effect is related to the large dependency on the temperature of the p(D,$\gamma$)$^3$He burning channel (about $T^{12}$); if $S(0)$ is artificially varied (e.g. reduced) by 7\% (independently of the temperature), to obtain the same energy production rate, which sustains the star, an increase of the burning temperature is required. However, given the high temperature dependency of the rate, a very small temperature variation is enough to restore the energy production. Thus the models are essentially unaffected by the current uncertainty on the p(D,$\gamma$)$^3$He reaction rate. From this analysis we can conclude that the main uncertainty source on the D-burning phase in stellar models is the error on the initial deuterium abundance which can be as large as 50\% as discussed in Section~\ref{Dburning}.

Recently \citet{tumino14} (see also \citet{tumino2011}) measured the reaction rate for two additional channels involving the D-burning, namely the D(D,p)$^3$H and the D(D,n)$^3$He processes, using the THM; such burning channels could  potentially contribute to the D-burning in stars. 
\begin{figure}[t]
    \centering
    \includegraphics[width=0.485\linewidth]{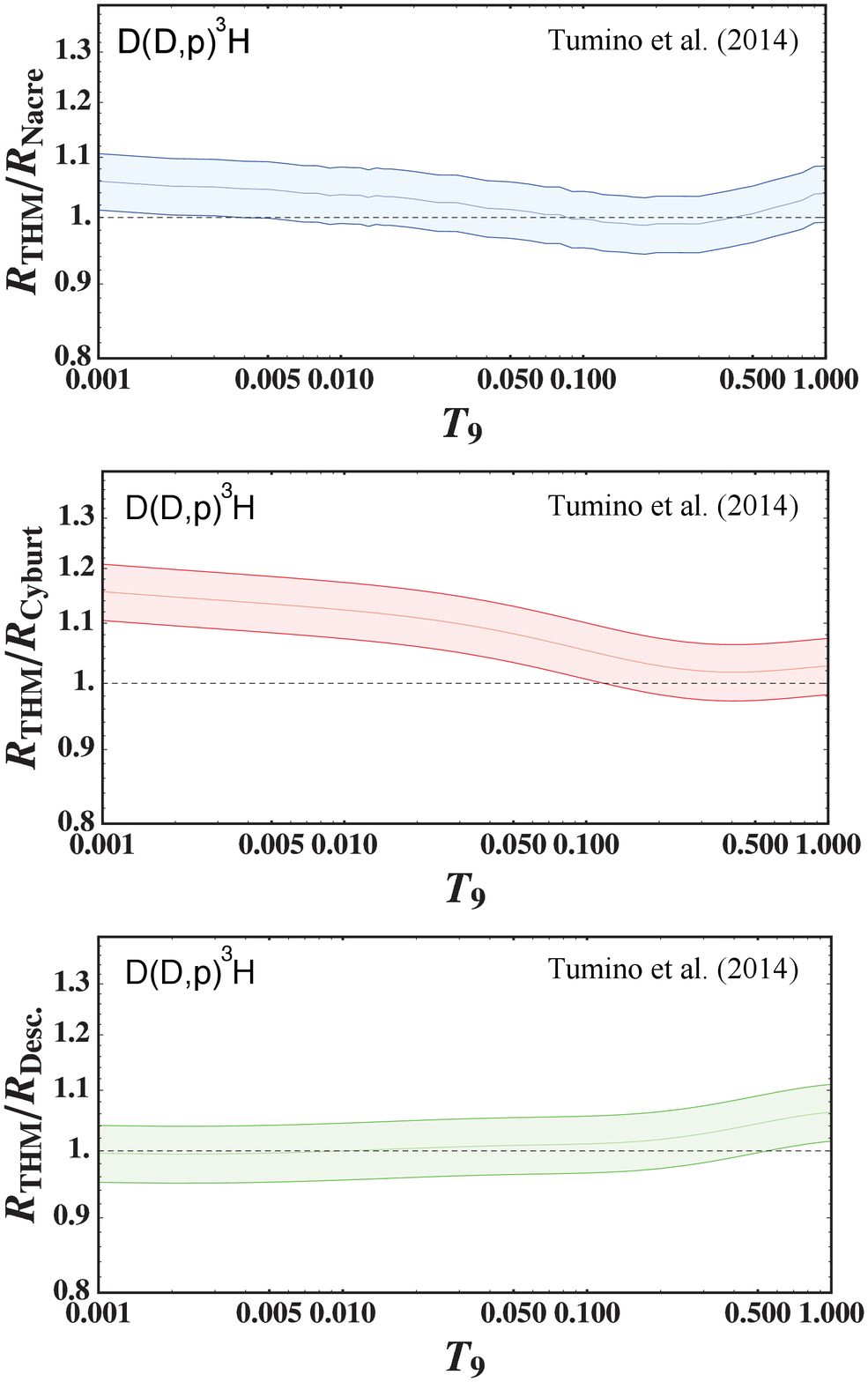}
    \includegraphics[width=0.48\linewidth]{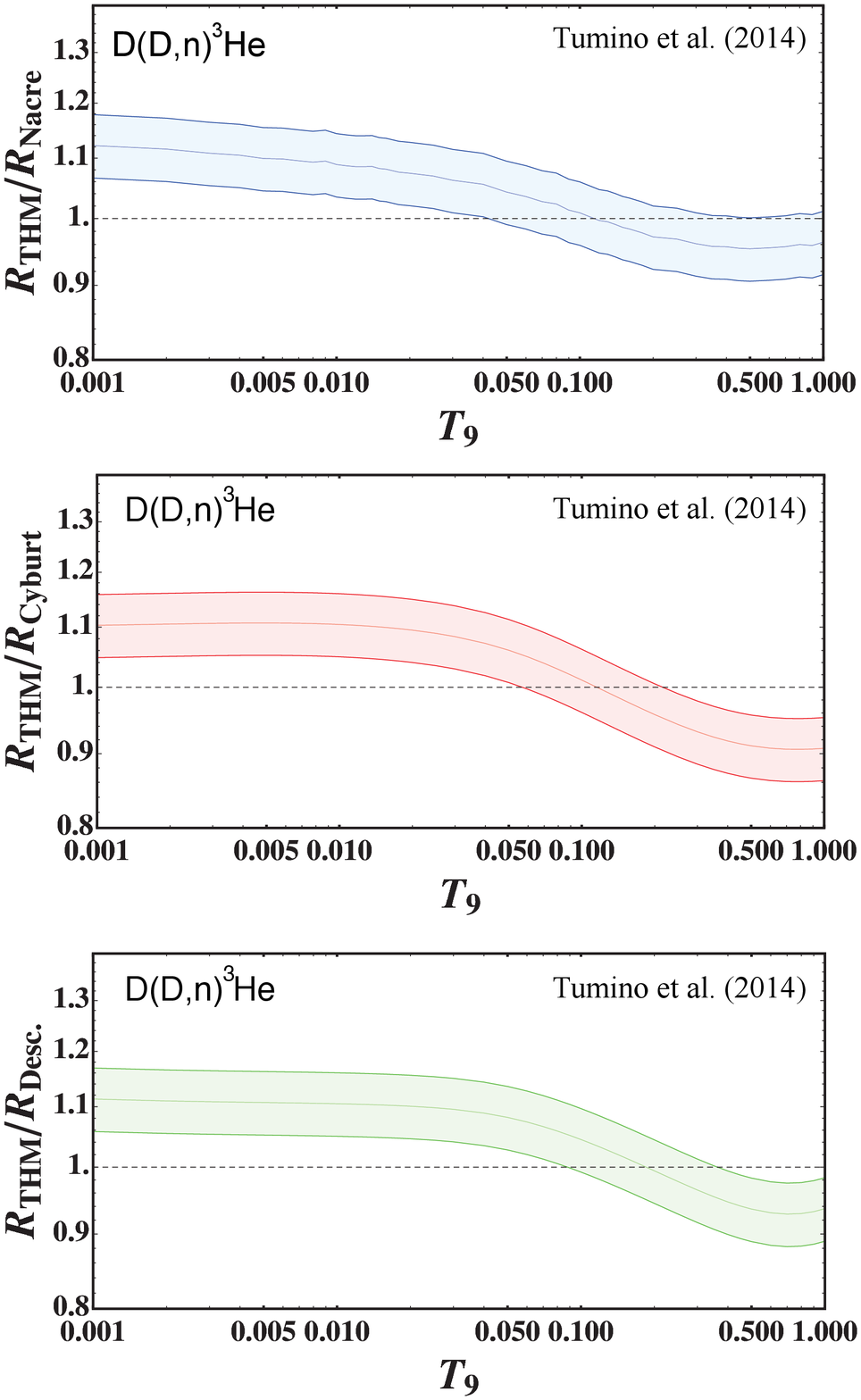}
    \caption{Comparison of the THM rates for the D(D,p)$^3$H (left panels) and D(D,n)$^3$He (right panels) reactions with the NACRE (\citet{nacre}), JINA REACLIB (\citet{cyburt04}) and \citet{descouvemont04} rates. The stripe corresponds to the THM estimated uncertainty. Figure adapted from \citet{tumino14}.}
    \label{fig:tumino}
\end{figure}
Figure~\ref{fig:tumino} shows, for the quoted reactions, the THM rates compared to the ones of still widely used NACRE compilation (\citet{nacre}), of the JINA REACLIB (\citet{cyburt04}) and  to the \citet{descouvemont04} rates. The estimated uncertainty on the analysed burning channels (of about 5\%) are also shown. At temperatures typical of stellar deuterium burning ($\sim 10^6$~K) the D(D,p)$^3$H is about 5\% larger than the NACRE, while it is much larger (about 15\%) than the value reported in \citet{cyburt04}. The differences sensitively reduce at larger temperatures, more important for cosmological calculations. If the \citet{descouvemont04} rate is considered, the difference with THM is very small at stellar temperatures (below 1\%), while it increases at larger temperatures, reaching about 5\% at $T\sim 10^9$~K. 

The new THM rate for the D(D,n)$^3$He reaction is $\sim$10\% larger than the others (NACRE, JINA and \citet{descouvemont04}) for temperatures smaller than about $5\times 10^7$~K. At larger temperatures the differences reduce and for $T\ga 2\times 10^8$~K the THM rate is smaller than the others by about 5-10\%. \citet{tumino14} evaluated the effect of the new rates in stellar models, showing that the change in the cross sections does not produce any effect on the stellar structure. The result was expected because such burning channels are quite negligible in stellar models, where D is mainly destroyed via the p(D,$\gamma$)$^3$He channel. On the contrary, these reactions could be more important for  BBN (\citet{coc10,cyburt04}). \citet{tumino14} estimated that the new reaction rates could result in a variation of the primordial deuterium abundance inferred from the BBN by about $2\%$, while an impact on the $^7$Li abundance up to about 10\% is expected. \\
\\
\\

\subsection{Stellar surface abundance of light elements and updated (p,$\alpha$) reaction rates}

The energy produced in the Li, Be and B nuclear reactions is negligible and  such reactions do not affect stellar structures evolution. However, the surface abundances of light elements strongly depend on the nuclear burning. The different fragility of Li, Be and B against (p,$\alpha$) destruction reactions potentially allows to investigate the characteristics of different depths of the stellar interior.

In Fig.s~\ref{fig:rate6Li7Li} and \ref{fig:rate9Be10B} the reaction rates for the most relevant light element burning (p,$\alpha$) reactions calculated with the THM are shown and compared with the JINA REACLIB and the less recent, but still widely used, NACRE rates. The results are discussed below.
\\
\\
\subsubsection{$^6$Li and $^7$Li surface abundance and (p,$\alpha$) reaction rates efficiency}
\begin{figure}[t]
\centering
\includegraphics[width=0.48\linewidth]{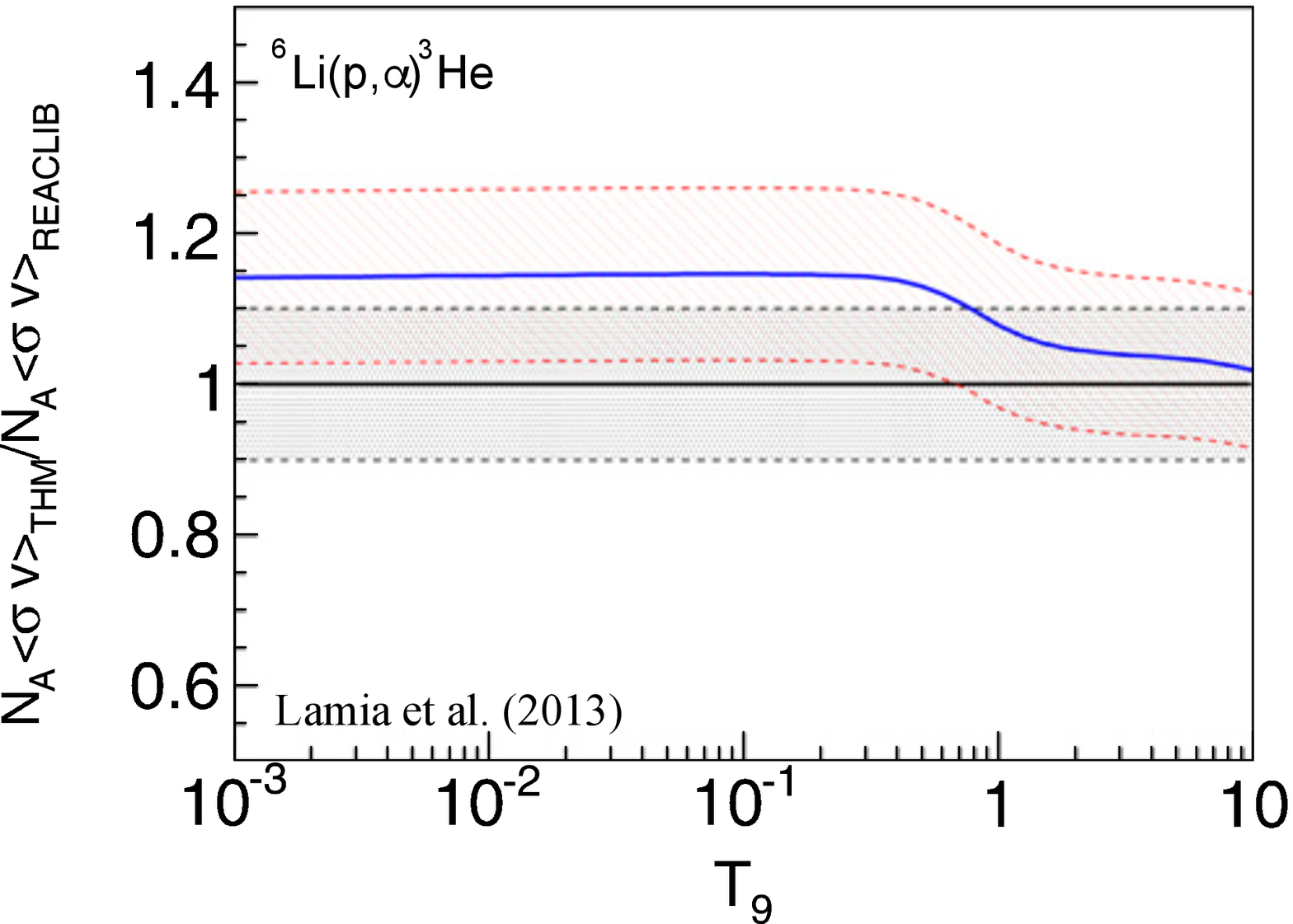}
\includegraphics[width=0.49\linewidth]{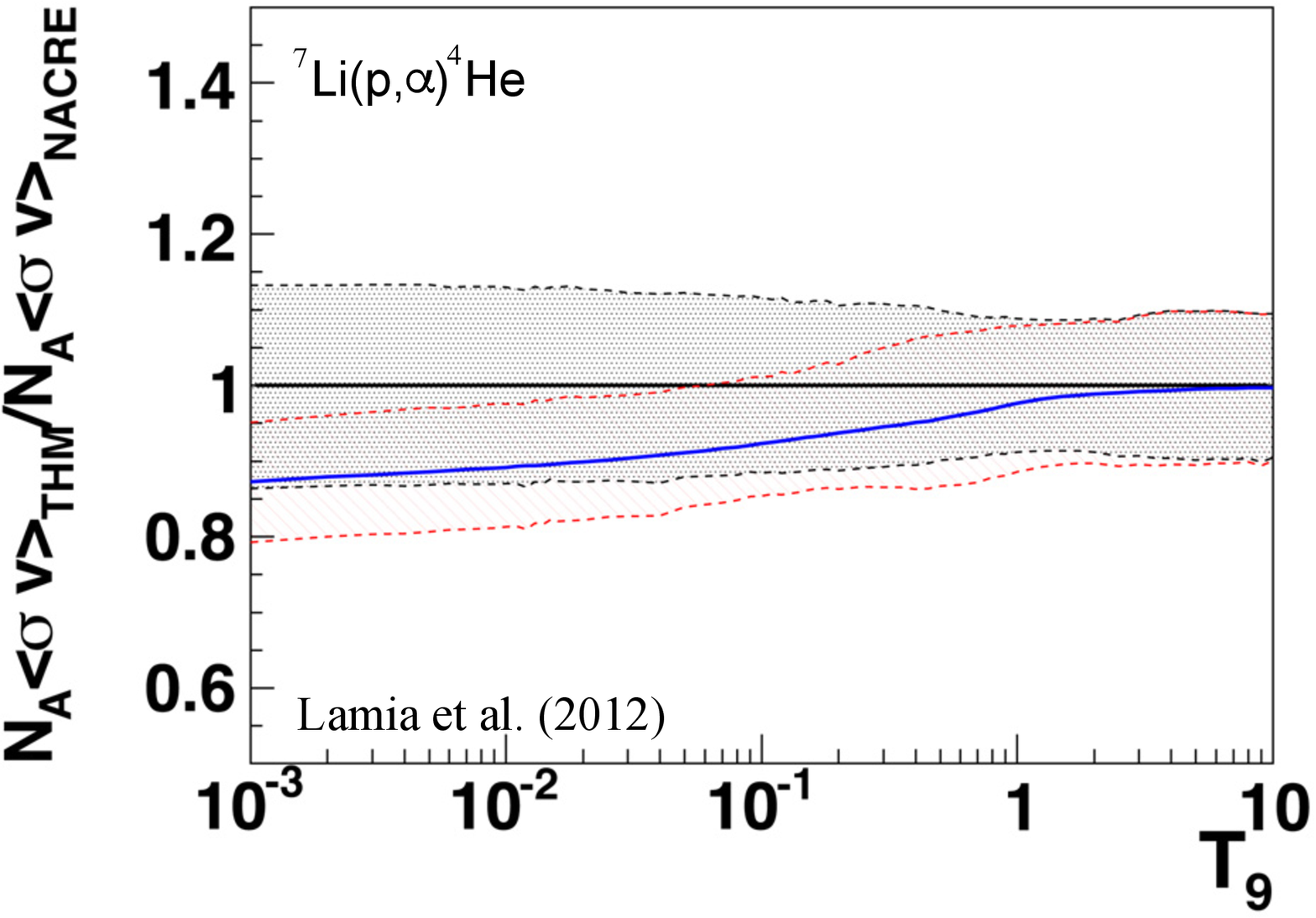}
\caption{Left panel: ratio between the THM $^6$Li(p,$\alpha$)$^3$He reaction rate and the \citep{pizzone05} (pt05) reported in the JINA REACLIB (solid blue line) together with THM upper and lower limits (red dashed line). The black dashed area represents the upper and lower limits of the JINA REACLIB rate assuming the same uncertainties given in the NACRE compilation. $T9$ indicates the temperature in units of 10$^9$ K. Figure adapted from \citet{lamia13}. Right panel: Ratio of the adopted THM $^7$Li(p, $\alpha$)$^4$He reaction rate to that evaluated by the still widely used NACRE compilation (solid blue line). The THM rate upper and lower limits (red dashed lines) are compared with the upper and lower values recommended by NACRE (black dashed lines). Figure adapted from \citet{lamia12}.}
\label{fig:rate6Li7Li}
\end{figure}
The first attempts to apply THM (p,$\alpha$) reaction rates to pre-MS lithium surface abundance calculations were performed by \citet{pizzone03} and \citep{pizzone05} (hereafter pt05) and successively updated, after re-normalization to recently released direct data, in \citet{lamia12,lamia13}. The left panel of Fig.~\ref{fig:rate6Li7Li} shows the $^6$Li(p,$\alpha$)$^3$He reaction rate obtained using the THM compared to the pt05 rate available on the JINA REACLIB page. The THM estimated rate deviates from the pt05 by about $15$\% at a temperature of T$\approx~10^6$~K, typical of $^6$Li burning in the pre-MS phase, a value that is larger than the current estimated uncertainty (about 10\%) on the rate itself. 

\begin{figure}
\centering
\includegraphics[width=0.48\linewidth]{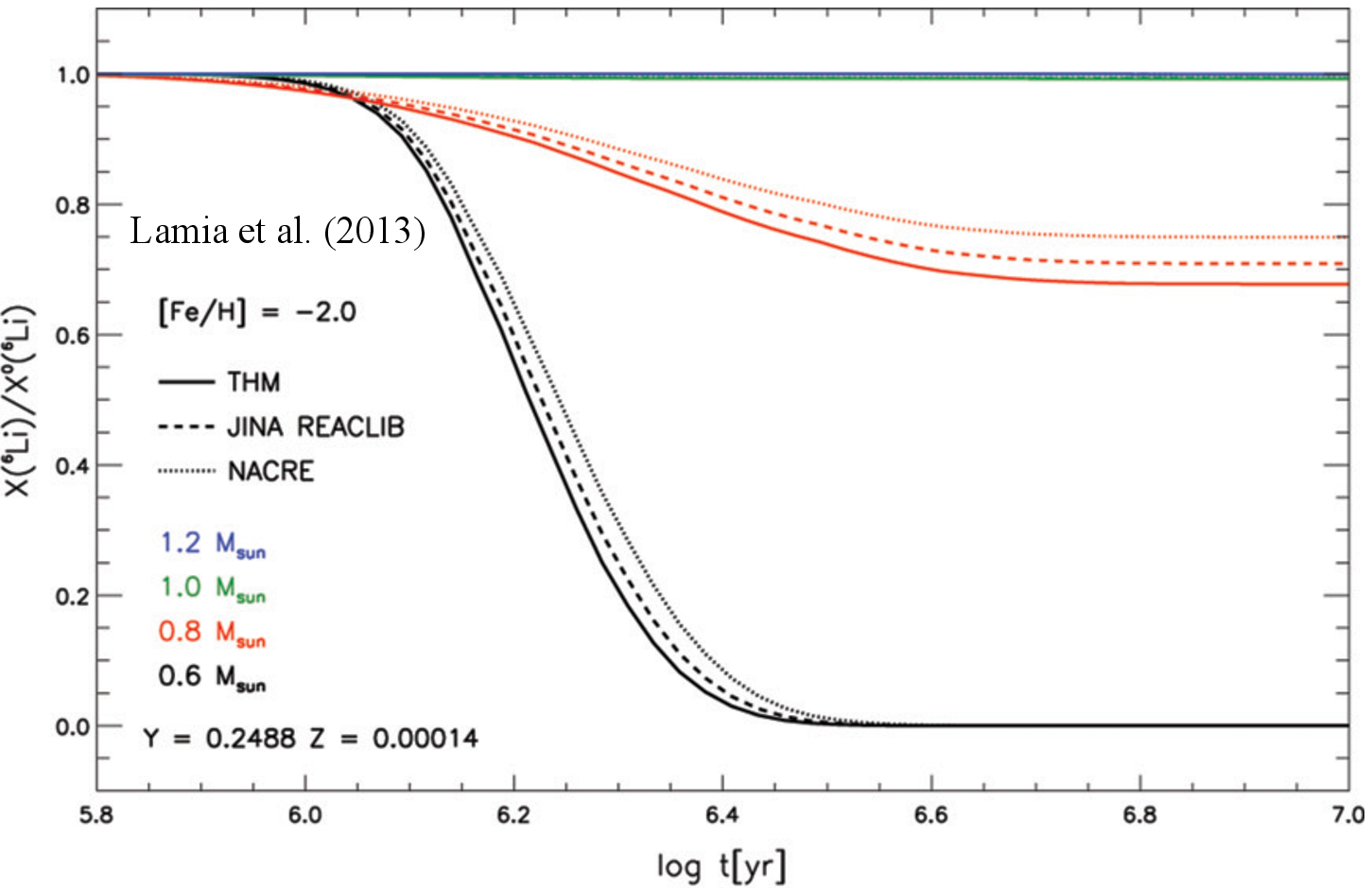}
\includegraphics[width=0.48\linewidth]{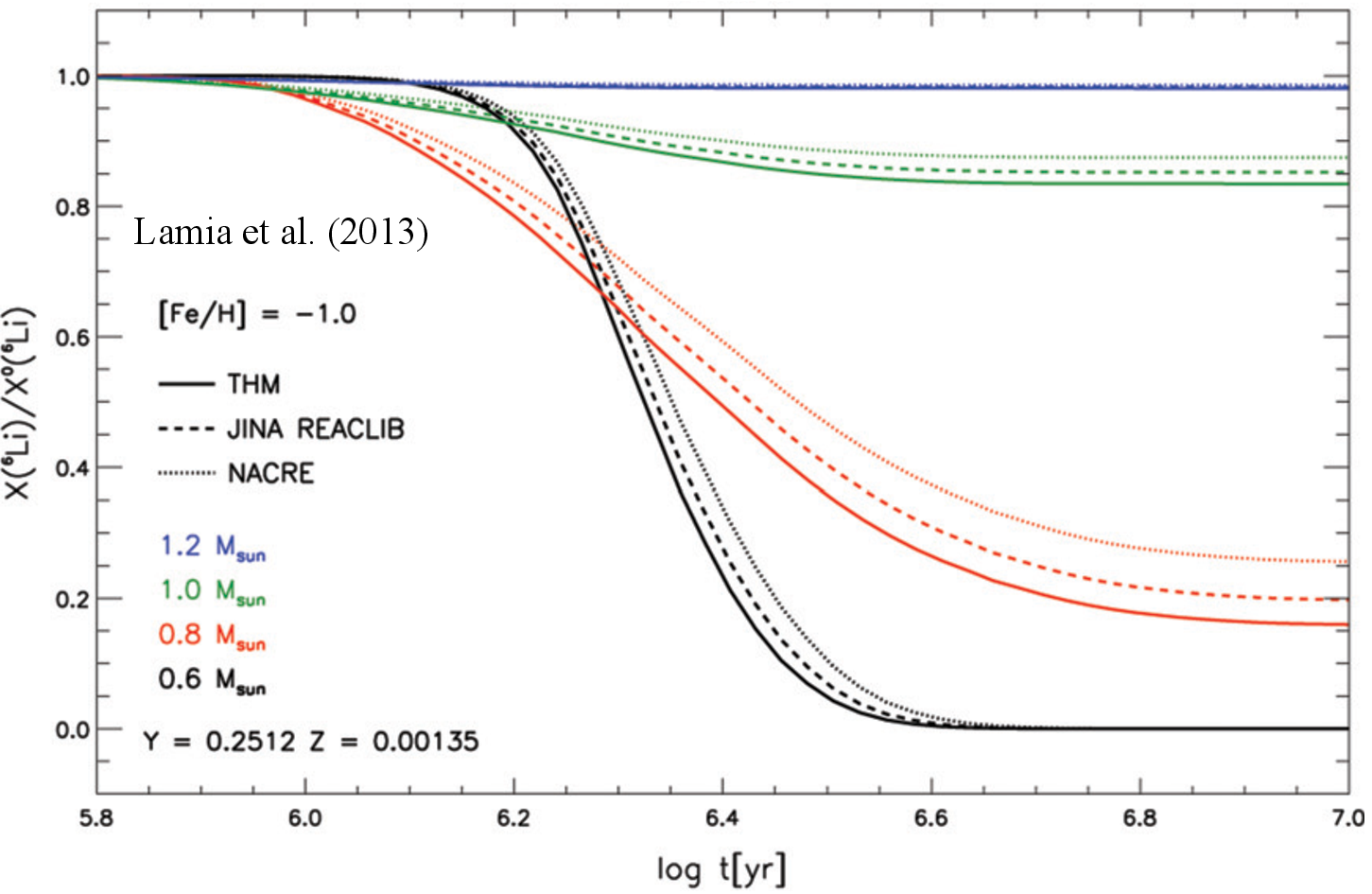}
\includegraphics[width=0.48\linewidth]{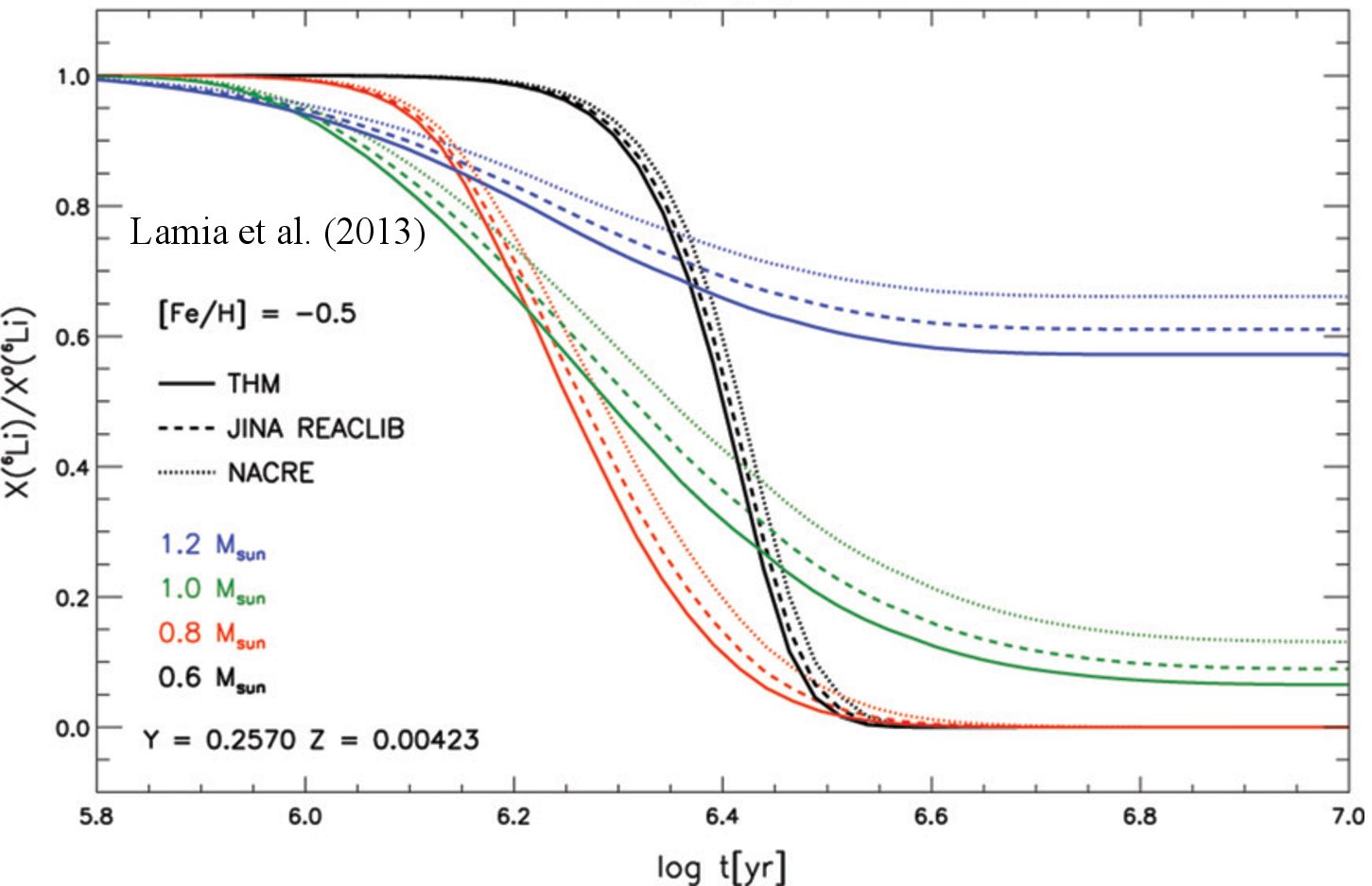}
\caption{Time evolution of the surface $^6$Li abundance, when the three different labelled $^6$Li(p,$\gamma$)$^3$He reaction rates are adopted in the models. Each panel corresponds, as indicated, to a different chemical composition for which four different masses are evolved. See text for details. Figure adapted from \citet{lamia13}.}
\label{fig:6Lievolution}
\end{figure}

\citet{lamia13} evaluated the effect on the surface $^6$Li abundance of the update of the $^6$Li+p reaction rate for a range of stellar masses at three different metallicities ([Fe/H]$=-0.5, -1.0$, and $-2.0$). Figure~\ref{fig:6Lievolution} shows the time evolution of the surface $^6$Li abundance -- normalised to the original value-- obtained adopting the three different $^6$Li(p,$\alpha$)$^3$He reaction rates -- THM, JINA (pt05), and NACRE. From Fig.~\ref{fig:6Lievolution} it is evident that $^6$Li depletion, at fixed burning reaction rate, varies significantly for different masses and metallicities. This can be understood recalling that the higher the metallicity (or the lower  the stellar mass) the deeper and hotter the base of the convective envelope. Note that among the most massive models (i.e., $M=1.2$~\msun), which have the thinnest external convective envelopes, only that with the highest metallicity (i.e., [Fe/H]$=-0.5$) efficiently depletes surface $^6$Li. In the selected [Fe/H] range, the difference in the $^6$Li depletion between the THM and NACRE models ranges from about 13\% (for $M=1.2$~\msun) to about 60\% (for $M=1.0$~\msun). The difference reduces if JINA rate is adopted, as expected due to the smaller differences between the two rates.

\begin{figure}
\centering
\includegraphics[width=0.48\linewidth]{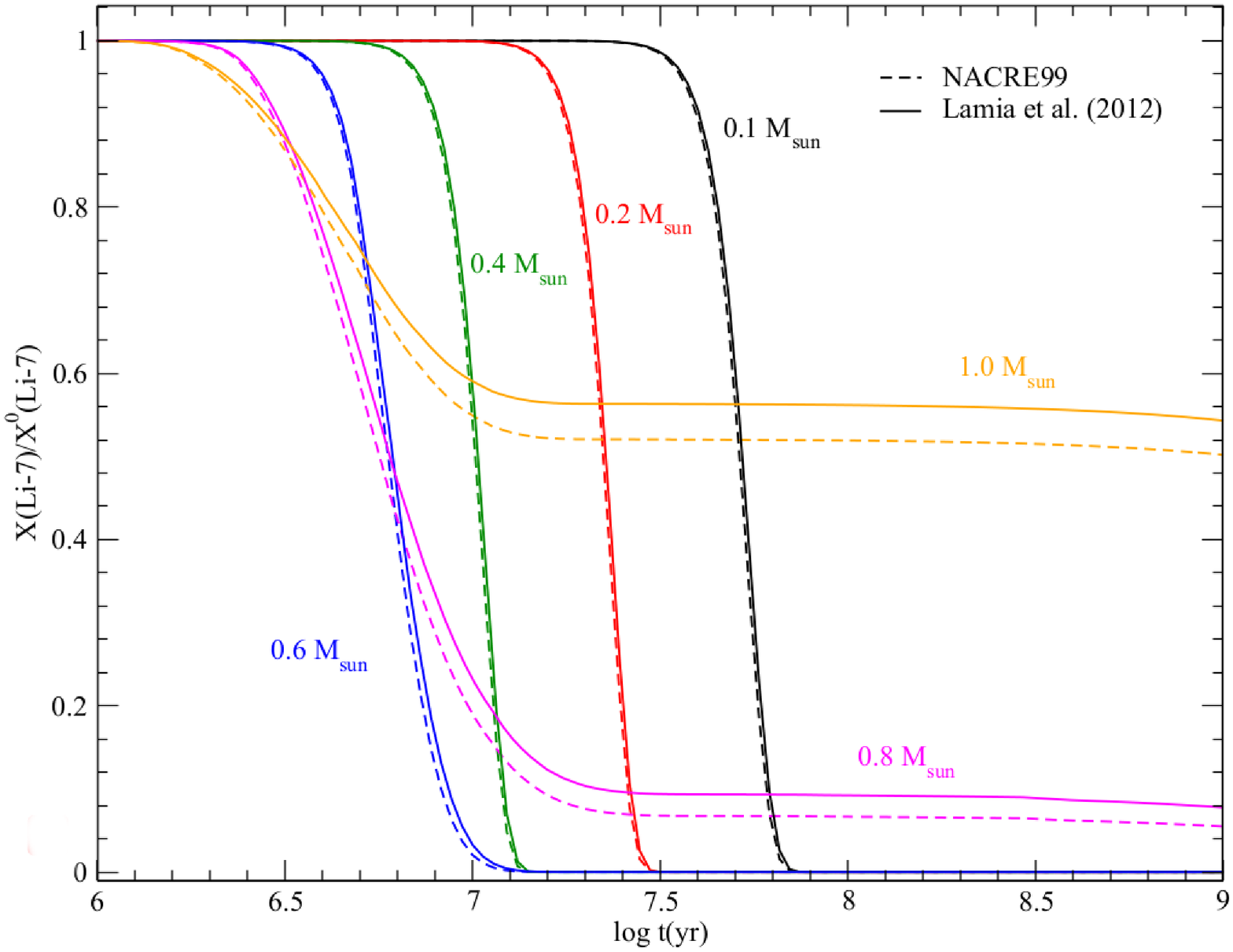}
\caption{Time evolution of surface $^7$Li abundance for PISA models with solar chemical composition and the labelled masses computed with THM, and NACRE reaction rates (figure adapted from \citet{nuclei15}).}
\label{fig:7Lievolution}
\end{figure}

Right panel of Fig.~\ref{fig:rate6Li7Li} shows the comparison between the THM and NACRE $^7$Li$+$p reaction rate; the difference is of about 13\% at a temperature of T$\approx~10^6$~K, not much larger than the current uncertainty on the rate (about 10\%). Figure~\ref{fig:7Lievolution} shows the time evolution of the surface $^7$Li abundance for different masses when THM and NACRE $^7$Li$+$p reaction rates are adopted. The differences between the two calculations range from about 7\% (for $M=1.0~$\msun) to about 25\% (for $M=0.6$~\msun).

In general, the effect of adopting different $^6$Li and $^7$Li burning reaction rates, although not negligible, is less important than the effects due to errors in other quantities used in the computation of a stellar model, such as the original chemical composition, external convection efficiency, or the uncertainties in some other physical inputs adopted in the calculations (e.g., opacity and equation of state, see Sec.~\ref{Liuncertainties} and the discussions in \citet{pizzone10} and \citet{tognelli12}). Thus, at the moment an uncertainty on the burning reaction rate of the order of $10$\% is not the main error source in the determination of the surface lithium abundance in stellar models (\citet{lamia12,lamia13}). \\
\\

\subsubsection{Lithium Depletion Boundary}
\label{LDB}
\begin{figure*}
\centering
\includegraphics[width=0.60\columnwidth]{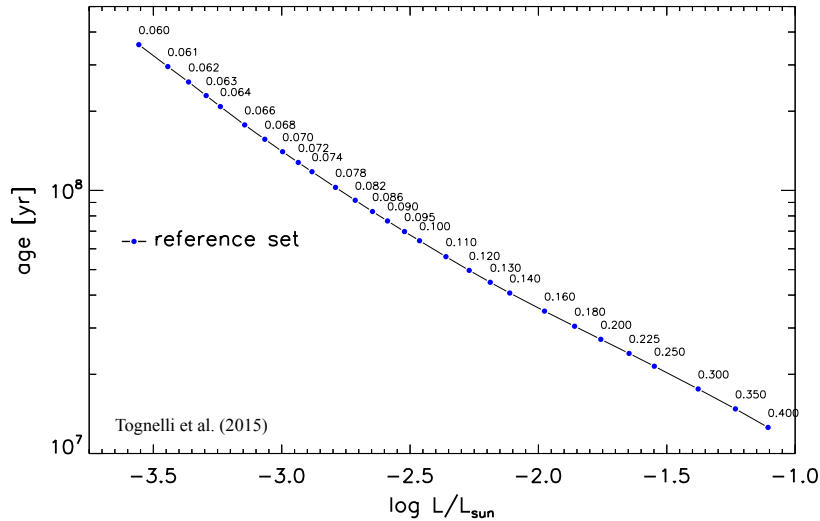}
\caption{Age vs luminosity relation for stars at the Lithium Depletion Boundary for solar chemical composition and \ml=1. The labels along the sequence indicate the stellar mass in \msun. Figure adapted from \citet{tognelli15b}.}
\label{fig:ldb}
\end{figure*}

In the mass range $M\approx$0.06-0.4~\msun{} (the exact values depending on the chemical composition), $^7$Li is completely destroyed in pre-MS in fully convective structures. The larger is the mass the higher is the temperature inside the star, and consequently the earlier is the onset of lithium burning. The larger temperature in more massive stars produces also a more efficient lithium burning, and consequently the age at which lithium is fully depleted in such convective stars strongly depends on the stellar mass. In a coeval population of stars with ages between about 15 and 350~Myr, one would expect to observe a sharp transition in very low-mass regime between stars with and without surface lithium at a given mass (corresponding to the higher stellar mass that, at the cluster age, fully destroy lithium). Such a transition, usually called the Lithium Depletion Boundary (LDB), is a powerful age indicator (see e.g. \citet{dantona94}) thanks to the connection between luminosity, mass and age. Figure~\ref{fig:ldb} shows an example of the age vs luminosity relation for stars located at the LDB in the mass range [0.06 $\div$ 0.4]~\msun.

\begin{figure*}
\centering
\includegraphics[width=0.60\columnwidth]{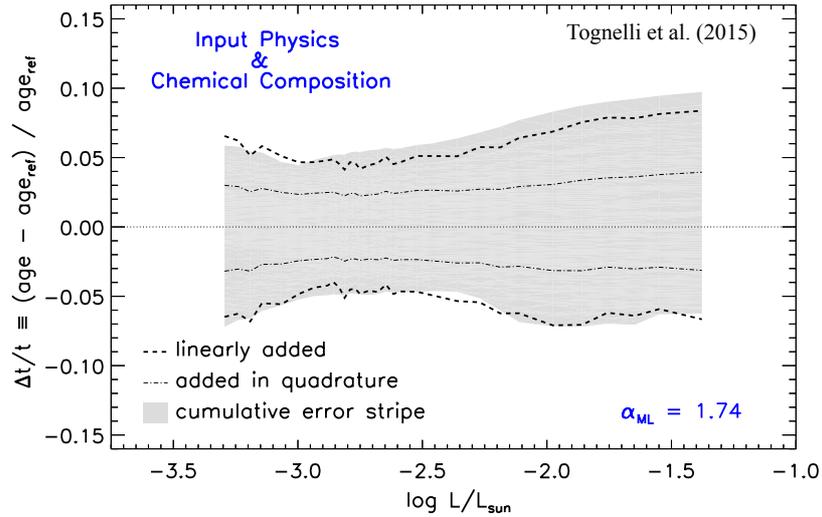}
\caption{Relative uncertainty on the LDB age as a function of the luminosity of the star at the LDB. The shaded area represents the cumulative error stripe (see text). The lines are estimations of the uncertainty obtained by a linear or quadratic combination of the individual effect of the variation of the input physics and chemical composition. Figure adapted from \citet{tognelli15b}.}
\label{fig:ldb_err}
\end{figure*}

The method of LDB has been successfully adopted to assign ages to young clusters as a competitive method to the use of isochrone fitting (e.g. \citet{barrado99,oliveira03,jeffries05,manzi08,dobbie10,jeffries13,binks14,juarez14,dahm15,dupuy16,martin18,martin20} and references therein). The uncertainties on age determination through the LDB technique have been analysed in \citet{burke04} and, more recently, in \citet{tognelli15b}; the main uncertainties that potentially affect the LDB age determination are those already discussed in Section \ref{Liuncertainties}. Figure~\ref{fig:ldb_err} shows the relative age uncertainty on LDB age determination obtained by \citet{tognelli15b} taking into account the errors on the adopted input physics and chemical composition. The shaded area has been obtained by a cumulative variation of all the considered input physics and chemical abundances within their uncertainties  (going into more detail would require a too long discussion out of the purposes of this review, see the quoted paper for additional information). The uncertainty of the method depends on the stellar luminosity (or mass) at the LDB, which, in turn, translates in an age, but it is in any case lower than about 10\%. As a general comment, faint stars that correspond to LDB ages of the order of 50-60 Myr ($0.06\la M$/\msun$\la 0.1$) have errors of the order of about 5\%, an uncertainty that increases increasing the stellar luminosity at the LDB and thus the derived age. \citet{tognelli15b} showed that a large part of the uncertainty on the LDB ($\sim40\%$) comes from the chemical composition indetermination while the lithium burning rate causes a variation of the LDB age of about 1\%.
\\
\\

\subsubsection{$^9$Be and $^{10}$B surface abundance and (p,$\alpha$) reaction rates efficiency}
\begin{figure}[t]
\centering
\includegraphics[width=0.48\linewidth]{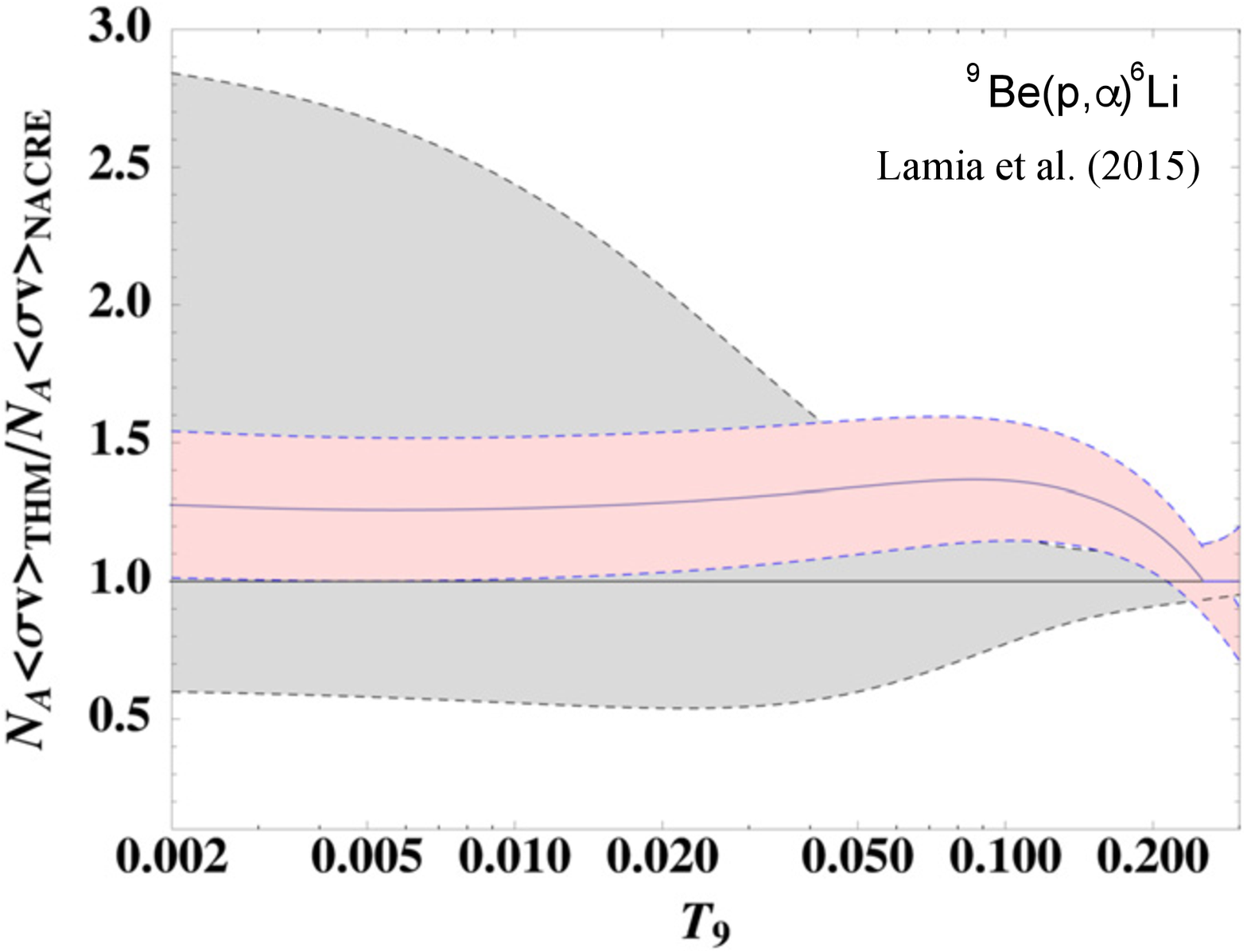}
\includegraphics[width=0.48\linewidth]{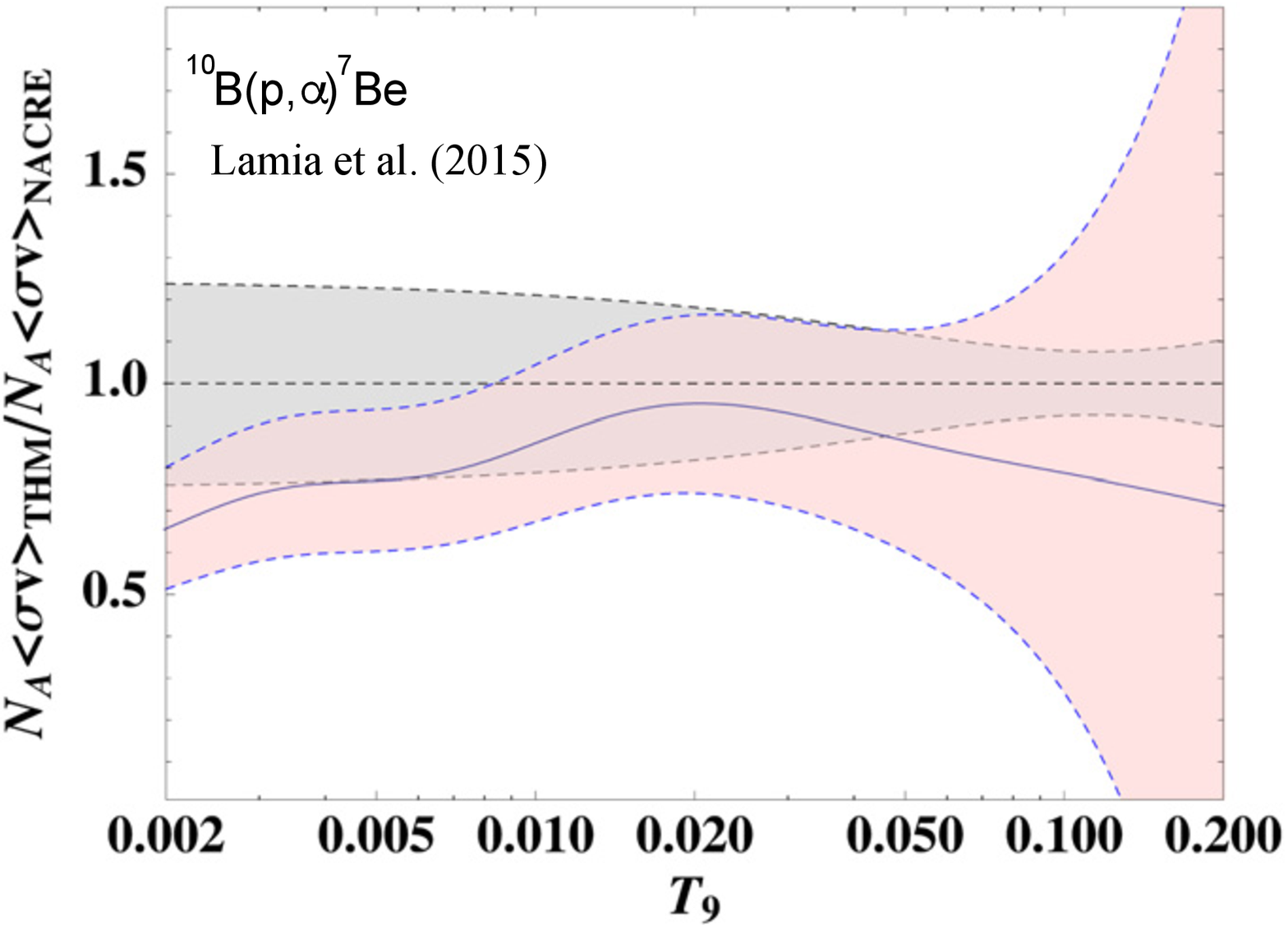}
\caption{Left panel: ratio between the $^9$Be(p,$\alpha$)$^6$Li THM reaction rate and the one reported in the NACRE compilation. The red filled area marks the THM reaction rate
uncertainty compared to the NACRE one (grey region). Right panel: ratio between the $^{10}$B(p,$\alpha$)$^7$Be THM reaction rate and that obtained by the NACRE compilation (blue line). The red filled area marks the THM reaction rate uncertainty compared to the NACRE one (grey region). Figure adapted from \citet{lamia15}.} 
\label{fig:rate9Be10B}
\end{figure}

\begin{figure}
\centering
\includegraphics[width=0.48\linewidth]{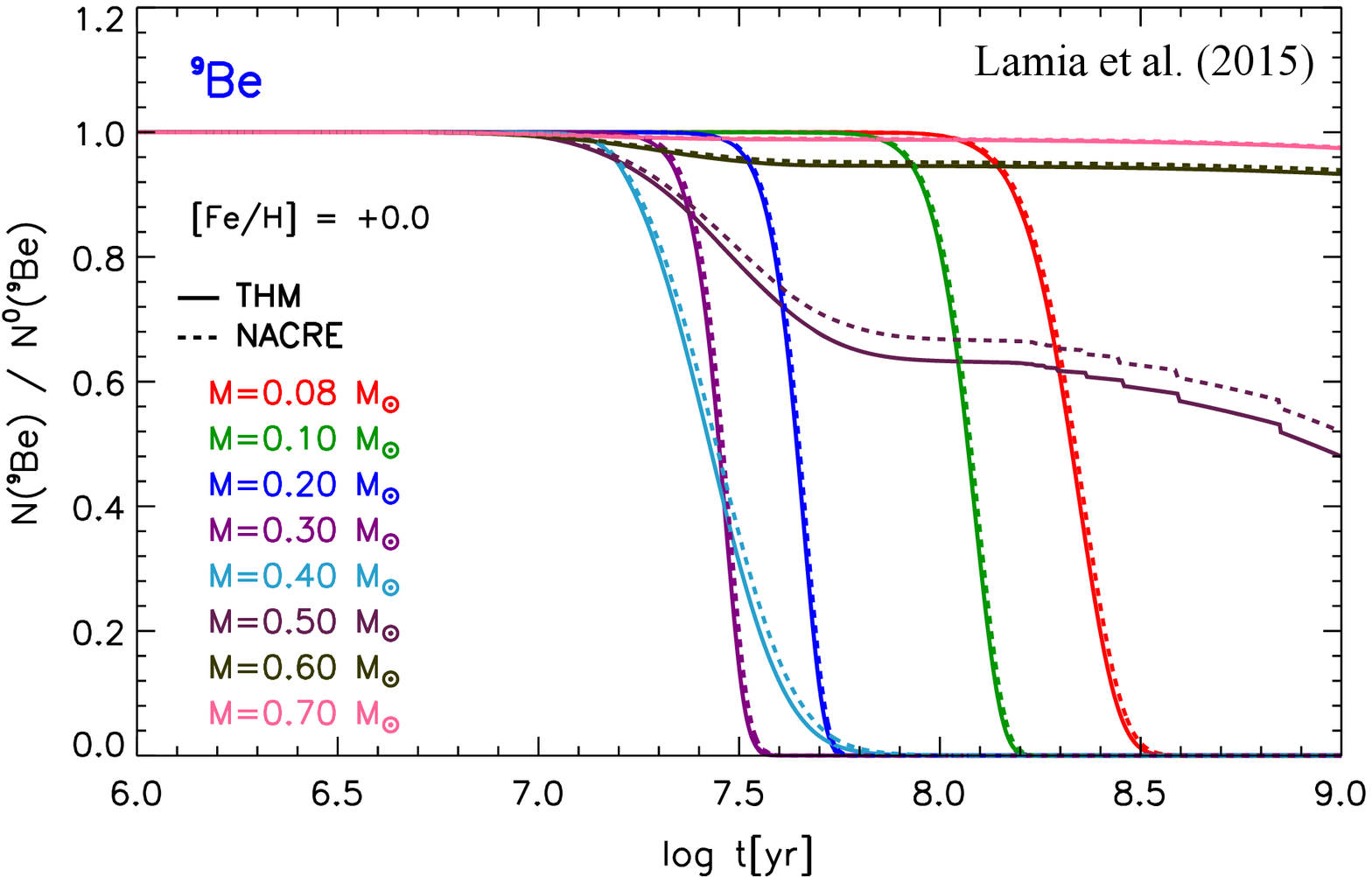}
\includegraphics[width=0.48\linewidth]{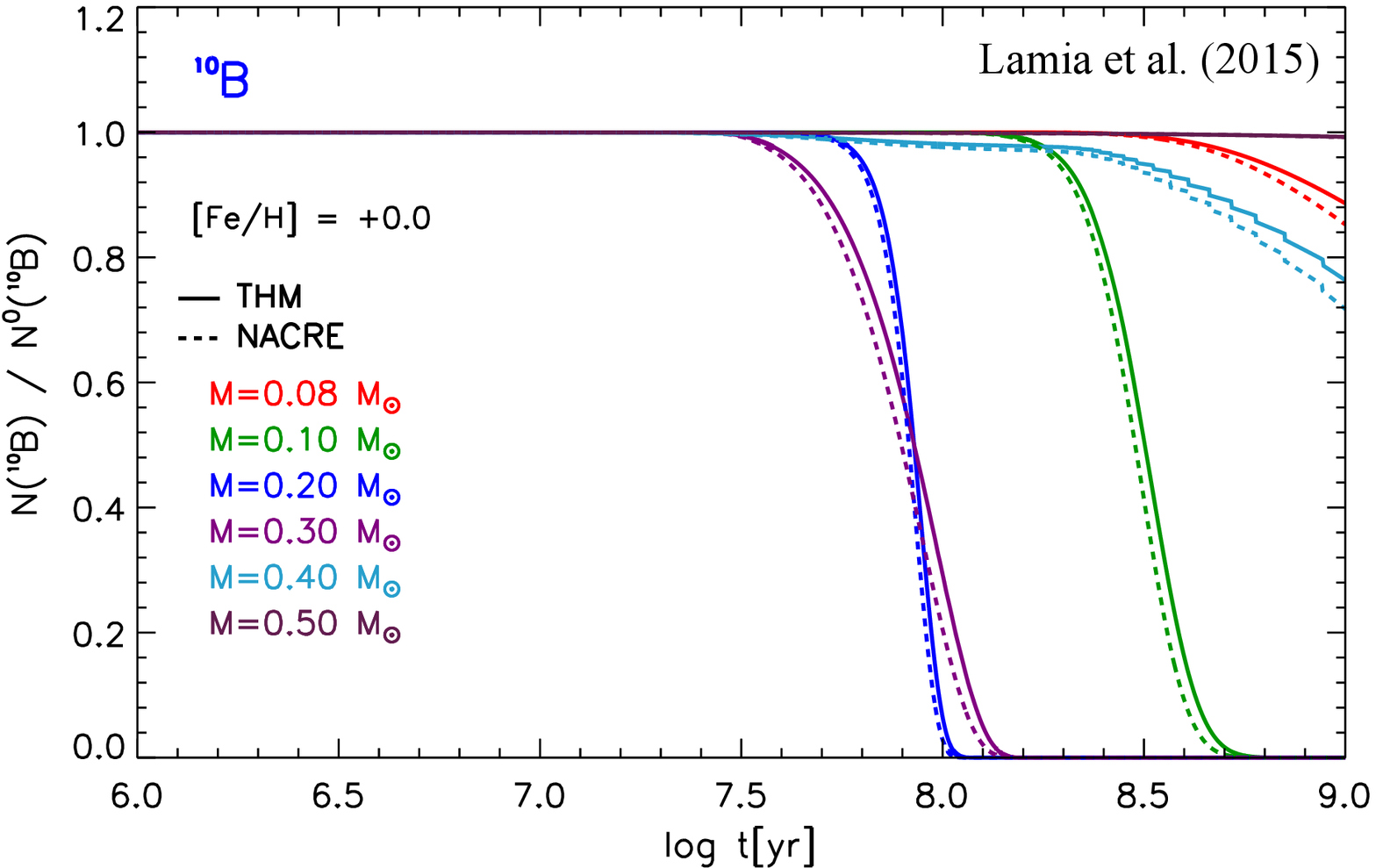}
\caption{Temporal evolution of surface $^9$Be (left panel) and $^{10}$B (right
panel) abundances (normalised to one) for solar chemical composition models with the labelled stellar mass. Models computed using the THM (thick solid line) and the NACRE (dashed line) reaction rates are shown. Figure adapted from \citet{lamia15}.}
\label{fig:9Be10Bevolution}
\end{figure}

Figure~\ref{fig:rate9Be10B} shows the comparison between the recent THM reaction rates and other reaction rates used in the literature. The THM $^9$Be(p,$\alpha$)$^6$Li rate (at temperatures of few million degrees) is about 25\% larger than NACRE and the uncertainty on the THM $^9$Be burning rate is about 25\% (blue dashed area in the figure). The THM reaction rate is quite similar to that in the recent NACRE II \citet{xu13}, even if the THM uncertainty region is sensibly smaller than that of the NACRE II rate, see \citet{lamia15}. 
Left panel of Fig.~\ref{fig:9Be10Bevolution} shows the comparison between the predicted surface Be abundances computed using the THM reaction rate and the NACRE reaction rate, for solar metallicity stars. The higher THM rate leads to a faster $^9$Be destruction consequently, at the same age, models with the THM rate show a lower $^9$Be surface abundance with respect to models with the NACRE rate. The differences in the predicted surface abundances are significant in those models where $^9$Be is efficiently destroyed (i.e. for M~$\la 0.5$~\msun). 

We recall that in stars, $^9$Be is destroyed through two channels: $^9$Be(p,$\alpha$)$^6$Li (R$_1$, the rate analysed by \citet{lamia15}) and $^9$Be(p, 2$\alpha$)$^2$H (R$_2$). The ratio R$_1$/R$_2\approx 1.2$, for stellar conditions at the temperature of interest; thus the R$_2$ contribution to beryllium destruction is not negligible. Changing only R$_1$, as done in \citet{lamia15} affects the final beryllium abundance by a factor that is given by the R$_1$ reaction rate change (about 25\%) multiplied by the probability that the $^9$Be burning occurs in that channel, i.e., 25\%~$\times$~R$_1$/(R$_1$ + R$_2$ )~$\approx$~14\%, thus leading to a change in the predicted Beryllium abundance lower than what expected if only R$_1$ channel was active in stars. The effect of the THM uncertainty on the $^9$Be(p,$\alpha$)$^6$Li rate -- approximately 25\% which is equal to the difference with respect to NACRE rate -- is expected to produce a variation of the predicted $^9$Be surface abundance of the same order of magnitude discussed above.

\begin{figure}
\centering
\includegraphics[width=0.96\linewidth]{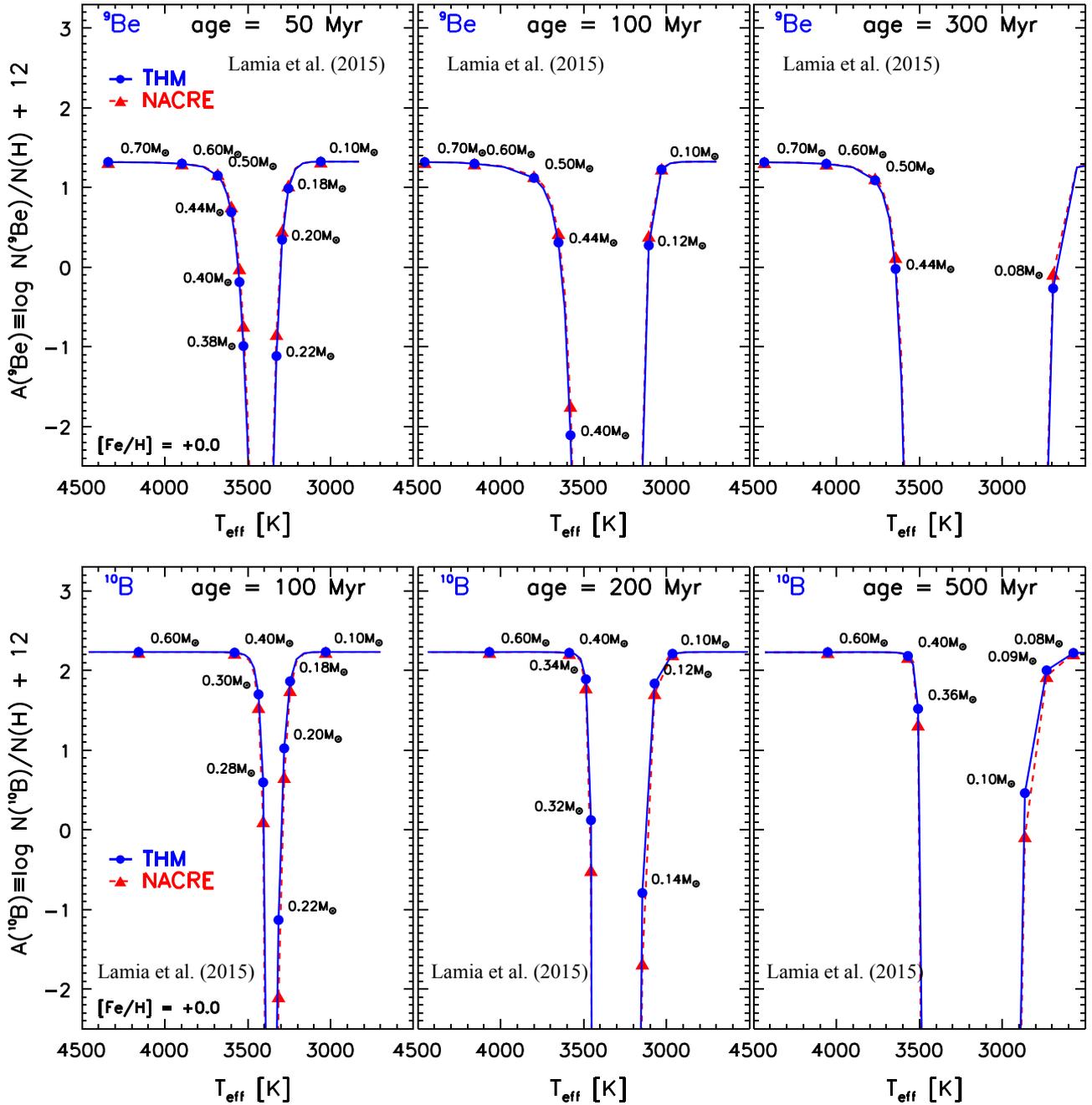}
\caption{Surface logarithmic abundances of $^9$Be (upper panel) and $^{10}$ B (lower panel) as a function of the star effective temperature for different stellar masses in the range [0.06, 0.8]~\msun{} and for three ages typical of young/intermediate open clusters. Models are calculated for [Fe/H]$ = +0.0$ adopting both the THM (solid blue line) and the NACRE (dashed red line) reaction rates. Figure adapted from \citet{lamia15}.}
\label{fig:ATeff}
\end{figure}

Right panel of Fig.~\ref{fig:rate9Be10B} shows the comparison between the THM and the NACRE rate for the $^{10}$B(p,$\alpha$)$^7$Be reaction: at temperatures of a few millions of kelvin, the THM rate is $\sim$30\% lower than the NACRE one. The error of the THM rate at the temperatures of interest is about 20\%. However if the THM rate is compared with the one of the updated NACRE II Compilation \citet{xu13}, the differences are significantly reduced, see \citet{lamia15}. 

The effect of the different $^{10}$B(p,$\alpha$)$^7$Be reaction rates on surface B abundance in low mass stars (at solar metallicity) is shown in the right panel of Fig.~\ref{fig:9Be10Bevolution}. The lower THM $^{10}$B(p,$\alpha$)$^7$Be cross section leads to a smaller $^{10}$B destruction and thus to a larger surface $^{10}$B abundance at a fixed age. Due to the higher $^{10}$B burning temperature with respect to $^9$Be, the effect of reaction rate change is significant only for masses  $M~\la 0.4$~\msun. Also notice that the typical timescale at a fixed mass where $^{10}$B is destroyed is longer than that of $^9$Be. 

For completeness in Fig.~\ref{fig:9Be10Bevolution} we point out for ages typical of MS  evolution (e.g. log~t$\ga 8.5$ for $M=0.5$~\msun ~for the $^9$Be abundance behaviour and for $M=0.4$~\msun ~for $^{10}$B) the effect of microscopic diffusion, which leads to the settling of light elements toward the stellar interior.

Figure~\ref{fig:ATeff} shows the surface logarithmic abundances of $^9$Be and $^{10}$B as a function of the effective temperature, \teff. Although observational $^9$Be and $^{10}$B abundances are still not available for the low temperature/mass regimes typical of efficient $^9$Be and/or $^{10}$B burning, it is worth to estimate the role of the improvements in nuclear physics in surface abundance predictions. The models are computed using the THM and NACRE reaction rates discussed above; we remind that \teff~ is not affected by the change of the (p,$\alpha$) rates. Each curve represents the abundance isochrone, i.e., the locus of models with the same age but different masses. For those models where $^9$Be ($^{10}$B) burns the differences between the adoption of the NACRE and THM reaction rates can be as large as about 0.2-0.3~dex for $^9$Be and almost 1 dex for $^{10}$B. 

To our knowledge, an analysis of the dependence of surface $^9$Be or $^{10}$B abundances on the errors in input physics and chemical composition adopted in the calculations is not available in the literature. However, it is a good approximation assuming a Be and B burning sensitivity to  the input physics similar to that obtained for $^7$Li, as the burning temperatures are not much different. Under this assumption, the effects of the uncertainty on (p,$\alpha$) Be and B burning reaction rates is not the dominant error source for the surface abundance predictions of such elements. \\
\\

\section{Summary and conclusions}

Surface light elements abundances prediction in stellar models is a difficult task because it is affected by several errors in the adopted input physics and uncertainties in the efficiency of some physical mechanisms as e.g. convection in the stellar envelope. Moreover, pre-MS characteristics and surface light element abundances depend on the previous protostellar phase,  which is the phase when the star forms. Analysis of the effects of different choices of accretion parameters (accretion rate, radius and mass of the first stable hydrostatic protostar, accretion history, accretion geometry, the amount of energy transferred from the accreted matter to the accreting star, etc..) on the subsequent pre-MS evolution have been performed in the literature. The results show that if the accretion process leads to bright and extended structures at the end of the protostellar phase the stellar characteristics (including the surface light element abundances) are very similar to those of standard (non-accreting) pre-MS models with the same final mass. The structure of a pre-MS star at the end of the accretion phase is affected by the inclusion of the protostellar accretion only for a restricted range of accretion parameters, mainly in the so called "cold accretion scenario". In these cases a significant reduction of the surface light element abundances during the protostellar phase (in contrast to standard models) has been obtained; however the position of the stars in the HR diagram is in disagreement with observations for disk stars, rising doubts about the validity of the adopted accretion parameters.

Protostellar accretion in low mass halo stars was suggested in the literature as one of the possible solutions for the cosmological lithium problem. However, theoretical calculations show that the reproduction of the Spite Plateau would require a fine tuning of the parameters that govern the protostellar phase and, more important, the models with the required Li depletion follow a pre-MS evolution in the HR diagram which is quite different to the one observed for high metallicity pre-MS stars. Comparison between theoretical predictions and observations for surface lithium abundance in young open clusters still show discrepancies. During the pre-MS phase surface Li abundance is strongly influenced by the nuclear burning as well as by the extension towards the interior of the convective envelope and by the temperature at its bottom. These last two quantities depend on the input physics adopted in the calculations (radiative opacity, atmospheric models etc..), on the assumed stellar chemical composition and on the convection efficiency in superadiabatic regions, whose precise physical treatment is not still fully available. 

Comparison between predictions and observations for pre-MS stars in open clusters suggest a less efficient convection during the pre-MS phase with respect to the Main Sequence. This is true even if one takes into account the uncertainties on the results due to the errors in the adopted input physics and assumed chemical composition. A possible explanation of this result could be the fact that in 1D evolutionary codes a reduced convection efficiency could mimic the main effects of some non-standard mechanisms active in young stars, such as the presence of a not negligible magnetic field and/or surface spot coverage. 

The energy produced by the Li, Be, B burning reactions is negligible, thus their effects on stellar structures are irrelevant. However, the surface abundances of light elements strongly depends on the nuclear burning and thus on the adopted reaction rates. The only nuclear burning that during the pre-MS or protostellar accretion phase affects stellar evolution is the deuterium burning. The impact on pre-MS evolution of a variation of the p(D,$\gamma$)$^3$He reaction rate by its present uncertainty ($\pm$ 7\%) has been analysed in the literature, finding a negligible effect on stellar models. Two other D-burning channels have been considered, namely the D(D,p)$^3$H and the D(D,n)$^3$He. However, as expected, the inclusion of such channels does not produce relevant effects on pre-MS evolution as the largest part of deuterium is destroyed via p(D,$\gamma$)$^3$He.

The effects on the other light elements surface abundance predictions of the still present (even if greatly reduced) uncertainty on (p,$\alpha$) cross sections have been evaluated in detail and compared to the influence on the results of the errors in the other physics ingredients and in the stellar chemical composition.  Light element (p,$\alpha$) reaction rates have been recently revised through the indirect Trojan Horse method (THM), sensibly reducing their estimated uncertainty and finding differences with previous estimates at the energies of astrophysical interest. In general, differences in the predicted surface Li, Be, B abundances if the THM or the less recent but still widely used NACRE reaction rates are adopted are significant for stars in which light elements are efficiently burned. 

The current uncertainty on the $^6$Li and $^7$Li proton capture reaction rates is of the order of 10\%. Numerical calculations show that the effects on the $^6$Li and $^7$Li surface abundances due to this uncertainty, although not negligible, are less important than the influence of errors in other quantities used in the computation of a stellar model. 
The present errors on the $^9$Be(p,$\alpha$)$^6$Li and  $^{10}$B(p,$\alpha$)$^7$Be rates, at the temperatures of interest, are, respectively, of about 25\% and 20\%.  Due to the higher $^9$Be and $^{10}$B burning temperature, with respect to the Li burning, the effects of the reaction rate change/uncertainty are significant only for masses lower than M~$\la 0.5$~\msun~and M~$\la 0.4$~\msun, respectively. 

In conclusion, recent cross section measurements for light element (p,$\alpha$) burning reactions sensibly reduced their uncertainty, even if it is still not negligible. Pre-Main Sequence theoretical calculations and consequently prediction for light element surface abundances are affected by several uncertainty sources: the not precise knowledge of the protostellar evolution and the efficiency of superadiabatic convection, the still present errors on the input physics adopted in the calculations and in the assumed stellar chemical composition. On the other hand, the errors on light element nuclear cross sections do not constitute the main uncertainty source for the prediction of light elements surface abundances.
\\
\\

\section*{Author Contributions}
All the authors contributed to the writing of the paper. 
\\
\\

\section*{Funding}
This work has been partially supported by INFN (Iniziativa specifica TAsP) and the Pisa University project PRA 2018/2019 "Le stelle come laboratori cosmici di fisica fondamentale". 

\section*{Acknowledgements}
A.C. acknowledges the support of the "visiting fellow" program (BVF 2019) of the Pisa University. 
E.T. acknowledges INAF-OAAb for the fellowship "Analisi dell'influenza dell'evoluzione protostellare sull'abbondanza superficiale di elementi leggeri in stelle di piccola massa in fase di pre-sequenza principale". L.L. acknowledges the program "Starting Grant 2020" by University of Catania.  
\\
\\

\bibliographystyle{frontiersinSCNS_ENG_HUMS} 
\bibliography{bib_frontiers}


\end{document}